\documentclass[a4paper,12pt,numbers,sort&compress]{article}

\pdfoutput=1
\usepackage[paper=letterpaper,margin=1in]{geometry}
\usepackage{amsmath,array,amsfonts,graphicx,wrapfig,lscape,float,mathtools,multirow,longtable}
\usepackage[dvipsnames]{xcolor}
\usepackage{hyperref,caption,setspace,todonotes}
\usepackage[makeroom]{cancel}
\usepackage{cite}
\usepackage{tocloft}

\renewcommand{\baselinestretch}{1.2}
\newcommand{\eref}[1]{(\ref{#1})}
\newcommand{\fref}[1]{Figure~\ref{#1}}
\newcommand{\sref}[1]{\S\ref{#1}}
\newcommand{\tref}[1]{Table~\ref{#1}}
\newcommand{\nn}{\nonumber}

\newcommand{\beal}[1]{\begin{eqnarray}\label{#1}}
\newcommand{\eea}{\end{eqnarray}}
\newcommand{\ba}{\begin{array}}
\newcommand{\ea}{\end{array}}

\newcommand{\comment}[1]{}

\newcommand{\setall}{\setcounter{equation}{0}
        \setcounter{theorem}{0}}

\newtheorem{theorem}{\bf Theorem}

\newtheorem{conjecture}[theorem]{\bf Conjecture}

\newtheorem{proposition}[theorem]{Proposition}
\newtheorem{corollary}[theorem]{Corollary}
\newtheorem{definition}[theorem]{Definition}

\newcommand{\qed}{\nobreak \ifvmode \relax \else
      \ifdim\lastskip<1.5em \hskip-\lastskip
      \hskip1.5em plus0em minus0.5em \fi \nobreak
      \vrule height0.75em width0.5em depth0.25em\fi}

\begin{document}

\begin{flushright}
UNIST-MTH-24-RS-05
\end{flushright}

\vspace{1cm}

\begin{center}
{\Large \bf  Futaki Invariants and Reflexive Polygons}
\end{center}
\medskip

\vspace{.4cm}
\centerline{
{\large Jiakang Bao}$^1$, \
{\large Eugene Choi}$^2$, \
{\large Yang-Hui He}$^{3,4}$, 
}

\centerline{
{\large Rak-Kyeong Seong}$^{2}$, \
{\large Shing-Tung Yau}$^{5}$
}

\renewcommand{\baselinestretch}{0.5}
\begin{center}
{\it
  {\small
    \begin{tabular}{cl}
      ${}^{1}$
      & Kavli Institute for the Physics and Mathematics of the Universe, \\
      & University of Tokyo, Kashiwa, Chiba 277-8583, Japan\\
      ${}^{2}$ 
      & Department of Mathematical Sciences, Ulsan National Institute of Science and Technology, \\
      & 50 UNIST-gil, Ulsan 44919, South Korea\\
      ${}^{3}$ 
      & London Institute for Mathematical Sciences, Royal Institution, London, W1S 4BS, UK\\
      ${}^{4}$ 
      & Merton College, University of Oxford, OX1 4JD, UK\\
      ${}^{5}$
      & Yau Mathematical Sciences Center, Tsinghua University, Beijing 100084, China\\
      &\\
      & \qquad
      {\rm jiakang.bao@ipmu.jp \ xeugenechoi@gmail.com \ hey@maths.ox.ac.uk}\\
      & \qquad
       {\rm seong@unist.ac.kr \ styau@tsinghua.edu.cn}
    \end{tabular}
  }
}
\end{center}

\renewcommand{\baselinestretch}{1.2}

\vspace*{3.0ex}
\centerline{\textbf{Abstract}} \bigskip

Futaki invariants of the classical moduli space of $4d$ $\mathcal{N}=1$ supersymmetric gauge theories determine whether they have a conformal fixed point in the IR. 
We systematically compute the Futaki invariants for a large family of $4d$ $\mathcal{N}=1$ supersymmetric gauge theories coming from D3-branes probing Calabi-Yau 3-fold singularities whose bases are Gorenstein Fano surfaces.
In particular, we focus on the toric case where the Fano surfaces are given by the 16 reflexive convex polygons and the moduli spaces are given by the corresponding toric Calabi-Yau 3-folds.
We study the distribution of and conjecture new bounds on the Futaki invariants with respect to various topological and geometric quantities.
These include the minimum volume of the Sasaki-Einstein base manifolds as well as the Chern and Euler numbers of the toric Fano surfaces.
Even though the moduli spaces for the family of theories studied are known to be K-stable, our work sheds new light on how the topological and geometric quantities restrict the Futaki invariants for a plethora of moduli spaces.
\\

\newpage

\begin{singlespace}
\tableofcontents
\end{singlespace}


\newpage

\section{Introduction}\setall

The chiral ring encapsulates many of the fundamental features of a $4d$ $\mathcal{N}=1$ supersymmetric gauge theory \cite{Seiberg:2002jq}. 
Computing it exactly allows us to study various algebro-geometric and dynamical properties of $4d$ $\mathcal{N}=1$ theories. 
At the heart of the computation lies the counting of gauge invariant operators that carry charges under the global symmetry of the $4d$ $\mathcal{N}=1$ theory and form what is known as the classical moduli space \cite{Benvenuti:2006qr, Feng:2007ur, Butti:2006au, Butti:2007jv, Hanany:2007zz, Gray:2008yu}. 
The coordinate ring of the algebraic variety describing the moduli space is what we refer to as the chiral ring of the $4d$ $\mathcal{N}=1$ theory.

A natural question to ask is if the chiral ring indicates whether the $4d$ $\mathcal{N}=1$ supersymmetric gauge theory flows to some $4d$ superconformal field theory in the IR. 
Recently, initiated by the study in \cite{Collins:2016icw}, the question was intricately linked to a separate question of whether a chiral ring satisfies the conditions for \textit{K-stability}.

Originally, K-stability was introduced in mathematics in order to study certain algebro-geometric properties of varieties \cite{futaki1983obstruction,ding1992kahler,donaldson2002scalar, yau1993open, chen2012kahler, chen2015kahler, chen2015kahler2}. 
Given an algebraic variety, its K-stability can be determined by the computation of \textit{(Donaldson-)Futaki invariants}\footnote{Following the differential geometric definition, the Futaki invariant for some holomorphic vector field on the algebraic variety is a holomorphic invariant since it is a characteristic of the Lie algebra of the vector field and is independent of the choice of the K\"ahler form.} \cite{szekelyhidi2014introduction}. 
In \cite{Collins:2012dh,collins2019sasaki}, K-stability was studied in the context of Fano cone singularities, which are $\mathbb{Q}$-Gorenstein and have log-terminal singularities\footnote{A priori, K-stability also depends on the polarization and Reeb vector field of the algebraic variety. We will make this more precise and explicit in \S\ref{preliminaries}.}. 
It was shown that a Fano cone singularity is K-stable if and only if it admits a Ricci-flat K\"ahler cone metric. 
Moreover, by associating these Fano cone singularities with a family of $4d$ $\mathcal{N}=1$ supersymmetric gauge theories, the work in \cite{Collins:2016icw} conjectured, based on results in \cite{Collins:2012dh,collins2019sasaki}, that if the chiral ring of these $4d$ $\mathcal{N}=1$ theories is K-stable, then the theories are associated to a $4d$ superconformal field theory in the IR.
A physical interpretation of the case when the K-stability conditions are not satisfied by the Fano cone singularities is when the corresponding $4d$ $\mathcal{N}=1$ theories have gauge invariant operators that violate the unitarity bound with their $U(1)_R$ charges, as suggested in \cite{Bao:2020ugf}. 

In the following work, we concentrate on a family of $4d$ $\mathcal{N}=1$ supersymmetric gauge theories that are worldvolume theories of D3-branes probing a Calabi-Yau 3-fold singularity $\mathcal{X}$.
Following the arguments above, we consider these $4d$ $\mathcal{N}=1$ theories to flow to $4d$ superconformal field theories in the IR if their classical moduli spaces, also referred to as mesonic moduli spaces, are K-stable.
Under the AdS/CFT correspondence \cite{Maldacena:1997re, Witten:1998qj, Gubser:1998bc},
the IR superconformal field theories are dual to type IIB string theory on $\text{AdS}_5 \times Y_5$ in the large $N$ limit \cite{Klebanov:1998hh, Morrison:1998cs}, where $Y_5$ is the Sasaki-Einstein base manifold of the Calabi-Yau cone $\mathcal{X}$.  
In order to check K-stability, we need to calculate the Futaki invariants corresponding to the generators of the mesonic moduli space as well as the \textit{Hilbert series} \cite{Feng:2001xr, Feng:2000mi}.

We have a projective variety $X$ over which the toric Calabi-Yau 3-fold $\mathcal{X}$ is a complex cone.
It is realized as an affine variety in $\mathbb{C}^k$,
where the Hilbert series is the generating function for the dimension of the graded pieces of the coordinate ring
$\mathbb{C}[x_1, \dots, x_k] / \langle f_i \rangle$.
Here, $f_i$ are the defining polynomials of $X$.
We also note that the coordinates $x_1,\dots, x_k$ are the gauge invariant generators of the mesonic moduli space with defining relations given by $f_i$. 
The $U(1)_R$ symmetry associated to the Reeb vector field $\zeta$ on the Sasaki-Einstein base $Y_5$ introduces a natural positive grading of the coordinate ring, allowing the Hilbert series to be written in terms of a fugacity $t$ whose positive exponents refer to the $U(1)_R$ charges for the gauge invariant operators of the mesonic moduli space. 

By introducing a test $U(1)$ symmetry $\eta$, the Futaki invariants measure to what extent the mesonic moduli space of a $4d$ $\mathcal{N}=1$ theory can be destabilized along the RG flow.
Under the larger overall symmetry involving both the $U(1)_R$ symmetry given by $\zeta$ and the test symmetry $\eta$, the Hilbert series of the mesonic moduli space becomes perturbed under a new induced grading given by $\zeta + \epsilon \eta$.
The extent of the perturbation is measured by the resulting volume of the base manifold $Y_5$ under the perturbation given by $\zeta + \epsilon \eta$.
We note that the volume of the Sasaki-Einstein base manifold both in the perturbed and non-perturbed cases is obtained through the Laurent expansion of the Hilbert series \cite{Benvenuti:2006qr, Feng:2007ur, Butti:2006au, Butti:2007jv, Hanany:2007zz, Gray:2008yu}.
In the non-perturbed case, the volume of the Sasaki-Einstein manifold is inversely proportional to the central charge of the superconformal field theory via the AdS/CFT correspondence.
In the perturbed case, we identify the volume with Futaki invariants under the $U(1)$ test symmetry.
The process of introducing a perturbation under a test symmetry and the computation of the effect using the perturbed volume of the base manifold $Y_5$ has been interpreted in \cite{Collins:2016icw} as a generalized volume minimization \cite{Martelli:2005tp,Martelli:2006yb,He:2017gam} and a-maximization \cite{Intriligator:2003jj} procedure. 
Indeed, as stated above, the K-stability of the mesonic moduli space of the $4d$ $\mathcal{N}=1$ theory
is equivalent to 
the existence of a Ricci-flat conic metric on $\mathcal{X}$ 
as well as 
the existence of the Sasaki-Einstein metric on $Y_5$.

This fascinating role played by Futaki invariants in determining the K-stability of the mesonic moduli space of $4d$ $\mathcal{N}=1$ theories motivates us in this work to further investigate the connection between Futaki invariants and other geometric and topological features of the mesonic moduli space. 
In fact, a systematic study of the minimized volumes of a large family of Sasaki-Einstein manifolds corresponding to toric Calabi-Yau cones with reflexive polytopes as toric diagrams was conducted in \cite{He:2017gam}.
There, the minimized volumes were compared with topological quantities of the associated toric varieties, including the Chern numbers and the Euler number. 
By doing so, the work in \cite{He:2017gam} observed that the distribution of the volume minima is not at all random and satisfies bounds parameterized by the topological quantities of the associated toric varieties. 

\begin{figure}[H]
    \centering
    \resizebox{0.95\hsize}{!}{\includegraphics[width=13cm]{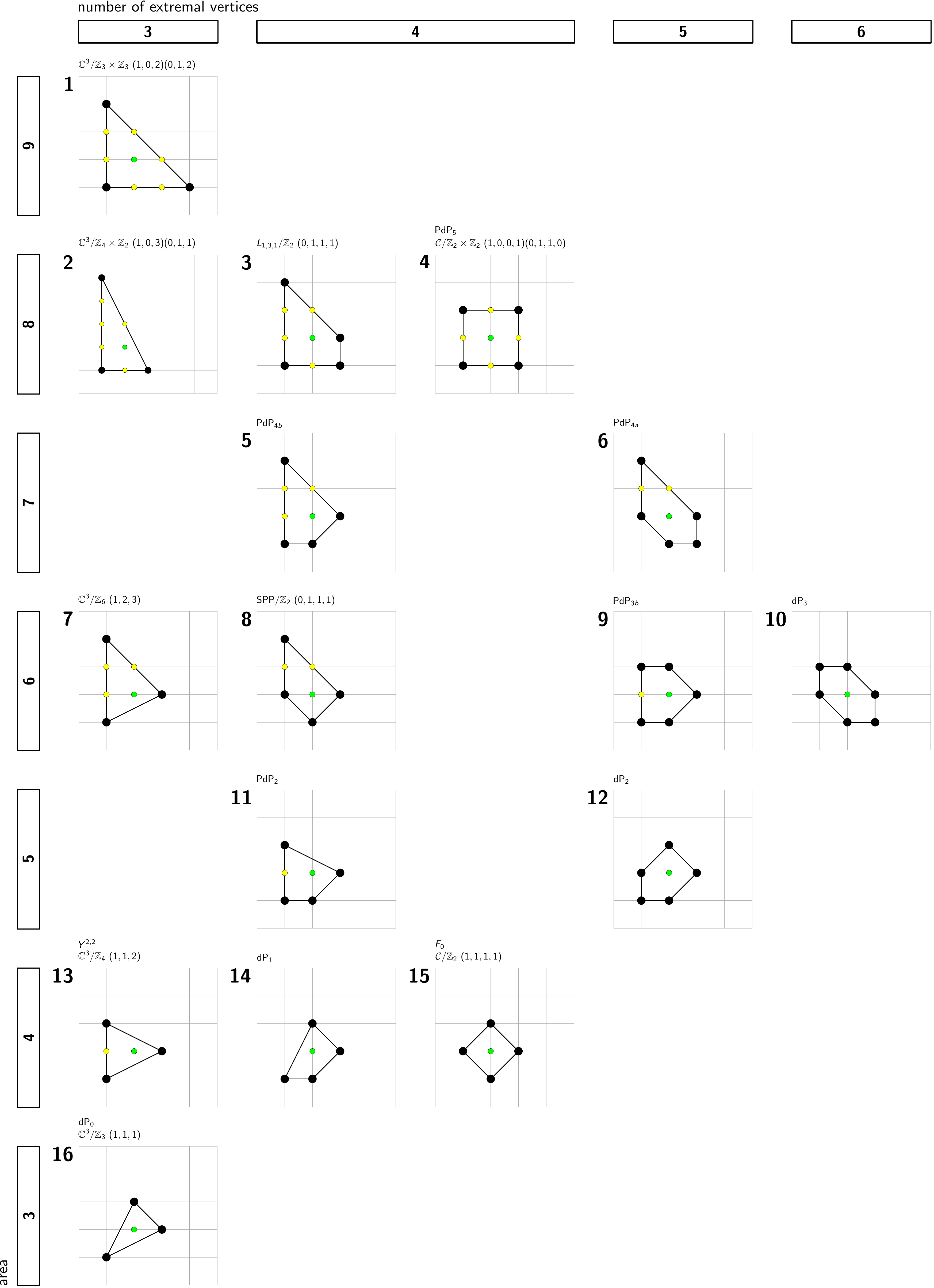}}
    \caption{The 16 reflexive polygons in $\mathbb{Z}^2$. The polygons are arranged in such a way that horizontally we have the number of extremal vertices in the polygons and vertically we have the normalized area of the polygons. Each reflexive polygon is gives rise to a toric Calabi-Yau 3-fold which is associated to at least one $4d$ $\mathcal{N}=1$ supersymmetric gauge theory \cite{Hanany:2012hi}. 
    }\label{fig_reflexive}
\end{figure}

In this work, we focus as in \cite{He:2017gam} on $4d$ $\mathcal{N}=1$ theories that are worldvolume theories of a D3-brane probing toric Calabi-Yau 3-folds, where the toric variety is given by one of the 16 convex reflexive polygons in $\mathbb{Z}^2$ as illustrated in \fref{fig_reflexive}.
This family of $4d$ $\mathcal{N}=1$ theories, fully classified in \cite{Hanany:2012hi}, is part of a wider family of $4d$ $\mathcal{N}=1$ theories realized by a type IIB brane configuration known as a brane tiling \cite{Franco:2005rj, Franco:2005sm, Hanany:2005ve}. 
We note that each of these $4d$ $\mathcal{N}=1$ theories have corresponding mesonic moduli spaces that are $K$-stable.
By concentrating on the values of Futaki invariants themselves, we calculate them for each of the generators of the mesonic moduli space by systematically introducing $U(1)$ test symmetries that are associated to these generators.
By calculating 
the topological invariants of the toric varieties such as Chern numbers and Euler number \cite{fulton1993introduction, cox2011toric}, 
the minimum volume of the associated Sasaki-Einstein 5-manifolds \cite{Martelli:2005tp,Martelli:2006yb,He:2017gam}, 
as well as the integrated curvature invariants \cite{Eager:2010dk}
and the minimum volumes associated to divisors in the toric Calabi-Yau 3-folds \cite{Martelli:2005tp,Martelli:2006yb,Butti:2006au}, 
we make a collection of fascinating observations that relate 
Futaki invariants to fixed topological and geometrical properties of mesonic moduli spaces of $4d$ $\mathcal{N}=1$ theories.
In fact, we show that Futaki invariants obey bounds like the minimum volumes of Sasaki-Einstein manifolds as observed in \cite{He:2017gam} giving us a measure of the rigidity of K-stable mesonic moduli spaces characterized by their geometric and topological properties. 
We expect that our work here on known K-stable mesonic moduli spaces will lead to further insights into more general $4d$ supersymmetric gauge theories and the K-stability of their moduli spaces.

The paper is organized as follows. In section \sref{preliminaries}, we give a quick overview on the relevant concepts that are used in this paper, including toric geometry, the computation of Hilbert series and minimum volumes of Sasaki-Einstein manifolds, and the calculation involved for Futaki invariants. 
In section \sref{Futforref}, we calculate the Futaki invariants for the family of $4d$ $\mathcal{N}=1$ theories associated to toric Calabi-Yau 3-folds whose toric diagrams are reflexive polygons. 
These Futaki invariants are then compared with other geometric and topological quantities of the associated toric Calabi-Yau 3-folds. 
We conclude with section \sref{KunstableSCFT}, where we 
discuss how K-stability of moduli spaces for more general $4d$ supersymmetric gauge theories
can be associated to the existence of corresponding superconformal field theories in the IR.
We preview possible avenues of generalizing Futaki invariants and how they could determine new notions of moduli space stability.
Appendices \sref{exactvalues} and \sref{plotsFp} give supplementary materials for the discussions in section \sref{Futforref}, including exact values for $U(1)_R$ charges and additional plots involving Futaki invariants. In appendix \sref{nonreflexive}, we compute the Futaki invariants and minimized volumes associated to toric Calabi-Yau 3-folds with non-reflexive toric diagrams
and comment on the generality of the bounds on the Futaki invariants that we discover in this work.

\paragraph{Nomenclature} 
\begin{align*}
    \Delta \quad &: \quad \text{a convex lattice polygon; $\Delta_{n-1} \subset \mathbb{Z}^{n-1}$};\\
    X \quad &: \quad \text{a (toric) variety constructed from $\Delta_{n-1}$, $\dim_{\mathbb{C}} X = n-1$};\\
    \mathcal{X} \quad &: \quad \text{affine Calabi-Yau cone over $X$, $\dim_{\mathbb{C}} \mathcal{X} = n$};\\
    & \qquad \text{here also called the mesonic moduli space $\mathcal{M}^{mes}$};\\
    Y \quad &: \quad \text{Sasaki-Einstein base manifold of $\mathcal{X}$, $\dim_{\mathbb{R}} Y = 2n-1$};\\
    n \quad &: \quad \text{(complex) dimension of $\mathcal{X}$, here also $\mathcal{M}^{mes}$};\\
    p_\alpha \quad &: \quad \text{(extremal) perfect matching/GLSM field};\\
    \zeta \quad &: \quad \text{$U(1)$ symmetry polarizing the mesonic moduli space (Reeb vector field)};\\
    b_i \quad &: \quad \text{components of the Reeb vector};\\
    g(t_i; \mathcal{X}) \quad &: \quad \text{Hilbert series (HS) of $\mathcal{X}$ in variables $t_i$;}\\
    \end{align*}
\begin{align*}
    V(b_i; Y) \quad &: \quad \text{volume function of $Y$};\\
    \int \text{Riem}^2 \quad &: \quad \text{integrated curvature of $Y$};\\
    D_\alpha \quad &: \quad \text{divisor in the Calabi-Yau cone $\mathcal{X}$ corresponding to $p_{\alpha}$, $\dim_{\mathbb{C}}(D_{\alpha}) = n - 1$};\\
    \Sigma_{\alpha} \quad &: \quad \text{submanifold of $Y$ corresponding to $D_{\alpha}$, $\dim_{\mathbb{R}}( \Sigma_{\alpha}) = 2n - 3$};\\
    V(b_i; \Sigma_{\alpha}) \quad &: \quad \text{divisor volume function of $\Sigma_{\alpha}$ };\\
    \chi \quad &: \quad \text{Euler number of $X$ (after complete resolution)};\\
    C_1 \quad &: \quad \text{first Chern number of the complete desingularization $\widetilde{X}$ of $X$}; \\
    & \qquad \text{this is the integral $\int_{\widetilde{X}} c_1\left(\widetilde{X}\right)$ of the first Chern class $c_1\left(\widetilde{X}\right)$};\\
    \eta \quad &: \quad \text{test symmetry with squared norm $||\eta||^2$};\\
    F \quad &: \quad \text{Futaki invariant}.
\end{align*}

\section{Background \label{preliminaries}} \setall

In the following section, we review some of the basic concepts regarding Gorenstein Fano varieties constructed from reflexive lattice polygons, non-compact toric Calabi-Yau 3-folds with Sasaki-Einstein base manifolds, as well as Hilbert series used to characterize them.
By reviewing the computation of the minimum volumes of Sasaki-Einstein 5-manifolds, we introduce the computation for Futaki invariants under a test symmetry -- the main subject of this work.
\\

\subsection{Toric Varieties and Reflexive Polytopes}\label{toric}

Let $\Delta$ be a convex lattice polytope in $\mathbb{Z}^m$. 
We define, 
\begin{definition}
A convex lattice polytope is reflexive if its dual polytope \cite{batyrev1982toroidal, Kreuzer:1995cd, Kreuzer:1998vb, Batyrev:1993oya, Kreuzer:2002uu}, given by
\beal{es02a01}
\Delta^{\circ} := \{ \textbf{u} \in \mathbb{Z}^{m} \mid \textbf{u}\cdot\textbf{v} \geq -1,\, \forall \textbf{v} \in \Delta \}
\eea
is also a lattice polytope in $\mathbb{Z}^m$. 
\end{definition}
In this paper, we shall mainly focus on $2d$ lattice polygons in $\mathbb{Z}^2$. 
A consequence of the reflexivity condition is that a reflexive polygon has only a single interior point, which can always be taken as the origin in $\mathbb{Z}^2$. 
There are $16$ reflexive polygons in $\mathbb{Z}^2$ up to $GL(2,\mathbb{Z})$ transformations as summarized in \fref{fig_reflexive}.

Given a lattice polytope $\Delta$, we can construct a compact toric variety $X(\Delta)$. 
When $\Delta$ is reflexive, we can take its unique internal point as the apex of a collection of cones that form an inner normal fan $\Sigma(\Delta)$.
These cones are bounded by rays extending from the origin to each of vertices of a face $\mathcal{F}$ of $\Delta$, such that
\beal{es02a05}
\Sigma(\Delta):=\left\{\text{pos}(\mathcal{F}):\mathcal{F}\in\text{Faces}(\Delta)\right\}
~,~
\eea
where
\beal{es02a06}
\text{pos}(\mathcal{F})=\left\{\sum_i\lambda_i\textbf{v}_i: \textbf{v}_i\in \mathcal{F},\lambda_i\geq0\right\}
~,~
\eea
is the positive hull of cones over face $\mathcal{F}$.
Using the fan $\Sigma(\Delta)$, we can construct a compact toric variety $X(\Delta)$ following the standard method in \cite{fulton1993introduction,cox2011toric}, where each cone gives an affine patch of $X(\Delta)$.

We can also think of the vertices in $\Delta$ as generators of a rational polyhedral cone $\sigma$ with the apex at the origin $(0,0,0)\in \mathbb{Z}^3:=M$.
Even though the reflexive polygon lives in $\mathbb{Z}^2$, we can consider the cone generated in $M$ by the vectors $\textbf{u}_i'=$($\textbf{u}_i$,1)$\in\mathbb{Z}^3$ as follows, 
\beal{es02a10}
\sigma=\left\{\sum_i\lambda_i\textbf{u}_i':\lambda_i\geq0\right\}\subset M\otimes_{\mathbb{Z}}\mathbb{R}=:M_{\mathbb{R}} ~.~
\eea
The dual cone $\sigma^\vee$ lives in the dual lattice $N_{\mathbb{R}}$, where $N:=\text{Hom}(M,\mathbb{Z})$. 
The dual cone takes the following form, 
\beal{es02a11}
\sigma^\vee=\left\{\textbf{w}\in N_{\mathbb{R}}:\textbf{w}\cdot\textbf{u}\geq0,\forall\textbf{u}\in\sigma\right\} ~.~
\eea
\begin{definition}
Given the dual cone $\sigma^\vee$, we can define 
an associated affine algebraic variety $\mathcal{X}$ as the maximal spectrum of the group algebra generated by the lattice points in $\sigma^\vee$, 
\beal{es02a12}
\mathcal{X} \cong {\rm Spec}_{\rm max}\mathbb{C}[\sigma^\vee\cap N]
~.~
\eea
\end{definition}

As an affine variety, we can explicitly define $\mathcal{X}$ as the vanishing locus of a set of multi-variate polynomials $f_i(x_1, \dots, x_k)$. Equivalently, the coordinate ring of $\mathcal{X}$ is
given by $\mathbb{C}[x_1, \dots, x_k] / \langle f_i \rangle$.
One can projectivize by letting $x_i$ be projective coordinates (with possible weights). Then, the base $X$ is also defined by $f_i$. In this sense, $\mathcal{X}$ is a complex affine cone over the toric variety $X(\Delta)$.
Now, given that the ewndpoints of the vector generators of the cone are all co-hyperplanar in $M$, $\mathcal{X}$ is a Gorenstein singularity \cite{fulton1993introduction,cox2011toric} and as a result, $\mathcal{X}$ admits a resolution to a Calabi-Yau manifold.

\subsection{Hilbert Series and the Mesonic Moduli Space \label{HS}}

Let us assume that we have a projective variety $X$ over which $\mathcal{X}$ is a Calabi-Yau cone.
Given this, we can define
\begin{definition}
The Hilbert series for $\mathcal{X}$ is a generating function for the graded pieces of its coordinate ring \beal{es02a15}
g(t; \mathcal{X}) = \sum_{i = 0}^{\infty} \dim_{\mathbb{C}}(X_{i}) ~t^{i},
\eea
where $X_{i}$ is the $i^\text{th}$ graded piece. 
\end{definition}
For multi-graded rings with pieces $X_{\vec{i}}$ and grading $\vec{i} = (i_1,\dots, i_k)$, the Hilbert series takes the following refined form, 
\beal{es02a16}
g(t_1, \dots, t_k; \mathcal{X}) = \sum_{\vec{i}=0}^\infty \dim_{\mathbb{C}}(X_{\vec{i}}) ~t_1^{i_1} \dots t_k^{i_k} ~.~
\eea

In this work,
we consider a family of abelian $4d$ $\mathcal{N}=1$ supersymmetric gauge theories
that are worldvolume theories of a D3-brane probing a toric Calabi-Yau 3-fold $\mathcal{X}$. 
Here, $\mathcal{X}$
is the mesonic moduli space $\mathcal{M}^{mes}$ 
of the $4d$ $\mathcal{N}=1$ supersymmetric gauge theory
and the grading given by $\vec{i} = (i_1,\dots, i_k)$ in \eref{es02a16}
can be interpreted as charges under a symmetry $\zeta$ in $\mathcal{M}^{mes}$, which usually is the global symmetry of the $4d$ $\mathcal{N}=1$ theory containing the $U(1)_R$ symmetry.
If $\zeta$ is chosen to be just the $U(1)_R$ symmetry, then $\vec{i} = (i_1,\dots, i_k)$ are the $U(1)_R$ charges $R(x_m)$ on the generators $(x_1, \dots, x_k)$ of the mesonic moduli space $\mathcal{M}^{mes}$.

\paragraph{Grading and Fugacities.}
In this work, we consider a family of $4d$ $\mathcal{N}=1$ supersymmetric gauge theories with $U(1)$ gauge groups whose mesonic moduli spaces $\mathcal{M}^{mes}$ are given by toric Calabi-Yau 3-folds $\mathcal{X}$. 
The Hilbert series of the coordinate ring $\mathbb{C}[x_1, \dots, x_k] / \langle f_i \rangle$ associated to $\mathcal{X}$ is the generating function of mesonic gauge invariant operators of the $4d$ $\mathcal{N}=1$ supersymmetric gauge theories \cite{Benvenuti:2006qr, Feng:2007ur, Butti:2006au, Butti:2007jv, Hanany:2007zz, Gray:2008yu}. 

For abelian $4d$ $\mathcal{N}=1$ theories where the mesonic moduli space $\mathcal{M}^{mes}$ is the toric Calabi-Yau 3-fold $\mathcal{X}$, 
we can make use of the forward algorithm \cite{Franco:2005rj, Feng:2000mi} to express the mesonic moduli space $\mathcal{M}^{mes}$ as the following symplectic quotient, 
\beal{es02a16a}
\mathcal{M}^{mes} = \text{Spec} ~ (\mathbb{C} [p_1, \dots, p_c] // Q_F) // Q_D ~,~
\eea
where $p_1, \dots, p_c$ are GLSM fields \cite{Witten:1993yc} that parameterize the toric Calabi-Yau 3-fold $\mathcal{X}$. 
$Q_F$ and $Q_D$ refer respectively to the $U(1)$ charges on the GLSM fields $p_\alpha$ under the $F$- and $D$-term of the $4d$ $\mathcal{N}=1$ supersymmetric gauge theory. 

In \eref{es02a16a}, the coordinates of $\mathcal{M}^{mes}$ are taken to be $(p_1,\dots, p_c)$ and by associating to each of the GLSM fields $p_\alpha$ a fugacity $t_\alpha$ that counts the degree of $p_\alpha$, the corresponding refined Hilbert series for \eref{es02a16a} can be calculated using the 
Molien-Weyl integral formula
\cite{Benvenuti:2006qr}, 
\beal{es02a17}
g(t_\alpha; \mathcal{X}) = \prod_{i = 1}^{c - 3} \oint_{|z_{i}| = 1} \frac{\text{d}z_{i}}{2\pi i z_{i}} \prod_{\alpha=1}^{c} \frac{1}{\left( 1-t_{\alpha} \prod\limits_{j=1}^{c-3}z_{j}^{(Q_{t})_{j\alpha}} \right)}
~.~
\eea
We refer to the above Hilbert series as the fully refined Hilbert series $g(t_\alpha; \mathcal{X}) $ of $\mathcal{M}^{mes}$ in terms of fugacities $t_\alpha$ corresponding to GLSM fields $p_\alpha$.

The global symmetry of the mesonic moduli space $\mathcal{M}^{mes}$ of the $4d$ $\mathcal{N}=1$ supersymmetric gauge theories that we are considering in this work includes the mesonic flavor symmetry of rank 2 and the $U(1)_R$ symmetry \cite{Forcella:2008bb, Forcella:2008eh, Hanany:2012hi, Hanany:2012vc}. 
It takes one of the following forms: 
\begin{itemize}
\item $U(1)_{f_1} \times U(1)_{f_2} \times U(1)_R$
\item $SU(2)_x \times U(1)_{f} \times U(1)_R$
\item $SU(2)_{x_1} \times SU(2)_{x_2} \times U(1)_R$
\item $SU(3)_{x_1,x_2} \times U(1)_R$. 
\end{itemize}
Above, $U(1)_R$ is the R-symmetry, whereas $U(1)_f$ corresponds to a global flavor symmetry, and $SU(2)_x$ and $SU(3)_x$ correspond to enhanced non-abelian global flavor symmetries. 

As we can see, the overall rank of the global symmetry group is 3.
The global symmetries originate from the isometry group of the toric Calabi-Yau 3-fold $\mathcal{X}$, which is of rank 3. 
The Hilbert series of the mesonic moduli space $\mathcal{M}^{mes}$ can be expressed in terms of a grading based on the global symmetry of $\mathcal{M}^{mes}$. 
In fact, any refinement of the Hilbert series for $\mathcal{M}^{mes}$ with more than 3 independent fugacities can be considered to be redundant due to the isometry group of the toric Calabi-Yau 3-folds $\mathcal{X}$.

As is standard, we will refer to multi-variable Hilbert series as refined and that of a single variable, the unrefined.
In the following work, we focus on two particular unrefinements of the Hilbert series $g(t_\alpha; \mathcal{X})$ of $\mathcal{M}^{mes}$.
These Hilbert series of $\mathcal{M}^{mes}$ are in terms of a single $U(1)$ inside the global symmetry of $\mathcal{M}^{mes}$. 
Let $\zeta$ refer to this $U(1)$ symmetry. 
In particular, we consider two choices for this $U(1)$ symmetry, the first being the 
$U(1)_R$ symmetry of the global symmetry. We refer to the $U(1)_R$ symmetry as $\zeta=\zeta_R$. 
The second choice for the $U(1)$ symmetry gives a grading of the coordinate ring for $\mathcal{M}^{mes}$ such that the fugacity of the Hilbert series counts degrees in GLSM fields for each of the mesonic gauge invariant operators. 
We refer to the symmetry resulting in this grading as $\zeta=\zeta_p$.
Below, we summarize these two choices for the unrefined Hilbert series of $\mathcal{M}^{mes}$:
\begin{enumerate}
\item \textbf{$U(1)_R$ Charges ($\zeta_R$).}
Each of the bifundamental chiral multiplets $X_{ij}$ of the $4d$ $\mathcal{N}=1$ supersymmetric gauge theory corresponding to toric Calabi-Yau 3-folds $\mathcal{X}$ can be expressed in terms of GLSM fields $p_\alpha$ associated to the extremal vertices of the toric diagram of $\mathcal{X}$, 
\beal{es02a19a}
X_{ij} = \prod_{X_{ij} \in p_\alpha} p_{\alpha} ~.~
\eea
The $U(1)_R$ charges $r(X_{ij})$ on bifundamental chiral fields $X_{ij}$, which can be obtained via $a$-maximization \cite{Intriligator:2003jj} for $4d$ $\mathcal{N}=1$ supersymmetric gauge theories, relate to the $U(1)_R$ charges $r_\alpha$ on GLSM fields $p_\alpha$ based on \eref{es02a19a} as follows, 
\beal{es02a19b}
r(X_{ij}) = \sum_{X_{ij} \in p_\alpha} r_{\alpha} ~.~
\eea
Accordingly, the fully refined Hilbert series $g(t_\alpha; \mathcal{X})$ defined in \eref{es02a17} can be unrefined in terms of the $U(1)_R$ symmetry of the global symmetry of $\mathcal{M}^{mes}$ by setting the GLSM field fugacities $t_\alpha = t^{r_\alpha}$, where now the fugacity $t$ refers to the $U(1)_R$ symmetry given by $\zeta_R$. 
We refer to this unrefined Hilbert series under $\zeta_R$ as follows, 

\beal{es02a19c}
g(t; \mathcal{X}, \zeta_R) =  g(t_\alpha = t^{r_\alpha}; \mathcal{X}) ~.~
\eea

\item \textbf{Degree in GLSM Fields ($\zeta_p$).}
The fully refined Hilbert series $g(t_\alpha; \mathcal{X})$ in \eref{es02a17} can be expressed in terms of a single fugacity $\bar{t}$, where now the exponent in $\bar{t}$ counts the degree in GLSM fields $p_\alpha$. 
Since the fugacities $t_\alpha$ in $g(t_\alpha; \mathcal{X})$ already correspond to each of the GLSM fields $p_\alpha$, respectively, this unrefinement can be achieved by setting 
\beal{es02a19d}
g(\bar{t}; \mathcal{X}, \zeta_p) = g(t_\alpha = \bar{t}; \mathcal{X}) ~.~
\eea
We refer to the $U(1)$ symmetry leading to the GLSM field grading as $\zeta_p$. 

\end{enumerate}
In the following work, we will use the above unrefined Hilbert series of the form $g(t; \mathcal{X}, \zeta_R)$ and $g(\bar{t}; \mathcal{X}, \zeta_p)$ in order to compute Futaki invariants $F_R$ and $F_p$, respectively.

\paragraph{Plethystics.}
The fully refined Hilbert series, as described in \eref{es02a17}, contains information about the algebraic structure of the toric Calabi-Yau $\mathcal{X}$. 
We can make use of the plethystic logarithm of the fully refined Hilbert series
\cite{Benvenuti:2006qr,Feng:2007ur}
in order to extract information about the generators and defining relations of the toric Calabi-Yau 3-fold $\mathcal{X}$ .
\begin{definition}
The plethystic logarithm of the fully refined Hilbert series $g(t_\alpha; \mathcal{X})$ is given by,
\beal{es02a20}
\text{PL}[g(t_1, \dots, t_c; \mathcal{X}) ] = \sum_{j = 1}^{\infty} \frac{\mu(j)}{j} \log(g(t_1^j, \dots, t_c^j; \mathcal{X})),
\eea
where $\mu(j)$ is the standard number-theoretic Möbius function. 
\end{definition}
The first positive terms of the plethystic logarithm are associated to the generators of $\mathcal{X}$, whereas the following negative terms relate to the defining relations amongst the generators. 
Any higher order terms in the expansion are associated to relations amongst relations, which are known as syzygies \cite{Gray:2006jb, Gray:2006gn}. 
A finite expansion indicates that $\mathcal{X}$ is a complete intersection \cite{Benvenuti:2006qr, Feng:2007ur, Butti:2006au, Butti:2007jv}.

\paragraph{Laurent Expansion.}
Let us consider a Hilbert series $g(t; \mathcal{X}, \zeta)$ for $\mathcal{X}$ under a $U(1)$ symmetry given by $\zeta$ with corresponding fugacity $t$.
The Laurent expansion of $g(t; \mathcal{X}, \zeta)$ around $s=0$ under the substitution $t=\text{e}^{-s}$ takes the following form,
\beal{es02a21}
g(t=\text{e}^{-s}; \mathcal{X}, \zeta)=\frac{a_0 (\zeta)}{s^n}+\frac{a_1(\zeta)}{s^{n-1}}+\dots ~,~
\eea
where $n$ corresponds to the complex dimension of the toric Calabi-Yau $n$-fold $\mathcal{X}$, in our case $n=3$.
We introduce, for $a_m$ the $m$-th coefficient in the expansion in \eref{es02a21},
\beal{es02a21b}
A_m=
\left\{
\begin{array}{rcl}
\frac{1}{(n-m-1)!} ~a_m  & \quad & m \leq n \\
a_m & \quad & m > n
\end{array}
\right. \ ,
\eea
and henceforth work primarily with $A_m$, in terms of which
the Laurent expansion of the Hilbert series $g(t; \mathcal{X}, \zeta)$ takes the form, 
\beal{es02a22}
g(t=\text{e}^{-s}; \mathcal{X}, \zeta)= 
\frac{(n-1)! A_0(\zeta)}{s^n}
+ \frac{(n-2)! A_1(\zeta)}{s^{n-1}} + \dots ~.~
\eea

\paragraph{The Sasaki-Einstein Base.}
We recall that the toric Calabi-Yau $3$-fold $\mathcal{X}$ has a Sasaki-Einstein 5-manifold $Y$ as its compact base manifold. 
We can consider the Calabi-Yau $3$-fold $\mathcal{X}$ as a real cone over the Sasaki-Einstein 5-manifold $Y$, where the metric of $\mathcal{X}$ is given by, 
\beal{es02a22}
\text{d}s^2(\mathcal{X})=\text{d}r^2+r^2\text{d}s^2(Y).
\eea
We emphasize that this is in parallel to and distinct from the fact that $\mathcal{X}$  is a complex cone over the toric Fano surface $X(\Delta)$.

The Laurent expansion of the Hilbert series around $s=0$ in \eref{es02a21} has coefficients that are directly related to topological invariants of $Y$. 
\begin{theorem}
The Hilbert series $g(t; \mathcal{X}, \zeta_R)$ in terms of a fugacity $t$ corresponding to the $U(1)_R$ symmetry $\zeta=\zeta_R$
has the following Laurent expansion \cite{Eager:2010dk},
\beal{es02a23}
\frac{8}{27} g(t=\text{e}^{-s}; \mathcal{X}, \zeta_R) &=& \frac{V_{min}}{s^{3}} + \frac{V_{min}}{s^{2}} 
\nn\\
&&
+ \left( \frac{91}{216} V_{min}+ \frac{1}{1728} \int_Y \text{Riem}^{2}(Y)\right) \frac{1}{s} 
+ \dots
~,~
\eea
where the coefficients directly relate to the integrated curvature 
$\int_Y\text{Riem}^2(Y)$
and the minimum volume $V_{min}$ of $Y$ \cite{Martelli:2005tp,Martelli:2006yb}. 
\end{theorem}
We note that in \eref{es02a23} the coefficients match the minimum volume $V_{min}$ only because the original Hilbert series is in terms of the $U(1)_R$ charge fugacity $t$.
In the following section, we discuss in detail the computation of the minimum volume $V_{\min}$.

\subsection{Minimized Volumes and Topological Invariants}\label{minimizedvolandtopinv}

\paragraph{Volume Function and Minimization.}
In our work, we require the Calabi-Yau cone $\mathcal{X}$ to be toric, which implies that we have a torus action 
$\mathbb{T}^3$ on $\mathcal{X}$ that leaves the K\"ahler form $\omega$
invariant. 
The generators of the torus action are given by $\partial/\partial\phi_i$, where $\phi_i$ are the angular coordinates with $\phi_i\sim\phi_i+2\pi$. 
Accordingly, the Sasaki-Einstein 5-manifold $Y=\mathcal{X} |_{r=1}$ has a Killing vector field called the Reeb vector, which can be expressed as 
\beal{es02a30}
\zeta=b_i\partial/\partial\phi_i ~,~
\eea 
where the Reeb vector components $b_i$ are algebraic numbers. 
\begin{definition}
The volume of the Sasaki-Einstein base $Y$ 
expressed in terms of 
Reeb vector components $b_i$
is given by
\beal{es02a31}
vol[Y] = \int_{r\leq 1} \omega^3 ~,~
\eea
where $\omega$ is the K\"ahler form and the integration of the $(3,3)$-form $\omega^3$ is over the Calabi-Yau threefold $\mathcal{X}$. 
The volume is normalized as follows, 
\beal{es02a32}
V(b_i; Y) = \frac{vol[Y]}{vol[S^5]} ~,~
\eea
where the volume of $S^5$ is given by he normalization $vol[S^5] = \pi^3$. 
\end{definition}
We note here that in the following work, we are going to use interchangeably the expression $V(b_i; Y)$ and $V(b_i; \mathcal{X})$ for the volume of the Sasaki-Einstein base manifold $Y$ associated to the Calabi-Yau cone $\mathcal{X}$.

The volume function $V(b_i; Y)$ in terms of the Reeb vector components $b_i$ can be obtained directly from the Hilbert series of $\mathcal{X}$.
\begin{theorem}
Using the Hilbert series $g(t_\alpha; \mathcal{X})$ refined under the extremal GLSM field fugacities $y_\alpha = t_\alpha$, the volume function for the Sasaki-Einstein manifold $Y$ corresponding to $\mathcal{X}$ is obtained as follows \cite{Bergman:2001qi,Martelli:2005tp,Martelli:2006yb}, 
\beal{es02a32}
V(b_\alpha; Y) = \frac{8}{27} \lim_{s \rightarrow 0} s^3 g(t_\alpha = \exp[- s b_\alpha]; \mathcal{X}) ~,~
\eea
where here $b_\alpha$ are the Reeb vector components now associated to GLSM fields corresponding to extremal points in the toric diagram of the toric Calabi-Yau 3-fold $\mathcal{X}$. 
\end{theorem}
In the above, the fugacities corresponding to GLSM fields associated to the non-extremal vertices in the toric diagram $\Delta$ of $\mathcal{X}$ are set to $y_\alpha=1$.
Based on the fact that the $U(1)_R$ charge of the superpotential $W$ of the associated $4d$ $\mathcal{N}=1$ supersymmetric gauge theories is $R(W)=2$, we set as a convention 
\beal{es02a32b}
\sum_\alpha b_\alpha = 2 ~.~ 
\eea

Recalling the Laurent expansion in \eref{es02a23}, we can see that the limit in \eref{es02a32} picks the leading order in $s$, which we identified with the minimum volume $V_{min}$ of the Sasaki-Einstein $5$-manifold $Y$, if the original Hilbert series is refined under the $U(1)_R$ symmetry.
In fact, under a global minimization of the volume function $V(b_\alpha; Y)$ in \eref{es02a32}, we indeed find the volume minimum $V_{min}$ in \eref{es02a23}, 
\beal{es02a33}
V_{min} = V(b_\alpha^*; Y) = \min_{b_\alpha} V(b_\alpha; Y) \big|_{\sum_{\alpha} b_\alpha = 2}~.~
\eea
We note here that the AdS/CFT correspondence relates the central charge $a$-function of the $4d$ $\mathcal{N}=1$ superconformal field theory with the volume of the Sasaki-Einstein 5-manifold $Y$ as follows \cite{Martelli:2005tp,Martelli:2006yb,He:2017gam}, 
\beal{es02a34}
a(R; Y) = \frac{\pi^3 N^2}{4 V(b_\alpha^*; Y)} ~.~
\eea
Under normalization, we can re-define the $a$-function to
\beal{es02a35}
A(R; Y) \equiv \frac{a(R; Y)}{a(R; S^5)} = \frac{vol[S^5]}{vol[Y]} = \frac{1}{V(b_\alpha^*; Y)}
~.~
\eea
This relationship between the central charge $a$-function and the volume function for the Sasaki-Einstein base manifold $Y$ implies that under volume minimization \cite{Martelli:2005tp,Martelli:2006yb,He:2017gam} in \eref{es02a33}, the $a$-function is maximized, which is known as $a$-maximization \cite{Intriligator:2003jj}. 
At the critical point of volume minimization, $V_{min} = V(b^*; Y)$, we can identify the critical Reeb vector $b^*$ and its components.
This is in line with the fact that the Reeb vector generates the $U(1)_R$ symmetry and the corresponding superconformal $U(1)_R$ charges at the critical point of the RG flow.

\paragraph{Topological Invariants and Volume Bounds.}
The distribution of minimum volumes of a large family of Sasaki-Einstein $(2n-1)$-manifolds corresponding to 
toric Calabi-Yau $n$-folds with reflexive toric diagrams $\Delta_{n-1}$ have been studied systematically in \cite{He:2017gam}. 
There, it was discovered that for this family of toric Calabi-Yau $n$-folds, with $n=3$ and $4$, the minimum volume satisfies lower and upper bounds parameterized by topological quantities of the corresponding toric varieties $X(\Delta_{n-1})$. 
\begin{conjecture}
According to \cite{He:2017gam}, 
there is a universal lower and upper bound on the 
minimum volume $V_{min}$ of Sasaki-Einstein $(2n-1)$-manifolds
corresponding to toric Calabi-Yau $n$-folds with reflexive toric diagrams $\Delta_{n-1}$
for any $n$, 
\beal{es02a40}
\frac{1}{\chi(\widetilde{X(\Delta_{n-1})})} 
\leq V_{min} \leq m_n \int c_1^{n-1} (\widetilde{X(\Delta_{n-1})}) ~,~
\eea
where $m_n > m_{n+1}$, and 
the lower and upper bounds are defined by two topological quantities of $X(\Delta_{n-1})$, the Euler number $\chi$ of $\widetilde{X(\Delta_{n-1}})$ and the first Chern number $\int c_1^{n-1}$ of $\widetilde{X(\Delta_{n-1}})$.
\end{conjecture}
It was shown in \cite{moraga2024bounding} that the lower bound is related to the non-symmetric Mahler conjecture in convex geometry. 
The upper bound was also studied in \cite{Manko:2022zfz}.

Here, the computations in \cite{He:2017gam}
gave values for
$m_3 \sim 3^{-3}$ and $m_4 \sim 4^{-4}$.
Furthermore, 
the complete resolution $\widetilde{X(\Delta_{n-1})}$ is achieved by the Fine Regular and Star (FRS) triangulation \cite{Altman:2014bfa} of the reflexive polytope given by $\Delta_{n-1}$.

As discussed in \cite{He:2017gam}, the Euler number $\chi(X(\Delta_{2}))$ for $n=3$ is given by,
\beal{es02a41}
\chi \equiv \chi(X(\Delta_{n-1})) = p ~,~
\eea
where $p$ is the number of perimeter lattice points of the 2-dimensional toric diagram $\Delta_2$. By Pick's formula, this number is related to the number of interior lattice points $i$ and the area $A$ of the convex lattice polygon $\Delta_2$, 
\beal{es02a42}
A(\Delta_2) = i + \frac{p}{2} - 1 ~.~
\eea
Similarly, the first Chern number $\int c_1^2(X(\Delta_{2}))$ is defined for a 2-dimensional toric diagram $\Delta_2$ as follows, 
\beal{es02a43}
C_1 \equiv \int c_1^2(X(\Delta_{2})) = p^\circ ~,~
\eea
where $p^\circ$ is the number of perimeter lattice points of the polar dual reflexive polygon $\Delta_2^\circ$ as defined in \eref{es02a01}.

In the following work, we are going to study the relationship between the Futaki invariants under a test $U(1)$ symmetry, the Euler number $\chi$ and the first Chern number $C_1$ of the related toric variety $X(\Delta_2)$, where $\Delta_2$ is one of the 16 reflexive polygons in $\mathbb{Z}^2$. 
Motivated by the findings in \cite{He:2017gam}, we discover interesting new relationships between the Futaki invariants and topological invariants of $X(\Delta_2)$, which we summarize in the following sections.
\\

\subsection{Divisor Volumes}\label{divisorvolumes}

The toric Calabi-Yau cone $\mathcal{X}$ has a Sasaki-Einstein 5-manifold $Y= \mathcal{X}|_{r=1}$ as its base, whose volume $V(b_i; Y)$ is related under minimization to the volumes of the divisors $D_\alpha$ in $\mathcal{X}$\cite{Martelli:2005tp,Martelli:2006yb,Butti:2006au}. 
The divisors $D_\alpha$ are associated to the extremal points of the toric diagram $\Delta$ of the toric Calabi-Yau cone $\mathcal{X}$ as well as the corresponding extremal GLSM fields $p_\alpha$.
In the following section, we discuss the volumes of the divisors, their connection to $U(1)_R$ charges on GLSM fields and the methods to compute them. 

\paragraph{Hilbert Series and Volume functions for Divisors.}
We recall from section \sref{HS} the definition of the Hilbert series of the mesonic moduli space in terms of GLSM fields $p_\alpha$, as given in \eref{es02a17}.
In our work, 
we study 
a family of abelian $4d$ $\mathcal{N}=1$ supersymmetric gauge theories 
where the mesonic moduli space is a toric Calabi-Yau 3-fold $\mathcal{X}$.
Here, we note that we can define a Hilbert series not just for the entire mesonic moduli space, but for one of the divisors $D_\alpha$ in $\mathcal{X}$:
\begin{theorem}
The Hilbert series \cite{Martelli:2005tp, Martelli:2006yb, Butti:2006au}
for the divisor $D_\alpha$ in the toric Calabi-Yau 3-fold $\mathcal{X}$
is given by, 
\beal{es02a50}
g(y_\alpha; D_\alpha) = \prod_{i = 1}^{c - 3} \oint_{|z_{i}| = 1} \frac{\text{d}z_{i}}{2\pi i z_{i}} \prod_{\beta = 1}^{c} \frac{\left[ y_{\alpha} \prod\limits_{k=1}^{c-3} z_{k}^{(Q_{t})_{k\alpha}} \right]^{-1}}{1-y_{\beta} \prod\limits_{j=1}^{c-3}z_{j}^{(Q_{t})_{j\beta}}}
~,~
\eea
where $Q_t=(Q_F,Q_D)$ is the total $U(1)$ charge matrix on GLSM fields $p_\alpha$.
\end{theorem}
In the above, $Q_t$ is obtained from the forward algorithm for the abelian $4d$ $\mathcal{N}=1$ supersymmetric gauge theory associated to the toric Calabi-Yau 3-fold $\mathcal{X}$ \cite{Franco:2005rj, Feng:2000mi}.
The $Q_t$-matrix encodes the $U(1)$ charges due to the $F$- and $D$-terms of the $4d$ $\mathcal{N}=1$ theory.
The number of GLSM fields is given by $c$ and the fugacities $y_\alpha$ are set to $y_\alpha=t_\alpha$ for GLSM fields $p_\alpha$ associated to extremal points of the toric diagram $\Delta$, whereas for all other GLSM fields we set $y_\alpha=1$. 

\begin{theorem}
The volume function $V(b_i; \Sigma_{\alpha})$ \cite{Martelli:2005tp, Martelli:2006yb, Butti:2006au} associated to the 
submanifold $\Sigma_\alpha$ of the Sasaki-Einstein manifold $Y$, 
which corresponds to the
divisor $D_\alpha$ in $\mathcal{X}$
and the associated GLSM field $p_\alpha$,
is given by
\beal{es02a51}
V(b_i; \Sigma_{\alpha}) = \frac{3}{2} V(b_i^*; Y) \lim_{ s \to 0 } \frac{1}{s} \left[ \frac{g(t_i = \text{e}^{-s b_i}; D_\alpha)}{g(t_i = \text{e}^{-s b_i}; \mathcal{X})} - 1 \right] ~,~
\eea
where $b_i$ are the Reeb vector components, 
$g(t; \mathcal{X})$ is the Hilbert series for $\mathcal{X}$ and 
$g(t; D_\alpha)$ is the Hilbert series for $D_\alpha$.
\end{theorem}
Here, we note that the Reeb vector components $b_i$ in \eref{es02a51} also appeared in the volume function for the Sasaki-Einstein 5-manifold $Y$ in \eref{es02a32}.

Following \cite{Martelli:2005tp},
the volume of the submanifold $\Sigma_\alpha$ of the Sasaki-Einstein 5-manifold $Y$
corresponding to the divisor $D_\alpha$,
which we refer to here simply as the divisor volume
$V(b_i; \Sigma_{\alpha})$,
 can also be obtained using a combinatorial formula based on the toric diagram $\Delta$ of the toric Calabi-Yau 3-fold $\mathcal{X}$. 
By identifying the extremal vertex $v_\alpha \in \Delta$ as the vertex corresponding to the extremal GLSM field $p_\alpha$ and divisor $D_\alpha$,  the normalized divisor volume $V(b_i; \Sigma_{\alpha})$ can be obtained using, 
\beal{es02a52}
V(b_i; \Sigma_{\alpha}) = \frac{\det(v_{\alpha-1}, v_{\alpha}, v_{\alpha + 1})}{\det(b, v_{\alpha-1}, v_{\alpha})\det(b, v_{\alpha}, v_{\alpha + 1})},
\eea
where the Reeb vector takes the form $b=(b_1,b_2,b_3)$.

\paragraph{$U(1)_R$ Charges and Divisor Volumes.}
When the Reeb vector components $b_i$ take critical values $b_i^*$ at which the volume $V(b_i; Y)$ of the Sasaki-Einstein 5-manifold $Y$ becomes a minimum $V_{min}$, as stated in \eref{es02a32}, then the volume $V(b_i; \Sigma_{\alpha})$ of the divisor $D_\alpha$ can be related to the $U(1)_R$ charge of the corresponding extremal GLSM field $p_\alpha$,
\beal{es02a53}
R(p_\alpha) \equiv 
\frac{2}{3}
\frac{V(b_i^*; \Sigma_{\alpha})}{V(b_i^*; Y)} 
=
\frac{2V(b_i^*; \Sigma_{\alpha})}{\sum_\alpha V(b_i^*; \Sigma_{\alpha})} 
~.~
\eea

\subsection{Futaki Invariants}\label{Fut}

As studied in \cite{Collins:2016icw, Bao:2020ugf}, Futaki invariants measure the K-stability of the mesonic moduli space $\mathcal{M}^{mes}$ of a $4d$ $\mathcal{N}=1$ supersymmetric gauge theory when the theory flows towards the IR. 
Knowing that the $U(1)_R$ charges of bifundamental chiral fields in the $4d$ $\mathcal{N}=1$ theories are determined via $a$-maximization \cite{Intriligator:2003jj}, 
we can consider the computation of Futaki invariants as a generalized version of $a$-maximization, where the original $U(1)_R$ symmetry $\zeta$ is modified by an additional test symmetry $\eta$ \cite{Collins:2016icw, Bao:2020ugf}.
The Futaki invariants measure the extent to which the mesonic moduli space $\mathcal{M}^{mes}$ becomes destabilized under the combined symmetry $\zeta+\epsilon \eta$ for small $\epsilon$.
In the following section, we review the computation of Futaki invariants under a test symmetry $\eta$.

\paragraph{Test Symmetries.}
We recall that the mesonic moduli spaces $\mathcal{M}^{mes}$ of the family of $4d$ $\mathcal{N}=1$ supersymmetric gauge theories that we consider in this work are non-compact toric Calabi-Yau 3-folds $\mathcal{X}$.
The Hilbert series for $\mathcal{X}$
is a generating function for the graded pieces of the coordinate ring given by $\mathbb{C}[x_1, \dots, x_k] / \langle f_i \rangle$, where we refer to $\langle f_i \rangle$ as the ideal $I$.

We also recall that the $U(1)_R$ symmetry is associated to the Reeb vector field $\zeta$ on the Sasaki-Einstein base $Y$ of the Calabi-Yau cone $\mathcal{X}$.
When we use a grading associated to the $U(1)_R$ symmetry and the Reeb vector field $\zeta$,
we refer to the toric Calabi-Yau 3-fold $\mathcal{X}$ and the associated affine variety as \textit{polarized}.
Following the discussion in section \sref{HS}, we note that we can also introduce a grading corresponding to the counting of degrees in GLSM fields associated to the toric Calabi-Yau 3-fold $\mathcal{X}$ \cite{Witten:1993yc}. 
In the following work, we will consider both gradings for the computation of Futaki invariants. We denote by $\zeta=\zeta_R$ the $U(1)_R$ symmetry and by $\zeta=\zeta_p$ the $U(1)$ symmetry for the grading associated to the degrees in GLSM fields, as discussed in section \sref{HS}.

For the computation of Futaki invariants, we add to the chosen $U(1)$ symmetry given by $\zeta$ a test symmetry $\eta$.
Here, we define the test symmetry as follows,
\begin{definition}
The test symmetry $\eta$ is defined as the following $\mathbb{C}^*$-action on the $k$ coordinates $x_1, \dots, x_k$ in the coordinate ring $\mathbb{C}[x_1, \dots, x_k]/\langle f_i \rangle$, 
\beal{es02a60}
\eta(\lambda):\mathbb{C}^*\hookrightarrow\text{GL}(k,\mathbb{C})
~,~
\eea
where $\lambda$ is the $\mathbb{C}^*$ parameter.
The above $\mathbb{C}^*$-action acts on the defining polynomials $f_i$ of the ideal $I$ as follows,
\beal{es02a61}
(\eta(\lambda)\cdot f_i)(x_1,\dots,x_k)=f_i(\eta(\lambda)x_1,\dots,\eta(\lambda)x_k)
~.~
\eea
\end{definition}

Under the test symmetry $\eta$, we obtain a test configuration, 
\beal{es02a62}
X_\lambda=\mathbb{C}[x_1,\dots,x_k]/I_\lambda
~,~
\eea
where the modified ideal takes the form, 
\beal{es02a63}
I_\lambda=\{\eta(\lambda)\cdot f_i | f_i \in I\}
~.~
\eea 
Following this, the central fibre $X_0$ of the ring is obtained by taking the flat limit \cite{eisenbud2013commutative}, 
\beal{es02a64}
    I_0=\lim\limits_{\lambda\rightarrow0}I_\lambda=\{\text{in}(f_i)|f_i\in I\} ~,~
\eea
where $\text{in}(f_i)$ is the lowest weight polynomial under $\eta$, which comes from the original defining polynomial $f_i$ in $I$ \cite{Collins:2016icw, Bao:2020ugf}. 

\noindent
\textit{Example.}
Let us consider here an example that illustrates the origin of $I_0$ based on $I$. Take the conifold $x_1^2+x_2^2+x_3^2+x_4^2=0$ and the $\mathbb{C}^*$ action giving $\eta(\lambda)\cdot(x_1, x_2, x_3, x_4)=(\lambda x_1,x_2,x_3,x_4)$. 
We have the test configuration and the central fibre cut out by $\lambda^2 x_1^2+x_2^2+x_3^2+x_4^2=0$ and $x_2^2+x_3^2+x_4^2=0$, respectively. 
However, when we consider the case when $\eta(\lambda)\cdot(x_1, x_2, x_3, x_4)=(\lambda^{-1} x_1,x_2,x_3,x_4)$ with weights under $\eta$ given by $(-1,0,0,0)$, then the test configuration results in $x_1^2 = 0$. 

We can introduce the following notation for a test symmetry $\eta$ that affects the $h$-th generator $x_h$ of $\mathcal{X}$ with weight $1$, 
\beal{es02a65}
\eta_h = (\delta_{h,1}, \dots, \delta_{h,k}) ~:~
(x_1,\dots,x_k) \mapsto (x_1,\dots, \lambda x_h, \dots,x_k) ~.~
\eea
In general, we can consider a test symmetry $\eta$ defined in terms of weights $(w_1,\dots,w_k) \in \mathbb{Z}^k_{\geq 0}$ such that,
\beal{es02a65b}
\eta_{(w_1,\dots,w_k)} ~:~ (x_1,\dots,x_k) \mapsto (\lambda^{w_1}x_1,\dots,\lambda^{w_k}x_k)~,~
\eea
where $(x_1,\dots,x_k)$ are the generators for $\mathcal{X}$.
The weights $(w_1,\dots,w_k)$ here parameterize a general $\mathbb{C}^*$-action on the generators of $\mathcal{X}$ given by $\eta=\eta_{(w_1,\dots,w_k)}$.

\paragraph{Futaki Invariants.}
Let us assume we have a toric Calabi-Yau cone $\mathcal{X}$
whose generators are weighted under a $U(1)$ symmetry given by $\zeta$.
Given that the generators of $\mathcal{X}$ are $(x_1, \dots x_k)$, let us denote the weights on the generators under $\zeta$ as $(q_1, \dots, q_k)$. 
By associating to $\zeta$ the fugacity $t$, we recall that the corresponding Hilbert series $g(t; \mathcal{X}, \zeta)$ 
has a Laurent expansion given in \eref{es02a22},
where $n$ is the complex dimension of the affine Calabi-Yau cone $\mathcal{X}$, which in our case has $n=3$.
We also recall that if $\zeta$ refers to the $U(1)_R$ symmetry, the coefficients $A_0(\zeta)$ and $A_1(\zeta)$ are proportional to the normalized minimum volume $V_{min}$ of the Sasaki-Einstein base manifold $Y$ of the toric Calabi-Yau cone $\mathcal{X}$ \cite{Martelli:2005tp,Martelli:2006yb,He:2017gam}, as discussed in section \sref{minimizedvolandtopinv}.

\begin{theorem}
Under a test symmetry $\eta_h$ that acts on the generator $x_h$ of $\mathcal{X}$ with weight $w_h \in \mathbb{Z}_{\geq 0}$, 
the Hilbert series under the grading given by $\zeta+ \epsilon \eta_h$ takes the form \cite{Bao:2020ugf}, 
\beal{es02a71b}
g(t; \mathcal{X}, \zeta + \epsilon \eta_h) 
= \frac{1-t^{q_h}}{1-t^{q_h + \epsilon w_h}}
g(t; \mathcal{X}, \zeta) 
~,~
\eea
where we associate the fugacity $t$ to $\zeta+\epsilon \eta_h$.
\end{theorem}
We also choose here $\eta_h$ such that the weight on $x_h$ under $\eta_h$ is $w_h=1$ as shown in \eref{es02a65}. 
The Laurent expansion of the new Hilbert series under $\zeta+ \epsilon \eta_h$ then takes the following new form, 
\beal{es02a71c}
g(t=e^{-s}; \mathcal{X}, \zeta + \epsilon \eta_h) 
&=&
\frac{(n-1)! A_0(\zeta) q_h}{(q_h+\epsilon) s^n}
\nn\\
&&
+ \frac{
\left((n-1)! A_0(\zeta) \epsilon + 2(n-2)! A_1(\zeta) \right)  q_h
}{
2(q_h + \epsilon) s^{n-1}
}
+ \dots 
\nn\\
&=&
\frac{(n-1)! A_0(\zeta+\epsilon \eta_h)}{s^n}
+ \frac{(n-2)! A_1(\zeta + \epsilon \eta_h)}{s^{n-1}}
+ \dots
~,~
\nn\\
\eea
where the new coefficients can be expressed in terms of the coefficients $A_0(\zeta)$ and $A_1(\zeta)$ as follows, 
\beal{es02a71d}
A_0(\zeta + \epsilon \eta_h) = 
\frac{A_0 (\zeta) q_h}{q_h + \epsilon} ~,~
A_1(\zeta + \epsilon \eta_h) = 
\frac{\left((n-1) \epsilon A_0 (\zeta) +2 A_1(\zeta)\right) q_h}{2 (q_h + \epsilon)} ~.~
\eea

The Futaki invariant is a measure on how these coefficients change under the introduction of a test symmetry $\eta$.
\begin{definition}
The Futaki invariant for a test symmetry $\eta$ is given by \cite{futaki1983obstruction,ding1992kahler,donaldson2002scalar, Collins:2012dh,collins2019sasaki,Collins:2016icw, Bao:2020ugf},
\beal{es02a71}
F(\mathcal{X}; \zeta, \eta) = \frac{A_{0}(\zeta)}{n-1} \text{D}_{\epsilon} \left[
\frac{A_{1}(\zeta + \epsilon \eta)}{A_{0}(\zeta + \epsilon \eta)} 
\right]
+ \frac{A_{1}(\zeta)}{n(n-1)A_{0}(\zeta)} \text{D}_{\epsilon} A_{0}(\zeta + \epsilon \eta) \bigg|_{\epsilon = 0}
~,~
\eea
where the leading coefficients
$A_0(\zeta + \epsilon \eta)$ and $A_1(\zeta + \epsilon \eta)$ are obtained from the Laurent expansion of the Hilbert series
$g(t; \mathcal{X}, \zeta + \epsilon \eta)$.
\end{definition}
The above definition for the Futaki invariant can be simplified to the following form, 
\beal{es02a75}
F(\mathcal{X}; \zeta, \eta) = \frac{A_1}{A_0} B_0 - B_1 
~,~
\eea
where $A_i = A_i(\zeta)$ and 
\beal{es02a76}
B_i = -\frac{1}{n-i} \text{D}_\epsilon A_i (\zeta + \epsilon \eta)\bigg|_{\epsilon = 0}~.~
\eea
By inserting the expressions for $A_i$ and $B_i$ above with the test symmetry given by $\eta=\eta_h$, we have
\begin{theorem}
Under a test symmetry $\eta_h$
giving a weight $1$ to the $h$-th generator $x_h$ of $\mathcal{X}$ and a weight $0$ to all other generators, 
the corresponding Futaki invariant for $\mathcal{X}$ takes the form,
\beal{es02a76}
F(\mathcal{X}; \zeta, \eta_h)
= 
\frac{A_0(\zeta)}{2} - \frac{A_1(\zeta)}{n(n-1) q_h} ~,~
\eea
where $q_h$ is the weight on $x_h$ under $\zeta$.
\end{theorem}

\noindent\textit{Proof.}
We recall the expressions for the coefficients $A_0(\zeta+\epsilon \eta_h)$ and $A_1(\zeta+\epsilon \eta_h)$ in \eref{es02a71d}.
Using the definition of the Futaki invariant in \eref{es02a71}, we have
\beal{es02a76b}
F(\mathcal{X}; \zeta, \eta_h) =
\frac{A_0(\zeta)}{n-1} \left[ \frac{n-1}{2} \right]
+ \frac{A_1(\zeta)}{n(n-1)A_0(\zeta)}
\left[
- \frac{A_0(\zeta)}{q_h}
\right]
~,~
\eea
which gives the expression for the Futaki invariant in \eref{es02a76}.
\hfill\qed

Here, we recall that the toric Calabi-Yau 3-fold $\mathcal{X}$ is the mesonic moduli space $\mathcal{M}^{mes}$ of the family of $4d$ $\mathcal{N}=1$ supersymmetric gauge theories that we study in this work.
We have $n=3$ as the complex dimension for $\mathcal{X}$ and
$\zeta$ as the $U(1)_R$ symmetry. 
Accordingly, the corresponding Futaki invariant for the $h$-th generator of $\mathcal{X}$ takes the following form, 
\beal{es02a77}
F(\mathcal{X}; \zeta_R, \eta_h) = 
\frac{A_0(\zeta_R)}{2}
- \frac{A_1(\zeta_R)}{6R_h}
~,~
\eea
where here $R_h$ denotes the $U(1)_R$ charge carried by the generator $x_h$ of $\mathcal{M}^{mes}$.
As discussed in section \sref{HS}, the coefficients $A_0(\zeta_R)$ and $A_1(\zeta_R)$ directly relate to the minimum volume $V_{min}$ of the Sasaki-Einstein base manifold of the toric Calabi-Yau 3-fold $\mathcal{X}$. 
We denote the Futaki invariant for the $h$-th generator $x_h$ of the mesonic moduli space $\mathcal{M}^{mes}$ under a test symmetry $\eta_h$ with the $U(1)_R$ symmetry as $F_{R,h} = F(\mathcal{X}; \zeta_R, \eta_h)$ in the following work. 

As discussed in section \sref{HS}, we can also choose $\zeta$ to correspond to a $U(1)$ symmetry that weights the generators of $\mathcal{X}$ according to their degrees in GLSM fields that parameterize the toric Calabi-Yau 3-fold $\mathcal{X}$. 
Denoting this symmetry as $\zeta=\zeta_p$, the resulting Futaki invariant of the $h$-th generator of the mesonic moduli space $\mathcal{M}^{mes}$ has the following form, 
\beal{es02a78}
F(\mathcal{X}; \zeta_p, \eta_h) = 
\frac{A_0(\zeta_p)}{2} - \frac{A_1(\zeta_p)}{6 d_h} ~,~
\eea
where $d_h$ refers to the number of GLSM fields that make up according to \eref{es02a19a} the $h$-th gauge-invariant generator $x_h$ of the mesonic moduli space $\mathcal{M}^{mes}$.
We denote the Futaki invariant for the $h$-th generator $x_h$ of $\mathcal{M}^{mes}$ under a test symmetry $\eta_h$ and a grading of the Hilbert series $g(\bar{t}; \mathcal{X}, \zeta_p)$ in terms of degrees in GLSM fields as $F_{p,h} = F(\mathcal{X}; \zeta_p, \eta_h)$ in the following work. 

\begin{corollary}
Under a general test symmetry $\eta=\eta_{(w_1, \dots, w_k)}$ giving weights $(w_1, \dots, w_k)\in \mathbb{Z}_{\geq 0}^k$ to generators $(x_1,\dots,x_k)$ of $\mathcal{X}$, 
the corresponding Futaki invariant takes the form,
\beal{es02a80}
F(\mathcal{X}; \zeta, \eta_{(w_1, \dots, w_k)}) = 
\sum_{m=1}^{k} 
\left(
\frac{A_0(\zeta)}{2}
- 
\frac{A_1(\zeta)}{n(n-1) q_m} 
\right)w_m
= 
\sum_{m=1}^{k}
F(\mathcal{X}; \zeta, \eta_m) ~w_m
~.~
\eea
\end{corollary}

\noindent\textit{Proof.}
For a general test symmetry given by $\eta=\eta_{(w_1, \dots, w_k)}$ as defined in \eref{es02a65b}, 
the Laurent expansion of the corresponding Hilbert series under $\zeta + \epsilon \eta$ gives the following leading order coefficients, 
\beal{es02a79}
A_0(\zeta + \epsilon \eta)
&=& 
\left(
\prod_{m=1}^{k}
\frac{q_m}{q_m + w_m \epsilon} 
\right)
A_0(\zeta)
~,~
\nn\\
A_1(\zeta + \epsilon \eta)
&=&
\left(
\prod_{m=1}^{k}
\frac{q_m}{q_m + w_m \epsilon} 
\right)
\left[
\frac{n-1}{2}
\left(
\sum_{m=1}^k
w_m
\right)
\epsilon A_0(\zeta)
+ A_1(\zeta)
\right]
~,~
\eea
where $q_m$ are the weights on the generators $x_m$ under the $U(1)$ symmetry given by $\zeta$.
Using the definition of the Futaki invariant in \eref{es02a71}, we obtain the general form in \eref{es02a80}.
\hfill\qed
\\

\paragraph{K-Stability.}
In \cite{Collins:2016icw}, it was conjectured that K-stability of the mesonic moduli space $\mathcal{M}^{mes}$, also known as the chiral ring of the $4d$ $\mathcal{N}=1$ supersymmetric gauge theories, can be associated to the existence of a corresponding $4d$ conformal field theory in the IR. 
This is certainly true for $4d$ $\mathcal{N}=1$ worldvolume theories of a D3-brane probing a toric Calabi-Yau 3-fold, 
where the mesonic moduli spaces $\mathcal{M}^{mes}$ 
of the $4d$ $\mathcal{N}=1$ theory is the probed toric Calabi-Yau 3-fold $\mathcal{X}$ itself. 
In the following work, we concentrate on this family of $4d$ $\mathcal{N}=1$ supersymmetric gauge theories corresponding to toric Calabi-Yau 3-folds,
with an additional restriction that the toric Calabi-Yau 3-folds have toric diagrams which are reflexive polygons in $\mathbb{Z}^2$ as originally studied in \cite{Hanany:2012hi}.
For this family of $4d$ $\mathcal{N}=1$ supersymmetric gauge theory, 
we determine the K-stability of their mesonic moduli spaces 
by the positivity of the Futaki invariants $F(\mathcal{X};\zeta, \eta)$ for a given test symmetry $\eta$. 

In principle, one needs to check the sign of the Futaki invariants $F(\mathcal{X};\zeta, \eta)$ corresponding to all possible test symmetries $\eta$ and associated test configurations in order to fully ensure that the toric Calabi-Yau 3-fold $\mathcal{X}$ is $K$-stable.
More generally, a test symmetry $\eta$ can lead to a Futaki invariant that is $F=0$, which may imply that $\eta$ is trivial for the particular toric Calabi-Yau 3-fold $\mathcal{X}$. 
In order to make sure that all non-trivial test symmetries $\eta$ are covered for K-stability of $\mathcal{X}$, we define
\begin{definition}
The norm for a test symmetry $\eta$ is defined as follows \cite{Collins:2016icw, Bao:2020ugf}, 
\beal{es02a95}
|| \eta ||^2 = \left\{ \ba{ll}
0, & I_0 \simeq I_{\lambda \neq 0}\\
C_0 - \frac{B_0^2}{A_0} & \text{otherwise}
\ea
\right.
~,~
\eea
where 
\beal{es02a96}
B_0 = -\frac{1}{n} ~\text{D}_\epsilon A_0(\zeta+\epsilon \eta) ~\bigg|_{\epsilon=0} ~,~
C_0  = \frac{1}{n(n+1)} ~\text{D}_\epsilon^2 A_0(\zeta+\epsilon \eta) ~\bigg|_{\epsilon = 0 } ~.~
\eea
\end{definition}
Here, $I_\lambda$ refers to the modified ideal in \eref{es02a63} under a test symmetry $\eta$.

Based on the definition of the norm $||\eta||^2$ in \eref{es02a95}, we can define $K$-stability as follows, 
\begin{definition}
Given $\mathcal{X}$ with symmetry $\zeta$, 
it is 
K-semistable if for any test symmetry $\eta$ the corresponding Futaki invariant $F(\mathcal{X}; \zeta, \eta) \geq 0$.
A K-semistable $\mathcal{X}$ is K-stable if $F(\mathcal{X}; \zeta, \eta)=0$ only when the norm of the test symmetry $|| \eta ||^2$ also vanishes. 
\end{definition}
In the following work, 
affine cone over $X$ is a toric Calabi-Yau 3-fold $\mathcal{X}$ whose toric diagram is one of the 16 reflexive polygons in \fref{fig_reflexive}.
The associated abelian
$4d$ $\mathcal{N}=1$ supersymmetric gauge theories have a mesonic moduli space $\mathcal{M}^{mes}$
which is given by $\mathcal{X}$.
Accordingly, when $\mathcal{X}$ is K-stable under the definition above, 
we call the corresponding mesonic moduli space $\mathcal{M}^{mes}$ to be K-stable. 

Given that the affine cone over $X$ is a toric Calabi-Yau 3-fold $\mathcal{X}$, 
we expect $\mathcal{X}$ to be always K-stable \cite{collins2019sasaki}.
In the following work, we focus on the actual values of the Futaki invariants $F(\mathcal{X}; \zeta, \eta_h)$ for test symmetries $\eta_h$ that affects individual generators $x_h$ of the mesonic moduli space $\mathcal{M}^{mes}$ as in \eref{es02a76}. 
By focusing on $4d$ $\mathcal{N}=1$ supersymmetric gauge theories with $U(1)$ gauge groups, whose mesonic moduli spaces $\mathcal{M}^{mes}$ are toric Calabi-Yau 3-folds associated to the 16 reflexive polygons in $\mathbb{Z}^2$ as summarized in \fref{fig_reflexive} and studied in \cite{Hanany:2012hi}, 
we discover that the values of the Futaki invariants $F(\mathcal{X}; \zeta, \eta_h)$ exhibit particularly interesting distributions that satisfy bounds parameterized by geometrical and topological invariants of the toric Calabi-Yau 3-folds $\mathcal{X}$ such as the minimum volume $V_{min}$ of the Sasaki-Einstein base manifold or volumes of divisors $V(b_i^*;\Sigma_\alpha)$ as discussed in sections \sref{minimizedvolandtopinv} and \sref{divisorvolumes}, respectively.

Before summarizing these discoveries in section \sref{Futforref}, we first review the computation of Futaki invariants $F(\mathcal{X}; \zeta, \eta_h)$ for test symmetries $\eta_h$ in \eref{es02a76} for a $4d$ $\mathcal{N}=1$ supersymmetric gauge theory corresponding to the toric Calabi-Yau cone over $L_{1,3,1}/\mathbb{Z}_2$ with orbifold action $(0,1,1,1)$ \cite{Hanany:2012hi}.
\\

\subsection{An Example: the $L_{1, 3, 1}/\mathbb{Z}_{2}$ $(0,1,1,1)$ Model }\label{example}

\begin{figure}[H]
    \centering
    \resizebox{0.3\hsize}{!}{\includegraphics[width=1\linewidth]{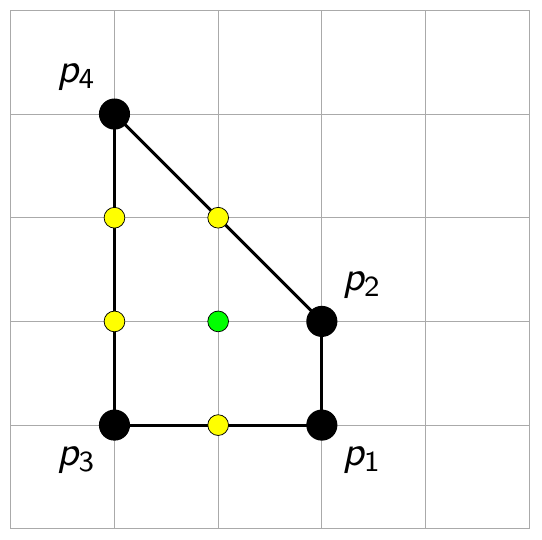}}
    \caption{The toric diagram for $L_{1, 3, 1}/\mathbb{Z}_{2}$ with orbifold action $(0,1,1,1)$ \cite{Hanany:2012hi}.}\label{fig_toric3a}
\end{figure}

\begin{figure}[H]
    \centering
    \resizebox{0.45\hsize}{!}{\includegraphics[width=1\linewidth]{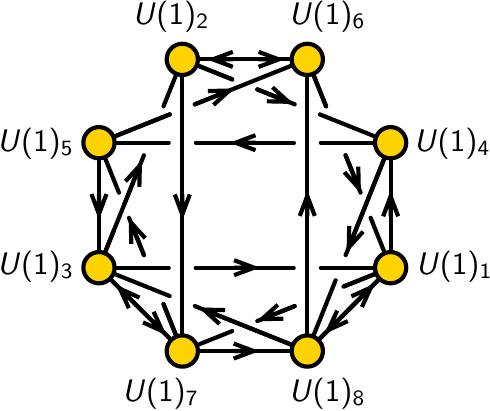}}
    \caption{The quiver for the $4d$ $\mathcal{N}=1$ supersymmetric gauge theory (phase a) corresponding to $L_{1, 3, 1}/\mathbb{Z}_{2}$ with orbifold action $(0,1,1,1)$ \cite{Hanany:2012hi}.}\label{fig_quiver3a}
\end{figure}

In the following section, we consider the $4d$ $\mathcal{N}=1$ supersymmetric gauge theory corresponding to $L_{1, 3, 1}/\mathbb{Z}_{2}$ with orbifold action $(0,1,1,1)$ \cite{Hanany:2012hi}.
The associated toric diagram is given in \fref{fig_toric3a}. 
The $4d$ $\mathcal{N}=1$ supersymmetric gauge theory is realized in terms of a brane tiling \cite{Franco:2005rj, Franco:2005sm, Hanany:2005ve} and has two two Seiberg dual phases which were both studied in detail in \cite{Hanany:2012hi}.
Here, we consider the first Seiberg dual phase, known as Model 3a in \cite{Hanany:2012hi}, whose superpotential takes the following form, 
\beal{es02a50}
W &=& X_{31} X_{18} X_{83} + X_{32} X_{27} X_{73} + X_{53} X_{37} X_{75}+ X_{78} X_{81} X_{17} 
\nn\\
&&
+X_{14} X_{45} X_{56}X_{61} + X_{62} X_{24} X_{48} X_{86}
\nn\\
&&
- X_{14} X_{48} X_{81} - X_{31} X_{17} X_{73} - X_{78} X_{83} X_{37} - X_{86} X_{61} X_{18}
\nn\\
&&
- X_{32} X_{24} X_{45} X_{53} - X_{62} X_{27} X_{75} X_{56}
~.~
\eea
The corresponding quiver diagram is shown in \fref{fig_quiver3a}.

Each of the extremal vertices in the toric diagram in \fref{fig_toric3a} are associated to GLSM fields $p_\alpha$. 
These GLSM fields can be used to parameterize the mesonic moduli space of the $4d$ $\mathcal{N}=1$ supersymmetric gauge theory.
\tref{tab_glsm} summarizes the GLSM fields $p_\alpha$ with their corresponding $U(1)_R$ charges as calculated in \cite{Hanany:2012hi}.

\begin{table}[H]
\centering
\begin{tabular}{|c|c|c|}
\hline
GLSM field & $U(1)_R$ charge & fugacity
\\
\hline
$p_1$ & $r_1 = \frac{1}{6} (5 - \sqrt{7})$ & $t_1 $
\\
$p_2$ & $r_2 = \frac{1}{6} (5 - \sqrt{7})$ & $t_2 $
\\
$p_3$ & $r_3 = \frac{1}{6} (1 + \sqrt{7})$ & $t_3$
\\
$p_4$ & $r_4 = \frac{1}{6} (1 + \sqrt{7})$ & $t_4$
\\
\hline
\end{tabular}
\caption{The GLSM fields $p_\alpha$ associated to extremal vertices of the toric diagram of $L_{1, 3, 1}/\mathbb{Z}_{2}$ with orbifold action $(0,1,1,1)$, with the corresponding $U(1)_R$ charges and fugacities $t_\alpha$ in the refined Hilbert series \cite{Hanany:2012hi}.}
\label{tab_glsm}
\end{table}

Using the refinement in terms of fugacities $t_\alpha$ associated to GLSM fields $p_\alpha$, the Hilbert series of the mesonic moduli space 
takes the following form \cite{Hanany:2012hi},  
\beal{es03a01}
g(t_{\alpha}; \mathcal{X})  = \frac{(1 - t_{1}^{2} t_{2}^{2} t_{3}^{2} t_{4}^{2})(1 - t_{1} t_{2} t_{3}^{3} t_{4}^{3})}{(1 - t_{1}^{2} t_{2}^{2})(1 - t_{1} t_{3}^{3})(1 - t_{3}^{2} t_{4}^{2})(1 - t_{2}t_{4}^{3})(1 - t_{1} t_{2} t_{3} t_{4})} 
~.~
\eea
The plethystics logarithm of the refined Hilbert series in \eref{es02a51} takes the following form, 
\beal{es02a82}
\text{PL}[g(t_{i}, \mathcal{X})] = t_{1}^{2} t_{2}^{2} + t_{1} t_{3}^{3} + t_{1} t_{2} t_{3} t_{4} + t_{3}^{2} t_{4}^{2} + t_{2} t_{4}^{3} - t_{1}^{2} t_{2}^{2} t_{3}^{2} t_{4}^{2} - t_{1} t_{2} t_{3}^{3} t_{4}^{3},
\eea
where we see that the expansion is finite, indicating that the mesonic moduli space here is a complete intersection.

The positive terms in the plethystic logarithm in \eref{es02a82} correspond to the generators of the mesonic moduli space, which are summarized in terms of the GLSM fields and their corresponding $U(1)_R$ charges in \tref{tab_generators}.
These generators form two binomial relations of the following form, 
\beal{es03a02}
x_1 x_4 = x_3^2 ~,~
x_3 x_4 = x_2 x_5~,~
\eea
which correspond to the two negative terms in the plethystic logarithm in \eref{es02a82}.
Accordingly, the mesonic moduli space of the $4d$ $\mathcal{N}=1$ supersymmetric gauge theory corresponding to $L_{1, 3, 1}/\mathbb{Z}_{2}$ $(0,1,1,1)$ can be expressed as, 
\beal{es03a03}
\mathcal{M}^{mes} = \text{Spec}~ 
\mathbb{C}[x_1,x_2,x_3,x_4,x_5] / \langle x_1 x_4 - x_3^2, x_3 x_4 - x_2 x_5\rangle ~.~
\eea

\paragraph{Grading under $\zeta_R$.}
If we take $\zeta=\zeta_R$ to be the $U(1)_R$ symmetry, we have the following fugacity assignment on the GLSM fields, 
\beal{es03a05}
t_{1} = t^{r_{1}}, \; t_{2} = t^{r_{2}}, \; t_{3} = t^{r_{3}}, \; t_{4} = t^{r_{4}},
\eea
where $r_\alpha$ denotes the $U(1)_R$ charge of the corresponding GLSM field $p_\alpha$, 
\beal{es03a06}
r_{1} = r_{2} = \frac{1}{6}(5 - \sqrt{ 7 }), \; 
r_{3} = r_{4} = \frac{1}{6}(1 + \sqrt{ 7 })~.~
\eea
Accordingly, the resulting Hilbert series $g(t; \mathcal{X}, \zeta_R)$ has the following Laurent expansion,
\beal{es03a07}
g(t=\text{e}^{-s}; \mathcal{X}, \zeta_R) = \frac{7 \sqrt{7}-10}{18 s^3} + \frac{7 \sqrt{7}-10}{18 s^2} + \frac{74+13 \sqrt{7}}{216 s} + \dots
~,~
\eea
where the coefficients are,
\beal{es03a08}
A_{0}(\zeta_R) = \frac{7 \sqrt{7}-10}{36}
~,~
A_{1}(\zeta_R) = \frac{7 \sqrt{7}-10}{18}
~.~
\eea

\begin{table}[H]
\centering
\begin{tabular}{|c|c|c|c|}
\hline
generator ($x_h$) & $U(1)_R$ charge ($R_h$) & GLSM fields ($d_h$) & fugacity
\\
\hline 
$p_1^2 p_2^2$ & 
$\frac{2}{3}(5-\sqrt{7})$ &
4 &
$t_{1}^{2} t_{2}^{2}$
\\
$p_1 p_3^3$ &
$\frac{1}{3}(4+\sqrt{7}) $ &
4 &
$t_{1} t_{3}^{3}$
\\
$p_1 p_2 p_3 p_4$ &
$2$ &
4 &
$t_{1} t_{2} t_{3} t_{4}$
\\
$p_3^2 p_4^2$ &
$\frac{2}{3}(1+\sqrt{7})$ &
4 &
$t_{3}^2 t_{4}^2$
\\
$p_2 p_4^3$ &
$\frac{1}{3}(4+\sqrt{7})$ &
4 &
$t_{2} t_{4}^{3}$
\\
\hline
\end{tabular}
\caption{
The generators $x_m$ for the mesonic moduli space of the $L_{1, 3, 1}/\mathbb{Z}_{2}$ $(0,1,1,1)$ (phase a) model with their corresponding $U(1)_R$ charges ($R_h$) and degrees in GLSM fields ($d_h$) \cite{Hanany:2012hi}.
}
\label{tab_generators}
\end{table}

We can now introduce test symmetries that adjust the grading on the generators in the Hilbert series under the fugacity $t$.
As introduced in \eref{es02a76}, the test symmetry takes the form $\eta_h = (\delta_{h,1}, \delta_{h,2}, \delta_{h,3}, \delta_{h,4}, \delta_{h,5})$ such that only the $h$-th generator is affected by the test symmetry and the grading is under  $\zeta_R + \epsilon \eta_h$.
Accordingly, using the resulting Hilbert series $g(t; \mathcal{X}, \zeta_R + \epsilon \eta_h)$ and the formula for the Futaki invariants $F(\mathcal{X}; \zeta_R, \eta_h)$ in \eref{es02a78}, we obtain
\beal{es03a10}
&
F_{R,1} = \frac{101 \sqrt{ 7 } - 179}{1296} ~,~
F_{R,2} = \frac{88 - 13\sqrt{ 7 }}{648}~,~ 
&
\nn\\
&
F_{R,3} = \frac{7\sqrt{ 7 } - 10}{108} ~,~
F_{R,4} = \frac{59 \sqrt{ 7 } - 119}{432} ~,~
F_{R,5} = \frac{88 - 13\sqrt{ 7 }}{648} ~.~
&
\eea
For a general test symmetry $\eta_{(w_{1}, w_{2}, w_{3}, w_{4}, w_{5})}$ with weights $w_m \geq 0$, the resulting Futaki invariant, which we call $F_R$, is a linear combination of the invariants in \eref{es03a10} as follows, 
\beal{es03a11}
F_R = \frac{101 \sqrt{ 7 } - 179}{1296} w_{1} + \frac{88 - 13\sqrt{ 7 }}{648} (w_{2} + w_{5}) + \frac{7\sqrt{ 7 } - 10}{108} w_{3} + \frac{59 \sqrt{ 7 } - 119}{432} w_{4}
~.~
\nn\\
\eea

\paragraph{Grading under $\zeta_p$.}
We can associate to $\zeta$ a $U(1)$ whose grading on the generators of the mesonic moduli space $\mathcal{M}^{mes}$ are given by the degree $d_h$ in GLSM fields $p_\alpha$ as summarized in \tref{tab_generators}.
Under $\zeta= \zeta_p$, the fugacities $t_\alpha$ corresponding to $p_\alpha$ are set to  
$t_\alpha = \overline{t}$ 
such that they count the degree in GLSM fields.
Under this unrefinement, the Hilbert series in \eref{es03a01} takes the following form
\beal{es03a16}
g(\overline{t}; \mathcal{X}, \zeta_p) = 
\frac{
(1 - \overline{t}^8 )^2
}{
(1 - \overline{t}^4 )^5
}
~.~
\eea
The corresponding Laurent expansion takes the following form, 
\beal{es03a17}
g(\overline{t}= e^{-s}; \mathcal{X}, \zeta_p) = 
\frac{1}{16 s^3} 
+ \frac{1}{8 s^2} 
+ \frac{1}{4 s}
+ \dots ~,~
\eea
where the leading order coefficients are, 
\beal{es03a17b}
A_0(\zeta_p) = \frac{1}{32}
~,~
A_1(\zeta_p) = \frac{1}{8}
~.~
\eea

By introducing a test symmetry of the form $\eta_h = (\delta_{h,1},\delta_{h,2},\delta_{h,3},\delta_{h,4},\delta_{h,5})$, 
the Futaki invariant $F_{p,h} = F(\mathcal{X}; \zeta_p, \eta_h)$ corresponding to the $h$-th generator of the mesonic moduli space $\mathcal{M}^{mes}$ takes the following form, 
\beal{es03a18}
F_{p,h} = \frac{1}{96}
~,~
\eea
where here $h=1, \dots, 5$. 
We can see that all generators $x_h$ of the mesonic moduli space $\mathcal{M}^{mes}$ of the $L_{1, 3, 1}/\mathbb{Z}_{2}$ $(0,1,1,1)$ (phase a) model have the same Futaki invariant under $\eta_h$ and $\zeta_p$.
For a general test symmetry of the form $\eta_{(w_{1}, w_{2}, w_{3}, w_{4}, w_{5})}$ with weights $w_m \geq 0$, the Futaki invariant is a linear combination of the invariants in \eref{es03a18} as follows, 
\beal{es03a18}
F_p = \frac{1}{96}(w_{1} + w_{2} + w_{3} + w_{4} + w_{5})
~.~
\eea
\\

\section{Futaki Invariants for Reflexive Polygons \label{Futforref}} \setall

In this section, we summarize the calculated Futaki invariants of the form $F(\mathcal{X}, \zeta_R, \eta)$ and $F(\mathcal{X}, \zeta_p, \eta)$
for toric Calabi-Yau 3-folds $\mathcal{X}$ associated to the 16 reflexive polygons in \fref{fig_reflexive}.
These are presented in Tables \ref{FRtableA} to \ref{FPtableB}.
Here, we notice that the Futaki invariants of the form $F(\mathcal{X}, \zeta_R, \eta)$ based on the $U(1)_R$ symmetry for $\zeta=\zeta_R$ are based on leading coefficients in the Laurent expansion of the Hilbert series that satisfy $A_1(\zeta_R)=2A_0(\zeta_R)$. This is not necessarily true for $F_p$ where we use the GLSM field degree as a grading under the $U(1)$ for $\zeta=\zeta_p$. Throughout, we shall always consider test symmetries of the form $\eta=\eta_h$ as defined in \eref{es02a65}.

\begin{table}[H]
    \centering
    \tiny
    \makebox[\textwidth][l]{\hspace*{-0.65cm}
    \resizebox{1.07\hsize}{!}{
    \begin{tabular}{|>{\centering\arraybackslash}p{1.5cm}|p{6.5cm}|p{1.5cm}|p{1.5cm}|p{2cm}|p{2.3cm}|p{2.5cm}|}
    \hline
    Model $a$
    & Global Symmetry 
    & Generators 
    & $t^R$
    & $A_0$
    & $F(\mathcal{X}_a, \zeta_R, \eta_h)$
    & $\|\eta\|^2_R$
    \\
    \hline
    Model 1
    &
        \hspace{-0.24cm}
        \begin{tabular}{c|c|c|c|c}
            & $U(1)_{f_1}$ 
            & $U(1)_{f_2}$ 
            & $U(1)_R$ 
            & fugacity
            \\
            \hline
            $p_1$ 
            & 1/3 
            & 0 
            & 2/3 
            & $t_1$
            \\
            $p_2$ 
            & -1/3 
            & -1/3 
            & 2/3 
            & $t_2$
            \\
            $p_3$ 
            & 0 
            & 1/3 
            & 2/3 
            & $t_3$
        \end{tabular}
    & 
        \begin{tabular}{@{}l}
        	$p_1^3$\\
        	$p_2^3$\\
        	$p_1 p_2 p_3$\\
        	$p_3^3$
        \end{tabular}
    & 
        \begin{tabular}{@{}l}
        	$t^2$\\
        	$t^2$\\
        	$t^2$\\
        	$t^2$
        \end{tabular}
    &  
        $3/16$
    & 
        \begin{tabular}{@{}l}
        	$1/16$\\
        	$1/16$\\
        	$1/16$\\
        	$1/16$
        \end{tabular}
    &
        \begin{tabular}{@{}l}
        	$1/384$\\
        	$1/384$\\
        	$1/384$\\
        	$1/384$
        \end{tabular}
    \\
    \hline
    Model 2
    &
        \hspace{-0.24cm}
        \begin{tabular}{c|c|c|c|c}
            & $U(1)_{f_1}$ 
            & $U(1)_{f_2}$ 
            & $U(1)_R$ 
            & fugacity
            \\
            \hline
            $p_1$ 
            & -1/4 
            & 1/4 
            & 2/3 
            & $t_1$
            \\
            $p_2$ 
            & -1/4 
            & -1/4 
            & 2/3 
            & $t_2$
            \\
            $p_3$ 
            & 1/2 
            & 0 
            & 2/3 
            & $t_3$
        \end{tabular}
    &
        \begin{tabular}{@{}l}
        	$p_3^2$\\
        	$p_1 p_2 p_3$\\
        	$p_1^4$\\
        	$p_1^2 p_2^2$\\
        	$p_2^4$
        \end{tabular}
    &  
        \begin{tabular}{@{}l}
        	$t^{4/3}$\\
        	$t^2$\\
        	$t^{8/3}$\\
        	$t^{8/3}$\\
        	$t^{8/3}$
        \end{tabular}
    &  
        $27/128$
    & 
        \begin{tabular}{@{}l}
        	$27/512$\\
        	$9/128$\\
        	$81/1024$\\
        	$81/1024$\\
        	$81/1024$
        \end{tabular}
    &
        \begin{tabular}{@{}l}
        	$27/4096$\\
        	$3/1024$\\
        	$27/16384$\\
        	$27/16384$\\
        	$27/16384$
        \end{tabular}
    \\
    \hline
    Model 3
    &  
        \hspace{-0.24cm}
        \begin{tabular}{c|c|c|c|c}
            & $U(1)_{f_1}$ 
            & $U(1)_{f_2}$ 
            & $U(1)_R$ 
            & fugacity
            \\      
            \hline
            $p_1$ 
            & 1/2 
            & 1/2 
            & $(5-\sqrt{7})/6$ 
            & $t_1$
            \\
            $p_2$ 
            & 0 
            & -1/2 
            & $(5-\sqrt{7})/6$ 
            & $t_2$
            \\
            $p_3$ 
            & -1/2 
            & -1/2 
            & $(1+\sqrt{7})/6$ 
            & $t_3$
            \\
            $p_4$ 
            & 0 
            & 1/2 
            & $(1+\sqrt{7})/6$ 
            & $t_4$
        \end{tabular}
    & 
        \begin{tabular}{@{}l}
        	$p_1^2 p_2^2$\\
        	$p_1 p_3^3$\\
        	$p_1 p_2 p_3 p_4$\\
        	$p_3^2 p_4^2$\\
        	$p_2 p_4^3$
        \end{tabular}
    &  
        \begin{tabular}{@{}l}
        	$t^{\frac{1}{3} \left(10-2 \sqrt{7}\right)}$\\
        	$t^{\frac{1}{3} \left(4+\sqrt{7}\right)}$\\
        	$t^2$\\
        	$t^{\frac{1}{3} \left(2+2 \sqrt{7}\right)}$\\
        	$t^{\frac{1}{3} \left(4+\sqrt{7}\right)}$
        \end{tabular}
    &  
        $\displaystyle \frac{7 \sqrt{7}-10}{36}$
    & 
        \begin{tabular}{@{}l}
        	$\frac{101 \sqrt{7}-179}{1296}$\\
        	$\frac{88-13 \sqrt{7}}{648}$\\
        	$\frac{7 \sqrt{7}-10}{108}$\\
        	$\frac{59 \sqrt{7}-119}{432}$\\
        	$\frac{88-13 \sqrt{7}}{648}$
        \end{tabular}
    &
        \begin{tabular}{@{}l}
        	$\frac{85+62 \sqrt{7}}{46656}$\\
        	$\frac{241 \sqrt{7}-622}{5832}$\\
        	$\frac{7 \sqrt{7}-10}{2592}$\\
        	$\frac{38 \sqrt{7}-89}{5184}$\\
        	$\frac{241 \sqrt{7}-622}{5832}$
        \end{tabular}
    \\
    \hline
    Model 4
    &
        \hspace{-0.24cm}
        \begin{tabular}{c|c|c|c|c}
            & $U(1)_{f_1}$ 
            & $U(1)_{f_2}$ 
            & $U(1)_R$ 
            & fugacity
            \\
            \hline
            $p_1$ 
            & 1/4
            & -1/4 
            & $1/2$ 
            & $t_1$
            \\
            $p_2$ 
            & 1/4
            & 1/4 
            & $1/2$ 
            & $t_2$
            \\
            $p_3$ 
            & -1/4 
            & -1/4 
            & $1/2$ 
            & $t_3$
            \\
            $p_4$ 
            & -1/4
            & 1/4
            & $1/2$ 
            & $t_4$
        \end{tabular}
    & 
        \begin{tabular}{@{}l}
        	$p_1^2 p_2^2$\\
        	$p_1^2 p_3^2$\\
        	$p_1 p_2 p_3 p_4$\\
        	$p_2^2 p_4^2$\\
        	$p_3^2 p_4^2$
        \end{tabular}
    &  
        \begin{tabular}{@{}l}
        	$t^2$\\
        	$t^2$\\
        	$t^2$\\
        	$t^2$\\
        	$t^2$
        \end{tabular}
    &  
        $1/4$
    & 
        \begin{tabular}{@{}l}
        	$1/12$\\
        	$1/12$\\
        	$1/12$\\
        	$1/12$\\
        	$1/12$
        \end{tabular}
    &
        \begin{tabular}{@{}l}
        	$1/288$\\
        	$1/288$\\
        	$1/288$\\
        	$1/288$\\
        	$1/288$
        \end{tabular}
    \\
    \hline
    Model 5
    &
        \hspace{-0.24cm}
        \begin{tabular}{c|c|c|c|c}
            & $U(1)_{f_1}$ 
            & $U(1)_{f_2}$ 
            & $U(1)_R$ 
            & fugacity
            \\
            \hline
            $p_1$ 
            & 0
            & -1/2 
            & $r_{5, 1}$ 
            & $t_1$
            \\
            $p_2$ 
            & 0
            & 1/2 
            & $r_{5, 2}$ 
            & $t_2$
            \\
            $p_3$ 
            & -1 
            & -1 
            & $r_{5, 3}$ 
            & $t_3$
            \\
            
            $p_4$ 
            & 1
            & 1
            & $r_{5, 4}$ 
            & $t_4$
        \end{tabular}
    & 
        \begin{tabular}{@{}l}
        	$p_1^2 p_4$\\
        	$p_1 p_2^3$\\
        	$p_1 p_2 p_3 p_4$\\
        	$p_2^4 p_3$\\
        	$p_2^2 p_3^2 p_4$\\
        	$p_3^3 p_4^2$
        \end{tabular}
    &  
        \begin{tabular}{@{}l}
        	$t^{R_{5, 1}}$\\
        	$t^{R_{5, 2}}$\\
        	$t^{R_{5, 3}}$\\
        	$t^{R_{5, 4}}$\\
        	$t^{R_{5, 5}}$\\
        	$t^{R_{5, 6}}$
        \end{tabular}
    &  
        $A_{5, 0}$
    & 
        \begin{tabular}{@{}l}
        	$F_{5, 1}$\\
        	$F_{5, 2}$\\
        	$F_{5, 3}$\\
        	$F_{5, 4}$\\
        	$F_{5, 5}$\\
        	$F_{5, 6}$
        \end{tabular}
    &
        \begin{tabular}{@{}l}
        	$\eta_{5, 1}$\\
        	$\eta_{5, 2}$\\
        	$\eta_{5, 3}$\\
        	$\eta_{5, 4}$\\
        	$\eta_{5, 5}$\\
        	$\eta_{5, 6}$
        \end{tabular}
    \\
    \hline
    Model 6
    &  
        \hspace{-0.24cm}
        \begin{tabular}{c|c|c|c|c}
            & $U(1)_{f_1}$ 
            & $U(1)_{f_2}$ 
            & $U(1)_R$ 
            & fugacity
            \\
            \hline
            $p_1$ 
            & -1
            & 0 
            & $r_{6, 1}$ 
            & $t_1$
            \\
            $p_2$ 
            & 1
            & 0
            & $r_{6, 2}$ 
            & $t_2$
            \\
            $p_3$ 
            & 0
            & 0
            & $r_{6, 3}$ 
            & $t_3$
            \\
            $p_4$ 
            & 0
            & 1 
            & $r_{6, 4}$ 
            & $t_4$
            \\
            $p_5$ 
            & 0 
            & -1 
            & $r_{6, 5}$ 
            & $t_5$
        \end{tabular}
    & 
        \begin{tabular}{@{}l}
        	$p_1^2 p_2 p_3^2$\\
        	$p_1^2 p_2^2 p_4$\\
        	$p_1 p_3^3 p_5$\\
        	$p_1 p_2 p_3 p_4 p_5$\\
        	$p_3^2 p_4 p_5^2$\\
        	$p_2 p_4^2 p_5^2$
        \end{tabular}
    &  
        \begin{tabular}{@{}l}
        	$t^{R_{6, 1}}$\\
        	$t^{R_{6, 2}}$\\
        	$t^{R_{6, 3}}$\\
        	$t^{R_{6, 4}}$\\
        	$t^{R_{6, 5}}$\\
        	$t^{R_{6, 6}}$
        \end{tabular}
    &  
        $A_{6, 0}$
    & 
        \begin{tabular}{@{}l}
        	$F_{6, 1}$\\
        	$F_{6, 2}$\\
        	$F_{6, 3}$\\
        	$F_{6, 4}$\\
        	$F_{6, 5}$\\
        	$F_{6, 6}$
        \end{tabular}
    &
        \begin{tabular}{@{}l}
        	$\eta_{6, 1}$\\
        	$\eta_{6, 2}$\\
        	$\eta_{6, 3}$\\
        	$\eta_{6, 4}$\\
        	$\eta_{6, 5}$\\
        	$\eta_{6, 6}$
        \end{tabular}
    \\
    \hline
        Model 7
    &  
        \hspace{-0.24cm}
        \begin{tabular}{c|c|c|c|c}
            & $U(1)_{f_1}$ 
            & $U(1)_{f_2}$ 
            & $U(1)_R$ 
            & fugacity
            \\
            \hline
            $p_1$ 
            & 1/2
            & 0 
            & $2/3$ 
            & $t_1$
            \\
            $p_2$ 
            & -1/6
            & 1/3
            & $2/3$ 
            & $t_2$
            \\
            $p_3$ 
            & -1/3
            & -1/3
            & $2/3$ 
            & $t_3$
        \end{tabular}
    & 
        \begin{tabular}{@{}l}
        	$p_1^2$\\
        	$p_1 p_2 p_3$\\
        	$p_3^3$\\
        	$p_1 p_2^3$\\
        	$p_2^2 p_3^2$\\
        	$p_2^4 p_3$\\
        	$p_2^6$
        \end{tabular}
    &  
        \begin{tabular}{@{}l}
        	$t^{4/3}$\\
        	$t^2$\\
        	$t^2$\\
        	$t^{8/3}$\\
        	$t^{8/3}$\\
        	$t^{10/3}$\\
        	$t^4$
        \end{tabular}
    &  
        $9/32$
    & 
        \begin{tabular}{@{}l}
        	$9/128$\\
        	$3/32$\\
        	$3/32$\\
        	$27/256$\\
        	$27/256$\\
        	$9/80$\\
        	$15/128$
        \end{tabular}
    &
        \begin{tabular}{@{}l}
        	$9/1024$\\
        	$1/256$\\
        	$1/256$\\
        	$9/4096$\\
        	$9/4096$\\
        	$9/6400$\\
        	$1/1024$
        \end{tabular}
    \\
    \hline
        Model 8
    &
        \hspace{-0.24cm}
        \begin{tabular}{c|c|c|c|c}
            & $U(1)_{f_1}$ 
            & $U(1)_{f_2}$ 
            & $U(1)_R$ 
            & fugacity
            \\
            \hline
            $p_1$ 
            & 1
            & 0 
            & $\frac{1}{\sqrt{3}}$
            & $t_1$
            \\
            $p_2$ 
            & -1/2
            & 1/2
            & $\frac{1}{\sqrt{3}}$
            & $t_2$
            \\
            $p_3$ 
            & -1
            & 0
            & $1-\frac{1}{\sqrt{3}}$
            & $t_3$
            \\
            $p_4$ 
            & 1/2
            & -1/2
            & $1-\frac{1}{\sqrt{3}}$
            & $t_4$
        \end{tabular}
    & 
        \begin{tabular}{@{}l}
        	$p_1^2 p_3$\\
        	$p_1^2 p_2^2$\\
        	$p_1 p_2 p_3 p_4$\\
        	$p_3^2 p_4^2$\\
        	$p_1 p_2^3 p_4$\\
        	$p_2^2 p_3 p_4^2$\\
        	$p_2^4 p_4^2$
        \end{tabular}
    &  
        \begin{tabular}{@{}l}
        	$t^{\frac{1}{3} \left(3+\sqrt{3}\right)}$\\
        	$t^{\frac{4}{\sqrt{3}}}$\\
        	$t^2$\\
        	$t^{\frac{1}{3} \left(12-4 \sqrt{3}\right)}$\\
        	$t^{1+\sqrt{3}}$\\
        	$t^{\frac{1}{3} \left(9-\sqrt{3}\right)}$\\
        	$t^{\frac{1}{3} \left(6+2 \sqrt{3}\right)}$
        \end{tabular}
    &  
        $\displaystyle \frac{3 \sqrt{3}}{16}$
    & 
        \begin{tabular}{@{}l}
        	$\frac{3}{32}$\\
        	$\frac{3(2 \sqrt{3}-1)}{64}$\\
        	$\frac{\sqrt{3}}{16}$\\
        	$\frac{3(3 \sqrt{3}-1)}{128}$\\
        	$\frac{4 \sqrt{3}-3}{32}$\\
        	$\frac{3(10 \sqrt{3}-1)}{416}$\\
        	$\frac{3(1+\sqrt{3})}{64}$
        \end{tabular}
    &
        \begin{tabular}{@{}l}
        	$\frac{2 \sqrt{3}-3}{64}$\\
        	$\frac{\sqrt{3}}{512}$\\
        	$\frac{1}{128 \sqrt{3}}$\\
        	$\frac{3+2 \sqrt{3}}{1024}$\\
        	$\frac{2 \sqrt{3}-3}{192}$\\
        	$\frac{9+14 \sqrt{3}}{10816}$\\
        	$\frac{2 \sqrt{3}-3}{256}$
        \end{tabular}
    \\
    \hline
    Model 9
    &  
        \hspace{-0.24cm}
        \begin{tabular}{c|c|c|c|c}
            & $U(1)_{f_1}$ 
            & $U(1)_{f_2}$ 
            & $U(1)_R$ 
            & fugacity
            \\
            \hline
            $p_1$ 
            & -2/5
            & 1/2
            & $2 \left(\sqrt{5}-2\right)$
            & $t_1$
            \\
            $p_2$ 
            & -1/5
            & -1/2
            & $2 \left(\sqrt{5}-2\right)$
            & $t_2$
            \\
            $p_3$ 
            & 2/5
            & 0
            & $2 \left(\sqrt{5}-2\right)$
            & $t_3$
            \\
            $p_4$ 
            & 1/5
            & 0
            & $7-3 \sqrt{5}$
            & $t_4$
            \\
            $p_5$ 
            & 0
            & 0
            & $7-3 \sqrt{5}$
            & $t_5$
        \end{tabular}
    & 
        \begin{tabular}{@{}l}
        	$p_3^2 p_4 p_5$\\
        	$p_1^2 p_3 p_4^2$\\
        	$p_1 p_2 p_3 p_4 p_5$\\
        	$p_2^2 p_3 p_5^2$\\
        	$p_1^3 p_2 p_4^2$\\
        	$p_1^2 p_2^2 p_4 p_5$\\
        	$p_1 p_2^3 p_5^2$
        \end{tabular}
    &  
        \begin{tabular}{@{}l}
        	$t^{17-7 \sqrt{5}}$\\
        	$t^{5 \sqrt{5}-9}$\\
        	$t^2$\\
        	$t^{13-5 \sqrt{5}}$\\
        	$t^{12 \sqrt{5}-24}$\\
        	$t^{7 \sqrt{5}-13}$\\
        	$t^{2 \sqrt{5}-2}$
        \end{tabular}
    &  
        $\displaystyle \frac{4119+1841 \sqrt{5}}{23232}$
    & 
        \begin{tabular}{@{}l}
        	$\frac{7 \left(4907+2192 \sqrt{5}\right)}{766656}$\\
        	$\frac{94379+42171 \sqrt{5}}{1533312}$\\
        	$\frac{4119+1841 \sqrt{5}}{69696}$\\
        	$\frac{7831+3499 \sqrt{5}}{139392}$\\
        	$\frac{56699+25337 \sqrt{5}}{836352}$\\
        	$\frac{87896+39277 \sqrt{5}}{1324224}$\\
        	$\frac{9026+4033 \sqrt{5}}{139392}$
        \end{tabular}
    &
        \begin{tabular}{@{}l}
        	$\frac{548792+245427 \sqrt{5}}{101198592}$\\
        	$\frac{419241+187489 \sqrt{5}}{202397184}$\\
        	$\frac{4119+1841 \sqrt{5}}{1672704}$\\
        	$\frac{54719+24471 \sqrt{5}}{18399744}$\\
        	$\frac{73891+33045 \sqrt{5}}{60217344}$\\
        	$\frac{422572+188979 \sqrt{5}}{301923072}$\\
        	$\frac{10781+4821 \sqrt{5}}{6690816}$
        \end{tabular}
    \\
     \hline
    Model 10
    &  
        \hspace{-0.24cm}
        \begin{tabular}{c|c|c|c|c}
            & $U(1)_{f_1}$ 
            & $U(1)_{f_2}$ 
            & $U(1)_R$ 
            & fugacity
            \\
            \hline
            $p_1$ 
            & -1
            & 0 
            & $1/3$
            & $t_1$
            \\
            $p_2$ 
            & -1
            & 1
            & $1/3$
            & $t_2$
            \\
            $p_3$ 
            & 1
            & 0
            & $1/3$
            & $t_3$
            \\
            $p_4$ 
            & 1
            & -1
            & $1/3$
            & $t_4$
            \\
            $p_5$ 
            & 0
            & 0
            & $1/3$
            & $t_5$
            \\
            $p_6$ 
            & 0
            & 0
            & $1/3$
            & $t_6$        
        \end{tabular}
    & 
        \begin{tabular}{@{}l}
        	$p_2^2 p_3^2 p_4 p_5$\\
        	$p_1 p_2 p_3^2 p_5^2$\\
        	$p_2^2 p_3 p_4^2 p_6$\\
        	$p_1 p_2 p_3 p_4 p_5 p_6$\\
        	$p_1^2 p_3 p_5^2 p_6$\\
        	$p_1 p_2 p_4^2 p_6^2$\\
        	$p_1^2 p_4 p_5 p_6^2$
        \end{tabular}
    &  
        \begin{tabular}{@{}l}
        	$t^2$\\
        	$t^2$\\
        	$t^2$\\
        	$t^2$\\
        	$t^2$\\
        	$t^2$\\
        	$t^2$
        \end{tabular}
    &  
        $3/8$
    & 
        \begin{tabular}{@{}l}
        	$1/8$\\
        	$1/8$\\
        	$1/8$\\
        	$1/8$\\
        	$1/8$\\
        	$1/8$\\
        	$1/8$
        \end{tabular}
    &
        \begin{tabular}{@{}l}
        	$1/192$\\
        	$1/192$\\
        	$1/192$\\
        	$1/192$\\
        	$1/192$\\
        	$1/192$\\
        	$1/192$
        \end{tabular}
    \\
    \hline
       \end{tabular}
    }
    }
\caption{The Futaki invariants $F(\mathcal{X}_a, \zeta_R, \eta_h)$ for the toric Calabi-Yau 3-folds $\mathcal{X}_a$ corresponding to the 16 reflexive polygons in $\mathbb{Z}^2$. 
The extremal perfect matchings $p_\alpha$ and the generators in terms of $p_\alpha$ are shown with their global symmetry charges. 
Exact values of certain Futaki invariants are given in Appendix \sref{exactvalues}. \textbf{(Part 1/2)}
}\label{FRtableA}
\end{table}

\begin{table}[H]
    \centering
    \tiny
    \makebox[\textwidth][l]{\hspace*{-0.65cm}
    \resizebox{1.07\hsize}{!}{
    \begin{tabular}{|>{\centering\arraybackslash}p{1.5cm}|p{6.5cm}|p{1.5cm}|p{1.5cm}|p{2cm}|p{2.3cm}|p{2.5cm}|}
    \hline
    Model $a$
    & Global Symmetry
    & Generators 
    & $t^R$
    & $A_0$
    & $F(\mathcal{X}_a, \zeta_R, \eta_h)$
    & $\|\eta\|^2_R$
    \\
       \hline
    Model 11
    &  
        \hspace{-0.24cm}
        \begin{tabular}{c|c|c|c|c}
            & $U(1)_{f_1}$ 
            & $U(1)_{f_2}$ 
            & $U(1)_R$ 
            & fugacity
            \\
            \hline
            $p_1$ 
            & -1/4
            & -1/3
            & $r_{11, 1}$
            & $t_1$
            \\
            $p_2$ 
            & -1/4
            & 0
            & $r_{11, 2}$
            & $t_2$
            \\
            $p_3$ 
            & 0
            & 2/3
            & $r_{11, 3}$
            & $t_3$
            \\
            $p_4$ 
            & 1/2
            & -1/3
            & $r_{11, 4}$
            & $t_4$
        \end{tabular}
    & 
        \begin{tabular}{@{}l}
        	$p_1^2 p_4$\\
        	$p_3 p_4^2$\\
        	$p_1^3 p_2$\\
        	$p_1 p_2 p_3 p_4$\\
        	$p_1^2 p_2^2 p_3$\\
        	$p_2^2 p_3^2 p_4$\\
        	$p_1 p_2^3 p_3^2$\\
        	$p_2^4 p_3^3$
        \end{tabular}
    &  
        \begin{tabular}{@{}l}
        	$t^{R_{11, 1}}$\\
        	$t^{R_{11, 2}}$\\
        	$t^{R_{11, 3}}$\\
        	$t^{R_{11, 4}}$\\
        	$t^{R_{11, 5}}$\\
        	$t^{R_{11, 6}}$\\
        	$t^{R_{11, 7}}$\\
        	$t^{R_{11, 8}}$
        \end{tabular}
    &  
        $A_{11, 0}$
    & 
        \begin{tabular}{@{}l}
        	$F_{11, 1}$\\
        	$F_{11, 2}$\\
        	$F_{11, 3}$\\
        	$F_{11, 4}$\\
        	$F_{11, 5}$\\
        	$F_{11, 6}$\\
        	$F_{11, 7}$\\
        	$F_{11, 8}$
        \end{tabular}
    &
        \begin{tabular}{@{}l}
        	$\eta_{11, 1}$\\
        	$\eta_{11, 2}$\\
        	$\eta_{11, 3}$\\
        	$\eta_{11, 4}$\\
        	$\eta_{11, 5}$\\
        	$\eta_{11, 6}$\\
        	$\eta_{11, 7}$\\
        	$\eta_{11, 8}$
        \end{tabular}
    \\
    \hline
    Model 12
    &  
        \hspace{-0.24cm}
        \begin{tabular}{c|c|c|c|c}
            & $U(1)_{f_1}$ 
            & $U(1)_{f_2}$ 
            & $U(1)_R$ 
            & fugacity
            \\        
            \hline
            $p_1$ 
            & 1/2
            & 0
            & $\frac{1}{16} \left(5 \sqrt{33}-21\right)$
            & $t_1$
            \\
            $p_2$ 
            & -1/2
            & 0
            & $\frac{1}{16} \left(57-9 \sqrt{33}\right)$
            & $t_2$
            \\
            $p_3$ 
            & 0
            & -1/2
            & $\frac{1}{16} \left(57-9 \sqrt{33}\right)$
            & $t_3$
            \\
            $p_4$ 
            & 0
            & 1/2
            & $\frac{1}{16} \left(5 \sqrt{33}-21\right)$
            & $t_4$
            \\
            $p_5$ 
            & 0
            & 0
            & $\frac{1}{2} \left(\sqrt{33}-5\right)$
            & $t_5$
        \end{tabular}
    & 
        \begin{tabular}{@{}l}
        	$p_1^2 p_3 p_4$\\
        	$p_1 p_2 p_4^2$\\
        	$p_1^2 p_3^2 p_5$\\
        	$p_1 p_2 p_3 p_4 p_5$\\
        	$p_2^2 p_4^2 p_5$\\
        	$p_1 p_2 p_3^2 p_5^2$\\
        	$p_2^2 p_3 p_4 p_5^2$\\
        	$p_2^2 p_3^2 p_5^3$
        \end{tabular}
    &  
        \begin{tabular}{@{}l}
        	$t^{\frac{1}{2} \left(9-\sqrt{33}\right)}$\\
        	$t^{\frac{1}{2} \left(9-\sqrt{33}\right)}$\\
        	$t^2$\\
        	$t^2$\\
        	$t^2$\\
        	$t^{\frac{1}{2} \left(\sqrt{33}-1\right)}$\\
        	$t^{\frac{1}{2} \left(\sqrt{33}-1\right)}$\\
        	$t^{\sqrt{33}-3}$
        \end{tabular}
    &  
        $\displaystyle \frac{79+15 \sqrt{33}}{384}$
    & 
        \begin{tabular}{@{}l}
        	$\frac{819+163 \sqrt{33}}{13824}$\\
        	$\frac{819+163 \sqrt{33}}{13824}$\\
        	$\frac{79+15 \sqrt{33}}{1152}$\\
        	$\frac{79+15 \sqrt{33}}{1152}$\\
        	$\frac{79+15 \sqrt{33}}{1152}$\\
        	$\frac{661+133 \sqrt{33}}{9216}$\\
        	$\frac{661+133 \sqrt{33}}{9216}$\\
        	$\frac{66+13 \sqrt{33}}{864}$
        \end{tabular}
    &
        \begin{tabular}{@{}l}
        	$\frac{1493+261 \sqrt{33}}{331776}$\\
        	$\frac{1493+261 \sqrt{33}}{331776}$\\
        	$\frac{79+15 \sqrt{33}}{27648}$\\
        	$\frac{79+15 \sqrt{33}}{27648}$\\
        	$\frac{79+15 \sqrt{33}}{27648}$\\
        	$\frac{919+167 \sqrt{33}}{442368}$\\
        	$\frac{919+167 \sqrt{33}}{442368}$\\
        	$\frac{131+23 \sqrt{33}}{82944}$
        \end{tabular}
    \\
    \hline
    Model 13
    &  
        \hspace{-0.24cm}
        \begin{tabular}{c|c|c|c|c}
            & $U(1)_{f}$ 
            & $SU(2)_{x}$ 
            & $U(1)_R$ 
            & fugacity
            \\
            \hline
            $p_1$ 
            & -1/4
            & 1/2
            & $2/3$
            & $t_1$
            \\
            $p_2$ 
            & -1/4
            & -1/2
            & $2/3$
            & $t_2$
            \\
            $p_3$ 
            & 1/2
            & 0
            & $2/3$
            & $t_3$
        \end{tabular}
    & 
        \begin{tabular}{@{}l}
        	$p_3^2$\\
        	$p_1^2 p_3$\\
        	$p_1 p_2 p_3$\\
        	$p_2^2 p_3$\\
        	$p_1^4$\\
        	$p_1^3 p_2$\\
        	$p_1^2 p_2^2$\\
        	$p_1 p_2^3$\\
        	$p_2^4$
        \end{tabular}
    &  
        \begin{tabular}{@{}l}
        	$t^{4/3}$\\
        	$t^2$\\
        	$t^2$\\
        	$t^2$\\
        	$t^{8/3}$\\
        	$t^{8/3}$\\
        	$t^{8/3}$\\
        	$t^{8/3}$\\
        	$t^{8/3}$
        \end{tabular}
    & 
        $27/64$
    & 
        \begin{tabular}{@{}l}
        	$27/256$\\
        	$9/64$\\
        	$9/64$\\
        	$9/64$\\
        	$81/512$\\
        	$81/512$\\
        	$81/512$\\
        	$81/512$\\
        	$81/512$
        \end{tabular}
    &
        \begin{tabular}{@{}l}
        	$27/2048$\\
        	$3/512$\\
        	$3/512$\\
        	$3/512$\\
        	$27/8192$\\
        	$27/8192$\\
        	$27/8192$\\
        	$27/8192$\\
        	$27/8192$
        \end{tabular}
    \\
    \hline
    Model 14
    &  
        \hspace{-0.24cm}
        \begin{tabular}{c|c|c|c|c}
            & $U(1)_{f_1}$ 
            & $U(1)_{f_2}$ 
            & $U(1)_R$ 
            & fugacity
            \\
            \hline
            $p_1$ 
            & 1
            & 0
            & $\sqrt{13}-3$
            & $t_1$
            \\
            $p_2$ 
            & 1
            & 1
            & $\frac{1}{3} \left(5 \sqrt{13}-17\right)$
            & $t_2$
            \\
            $p_3$ 
            & -1
            & -1
            & $-\frac{4}{3} \left(\sqrt{13}-4\right)$
            & $t_3$
            \\
            $p_4$ 
            & -1
            & 0
            & $-\frac{4}{3} \left(\sqrt{13}-4\right)$
            & $t_4$            
        \end{tabular}
    & 
        \begin{tabular}{@{}l}
        	$p_1^2 p_3$\\
        	$p_1^2 p_4$\\
        	$p_1 p_2 p_3^2$\\
        	$p_1 p_2 p_3 p_4$\\
        	$p_1 p_2 p_4^2$\\
        	$p_2^2 p_3^3$\\
        	$p_2^2 p_3^2 p_4$\\
        	$p_2^2 p_3 p_4^2$\\
        	$p_2^2 p_4^3$
        \end{tabular}
    &  
        \begin{tabular}{@{}l}
        	$t^{\frac{1}{3} \left(11 \sqrt{13}-35\right)}$\\
        	$t^{\frac{1}{3} \left(2 \sqrt{13}-2\right)}$\\
        	$t^{3 \sqrt{13}-9}$\\
        	$t^2$\\
        	$t^{13-3 \sqrt{13}}$\\
        	$t^{\frac{1}{3} \left(7 \sqrt{13}-19\right)}$\\
        	$t^{\frac{1}{3} \left(14-2 \sqrt{13}\right)}$\\
        	$t^{\frac{1}{3} \left(47-11 \sqrt{13}\right)}$\\
        	$t^{\frac{1}{3} \left(80-20 \sqrt{13}\right)}$
        \end{tabular}
    &  
        $\displaystyle \frac{52763+14609 \sqrt{13}}{213440}$
    & 
        \begin{tabular}{@{}l}
        	$\frac{2622485+725129 \sqrt{13}}{37138560}$\\
        	$\frac{97619+26984 \sqrt{13}}{1280640}$\\
        	$\frac{150382+41593 \sqrt{13}}{1920960}$\\
        	$\frac{52763+14609 \sqrt{13}}{640320}$\\
        	$\frac{714961+197824 \sqrt{13}}{8324160}$\\
        	$\frac{7 \left(353527+97795 \sqrt{13}\right)}{29454720}$\\
        	$\frac{29135+8063 \sqrt{13}}{334080}$\\
        	$\frac{6104843+1689323 \sqrt{13}}{67873920}$\\
        	$\frac{1181921+327071 \sqrt{13}}{12806400}$
        \end{tabular}
    &
        \begin{tabular}{@{}l}
        	$\frac{73466741+20375873 \sqrt{13}}{12924218880}$\\
        	$\frac{279629+77513 \sqrt{13}}{61470720}$\\
        	$\frac{287536+79747 \sqrt{13}}{69154560}$\\
        	$\frac{52763+14609 \sqrt{13}}{15367680}$\\
        	$\frac{287536+79747 \sqrt{13}}{99889920}$\\
        	$\frac{25793849+7153685 \sqrt{13}}{8129502720}$\\
        	$\frac{741268+205555 \sqrt{13}}{276618240}$\\
        	$\frac{98980961+27452045 \sqrt{13}}{43167813120}$\\
        	$\frac{3049463+845765 \sqrt{13}}{1536768000}$
        \end{tabular}
    \\
    \hline
        Model 15
    &  
        \hspace{-0.24cm}
        \begin{tabular}{c|c|c|c|c}
            & $SU(2)_{x_1}$ 
            & $SU(1)_{x_2}$ 
            & $U(1)_R$ 
            & fugacity
            \\
            \hline
            $p_1$ 
            & 1/2
            & 0
            & $1/2$
            & $t_1$
            \\
            $p_2$ 
            & -1/2
            & 0
            & $1/2$
            & $t_2$
            \\
            
            $p_3$ 
            & 0
            & 1/2
            & $1/2$
            & $t_3$
            \\

            $p_4$ 
            & 0
            & -1/2
            & $1/2$
            & $t_4$            
        \end{tabular}
    & 
        \begin{tabular}{@{}l}
        	$p_1^2 p_3^2$\\
        	$p_1 p_2 p_3^2$\\
        	$p_2^2 p_3^2$\\
        	$p_1^2 p_3 p_4$\\
        	$p_1 p_2 p_3 p_4$\\
        	$p_2^2 p_3 p_4$\\
        	$p_1^2 p_4^2$\\
        	$p_1 p_2 p_4^2$\\
        	$p_2^2 p_4^2$
        \end{tabular}
    &  
        \begin{tabular}{@{}l}
        	$t^2$\\
        	$t^2$\\
        	$t^2$\\
        	$t^2$\\
        	$t^2$\\
        	$t^2$\\
        	$t^2$\\
        	$t^2$\\
        	$t^2$
        \end{tabular}
    &  
        $1/2$
    & 
        \begin{tabular}{@{}l}
        	$1/6$\\
        	$1/6$\\
        	$1/6$\\
        	$1/6$\\
        	$1/6$\\
        	$1/6$\\
        	$1/6$\\
        	$1/6$\\
        	$1/6$
        \end{tabular}
    &
        \begin{tabular}{@{}l}
        	$1/144$\\
        	$1/144$\\
        	$1/144$\\
        	$1/144$\\
        	$1/144$\\
        	$1/144$\\
        	$1/144$\\
        	$1/144$\\
        	$1/144$
        \end{tabular}
    \\
    \hline
    Model 16
    &  
        \hspace{-0.24cm}
        \begin{tabular}{c|c|c|c}
            & $SU(3)_{(x_1, x_2)}$ 
            & $U(1)_R$ 
            & fugacity
            \\
            
            \hline
            
            $p_1$ 
            & (-1/3, -1/3)
            & $2/3$
            & $t_1$
            \\
            
            $p_2$ 
            & (2/3, -1/3)
            & $2/3$
            & $t_2$
            \\
            
            $p_3$ 
            & (-1/3, 2/3)
            & $2/3$
            & $t_3$     
        \end{tabular}
    & 
        \begin{tabular}{@{}l}
        	$p_1^3$\\
        	$p_1^2 p_2$\\
        	$p_1 p_2^2$\\
        	$p_2^3$\\
        	$p_1^2 p_3$\\
        	$p_1 p_2 p_3$\\
        	$p_2^2 p_3$\\
        	$p_1 p_3^2$\\
        	$p_2 p_3^2$\\
        	$p_3^3$
        \end{tabular}
    &  
        \begin{tabular}{@{}l}
        	$t^2$\\
        	$t^2$\\
        	$t^2$\\
        	$t^2$\\
        	$t^2$\\
        	$t^2$\\
        	$t^2$\\
        	$t^2$\\
        	$t^2$\\
        	$t^2$
        \end{tabular}
    &  
        $9/16$
    & 
        \begin{tabular}{@{}l}
        	$3/16$\\
        	$3/16$\\
        	$3/16$\\
        	$3/16$\\
        	$3/16$\\
        	$3/16$\\
        	$3/16$\\
        	$3/16$\\
        	$3/16$\\
        	$3/16$
        \end{tabular}
    &
        \begin{tabular}{@{}l}
        	$1/128$\\
        	$1/128$\\
        	$1/128$\\
        	$1/128$\\
        	$1/128$\\
        	$1/128$\\
        	$1/128$\\
        	$1/128$\\
        	$1/128$\\
        	$1/128$
        \end{tabular}
    \\
    \hline
    \end{tabular}
    }
    }
\caption{The Futaki invariants $F(\mathcal{X}_a, \zeta_R, \eta_h)$ for the toric Calabi-Yau 3-folds $\mathcal{X}_a$ corresponding to the 16 reflexive polygons in $\mathbb{Z}^2$. 
The extremal perfect matchings $p_\alpha$ and the generators in terms of $p_\alpha$ are shown with their global symmetry charges. 
Exact values of certain Futaki invariants are given in Appendix \sref{exactvalues}. \textbf{(Part 2/2)}
}\label{FRtableB}
\end{table}

\begin{table}[H]
    \centering
    \tiny
    \makebox[\textwidth][l]{\hspace*{-0.65cm}
    \resizebox{1.07\hsize}{!}{
    \begin{tabular}{|>{\centering\arraybackslash}p{1.5cm}|p{6.5cm}|p{1.5cm}|p{1.5cm}|p{2cm}|p{2.3cm}|p{2.5cm}|}
    \hline
    Model $a$
    & Global Symmetry 
    & Generators 
    & $t_\alpha = \bar{t}$
    & $A_0, A_1$
    & $F(\mathcal{X}_a, \zeta_p, \eta_h)$
    & $\|\eta\|^2_p$
    \\
    \hline
    Model 1
    &
        \hspace{-0.24cm}
        \begin{tabular}{c|c|c|c|c}
            & $U(1)_{f_1}$ 
            & $U(1)_{f_2}$ 
            & $U(1)_R$ 
            & fugacity
            \\
            \hline
            $p_1$ 
            & 1/3 
            & 0 
            & 2/3 
            & $t_1$
            \\
            $p_2$ 
            & -1/3 
            & -1/3 
            & 2/3 
            & $t_2$
            \\
            $p_3$ 
            & 0 
            & 1/3 
            & 2/3 
            & $t_3$
        \end{tabular}
    & 
        \begin{tabular}{@{}l}
        	$p_1^3$\\
        	$p_2^3$\\
        	$p_1 p_2 p_3$\\
        	$p_3^3$
        \end{tabular}
    & 
        \begin{tabular}{@{}l}
        	$\bar{t}^3$\\
        	$\bar{t}^3$\\
        	$\bar{t}^3$\\
        	$\bar{t}^3$
        \end{tabular}
    &  
        \begin{tabular}{@{}l}
        	$1/18$\\
        	$1/6$
        \end{tabular}
    & 
        \begin{tabular}{@{}l}
        	$1/54$\\
        	$1/54$\\
        	$1/54$\\
        	$1/54$
        \end{tabular}
    &
        \begin{tabular}{@{}l}
        	$1/2916$\\
        	$1/2916$\\
        	$1/2916$\\
        	$1/2916$
        \end{tabular}
    \\
    \hline
    Model 2
    &
        \hspace{-0.24cm}
        \begin{tabular}{c|c|c|c|c}
            & $U(1)_{f_1}$ 
            & $U(1)_{f_2}$ 
            & $U(1)_R$ 
            & fugacity
            \\
            \hline
            $p_1$ 
            & -1/4 
            & 1/4 
            & 2/3 
            & $t_1$
            \\
            $p_2$ 
            & -1/4 
            & -1/4 
            & 2/3 
            & $t_2$
            \\
            $p_3$ 
            & 1/2 
            & 0 
            & 2/3 
            & $t_3$
        \end{tabular}
    &
        \begin{tabular}{@{}l}
        	$p_3^2$\\
        	$p_1 p_2 p_3$\\
        	$p_1^4$\\
        	$p_1^2 p_2^2$\\
        	$p_2^4$
        \end{tabular}
    &  
        \begin{tabular}{@{}l}
        	$\bar{t}^2$\\
        	$\bar{t}^3$\\
        	$\bar{t}^4$\\
        	$\bar{t}^4$\\
        	$\bar{t}^4$
        \end{tabular}
    &  
        \begin{tabular}{@{}l}
        	$1/16$\\
        	$3/16$
        \end{tabular}
    & 
        \begin{tabular}{@{}l}
        	$1/64$\\
        	$1/48$\\
        	$3/128$\\
        	$3/128$\\
        	$3/128$
        \end{tabular}
    &
        \begin{tabular}{@{}l}
        	$1/1152$\\
        	$1/2592$\\
        	$1/4608$\\
        	$1/4608$\\
        	$1/4608$
        \end{tabular}
    \\
    \hline
    Model 3
    &  
        \hspace{-0.24cm}
        \begin{tabular}{c|c|c|c|c}
            & $U(1)_{f_1}$ 
            & $U(1)_{f_2}$ 
            & $U(1)_R$ 
            & fugacity
            \\
            \hline
            $p_1$ 
            & 1/2 
            & 1/2 
            & $(5-\sqrt{7})/6$ 
            & $t_1$
            \\
            $p_2$ 
            & 0 
            & -1/2 
            & $(5-\sqrt{7})/6$ 
            & $t_2$
            \\
            $p_3$ 
            & -1/2 
            & -1/2 
            & $(1+\sqrt{7})/6$ 
            & $t_3$
            \\
            $p_4$ 
            & 0 
            & 1/2 
            & $(1+\sqrt{7})/6$ 
            & $t_4$
        \end{tabular}
    & 
        \begin{tabular}{@{}l}
        	$p_1^2 p_2^2$\\
        	$p_1 p_3^3$\\
        	$p_1 p_2 p_3 p_4$\\
        	$p_3^2 p_4^2$\\
        	$p_2 p_4^3$
        \end{tabular}
    &  
        \begin{tabular}{@{}l}
        	$\bar{t}^4$\\
        	$\bar{t}^4$\\
        	$\bar{t}^4$\\
        	$\bar{t}^4$\\
        	$\bar{t}^4$
        \end{tabular}
    &  
        \begin{tabular}{@{}l}
        	$1/32$\\
        	$1/8$
        \end{tabular}
    & 
        \begin{tabular}{@{}l}
        	$1/96$\\
        	$1/96$\\
        	$1/96$\\
        	$1/96$\\
        	$1/96$
        \end{tabular}
    &
        \begin{tabular}{@{}l}
        	$1/9216$\\
        	$1/9216$\\
        	$1/9216$\\
        	$1/9216$\\
        	$1/9216$
        \end{tabular}
    \\
    \hline    
    Model 4
    &
        \hspace{-0.24cm}
        \begin{tabular}{c|c|c|c|c}
            & $U(1)_{f_1}$ 
            & $U(1)_{f_2}$ 
            & $U(1)_R$ 
            & fugacity
            \\
            \hline
            $p_1$ 
            & 1/4
            & -1/4 
            & $1/2$ 
            & $t_1$
            \\
            $p_2$ 
            & 1/4
            & 1/4 
            & $1/2$ 
            & $t_2$
            \\
            $p_3$ 
            & -1/4 
            & -1/4 
            & $1/2$ 
            & $t_3$
            \\
            $p_4$ 
            & -1/4
            & 1/4
            & $1/2$ 
            & $t_4$
        \end{tabular}
    & 
        \begin{tabular}{@{}l}
        	$p_1^2 p_2^2$\\
        	$p_1^2 p_3^2$\\
        	$p_1 p_2 p_3 p_4$\\
        	$p_2^2 p_4^2$\\
        	$p_3^2 p_4^2$
        \end{tabular}
    &  
        \begin{tabular}{@{}l}
        	$\bar{t}^4$\\
        	$\bar{t}^4$\\
        	$\bar{t}^4$\\
        	$\bar{t}^4$\\
        	$\bar{t}^4$
        \end{tabular}
    &  
        \begin{tabular}{@{}l}
        	$1/32$\\
        	$1/8$
        \end{tabular}
    & 
        \begin{tabular}{@{}l}
        	$1/96$\\
        	$1/96$\\
        	$1/96$\\
        	$1/96$\\
        	$1/96$
        \end{tabular}
    &
        \begin{tabular}{@{}l}
        	$1/9216$\\
        	$1/9216$\\
        	$1/9216$\\
        	$1/9216$\\
        	$1/9216$
        \end{tabular}
    \\
    \hline
    Model 5
    &
        \hspace{-0.24cm}
        \begin{tabular}{c|c|c|c|c}
            & $U(1)_{f_1}$ 
            & $U(1)_{f_2}$ 
            & $U(1)_R$ 
            & fugacity
            \\
            \hline
            $p_1$ 
            & 0
            & -1/2 
            & $R_{5, 1}$ 
            & $t_1$
            \\
            $p_2$ 
            & 0
            & 1/2 
            & $R_{5, 2}$ 
            & $t_2$
            \\
            $p_3$ 
            & -1 
            & -1 
            & $R_{5, 3}$ 
            & $t_3$
            \\
            $p_4$ 
            & 1
            & 1
            & $R_{5, 4}$ 
            & $t_4$
        \end{tabular}
    & 
        \begin{tabular}{@{}l}
        	$p_1^2 p_4$\\
        	$p_1 p_2^3$\\
        	$p_1 p_2 p_3 p_4$\\
        	$p_2^4 p_3$\\
        	$p_2^2 p_3^2 p_4$\\
        	$p_3^3 p_4^2$
        \end{tabular}
    &  
        \begin{tabular}{@{}l}
        	$\bar{t}^3$\\
        	$\bar{t}^4$\\
        	$\bar{t}^4$\\
        	$\bar{t}^5$\\
        	$\bar{t}^5$\\
        	$\bar{t}^5$
        \end{tabular}
    &  
        \begin{tabular}{@{}l}
        	$7/200$\\
        	$7/50$
        \end{tabular}
    & 
        \begin{tabular}{@{}l}
        	$7/720$\\
        	$7/600$\\
        	$7/600$\\
        	$77/6000$\\
        	$77/6000$\\
        	$77/6000$
        \end{tabular}
    &
        \begin{tabular}{@{}l}
        	$7/32400$\\
        	$7/57600$\\
        	$7/57600$\\
        	$7/90000$\\
        	$7/90000$\\
        	$7/90000$
        \end{tabular}
    \\
    \hline
    Model 6
    &  
        \hspace{-0.24cm}
        \begin{tabular}{c|c|c|c|c}
            & $U(1)_{f_1}$ 
            & $U(1)_{f_2}$ 
            & $U(1)_R$ 
            & fugacity
            \\
            \hline
            $p_1$ 
            & -1
            & 0 
            & $R_{6, 1}$ 
            & $t_1$
            \\
            $p_2$ 
            & 1
            & 0
            & $R_{6, 2}$ 
            & $t_2$
            \\
            $p_3$ 
            & 0
            & 0
            & $R_{6, 3}$ 
            & $t_3$
            \\
            $p_4$ 
            & 0
            & 1 
            & $R_{6, 2}$ 
            & $t_4$
            \\
            $p_5$ 
            & 0 
            & -1 
            & $R_{6, 1}$ 
            & $t_5$
        \end{tabular}
    & 
        \begin{tabular}{@{}l}
        	$p_1^2 p_2 p_3^2$\\
        	$p_1^2 p_2^2 p_4$\\
        	$p_1 p_3^3 p_5$\\
        	$p_1 p_2 p_3 p_4 p_5$\\
        	$p_3^2 p_4 p_5^2$\\
        	$p_2 p_4^2 p_5^2$
        \end{tabular}
    &  
        \begin{tabular}{@{}l}
        	$\bar{t}^5$\\
        	$\bar{t}^5$\\
        	$\bar{t}^5$\\
        	$\bar{t}^5$\\
        	$\bar{t}^5$\\
        	$\bar{t}^5$
        \end{tabular}
    &  
        \begin{tabular}{@{}l}
        	$1/50$\\
        	$1/10$
        \end{tabular}
    & 
        \begin{tabular}{@{}l}
        	$1/150$\\
        	$1/150$\\
        	$1/150$\\
        	$1/150$\\
        	$1/150$\\
        	$1/150$
        \end{tabular}
    &
        \begin{tabular}{@{}l}
        	$1/22500$\\
        	$1/22500$\\
        	$1/22500$\\
        	$1/22500$\\
        	$1/22500$\\
        	$1/22500$
        \end{tabular}
    \\
    \hline
    Model 7
    &  
        \hspace{-0.24cm}
        \begin{tabular}{c|c|c|c|c}
            & $U(1)_{f_1}$ 
            & $U(1)_{f_2}$ 
            & $U(1)_R$ 
            & fugacity
            \\
            \hline
            $p_1$ 
            & 1/2
            & 0 
            & $2/3$ 
            & $t_1$
            \\
            $p_2$ 
            & -1/6
            & 1/3
            & $2/3$ 
            & $t_2$
            \\
            $p_3$ 
            & -1/3
            & -1/3
            & $2/3$ 
            & $t_3$
        \end{tabular}
    & 
        \begin{tabular}{@{}l}
        	$p_1^2$\\
        	$p_1 p_2 p_3$\\
        	$p_3^3$\\
        	$p_1 p_2^3$\\
        	$p_2^2 p_3^2$\\
        	$p_2^4 p_3$\\
        	$p_2^6$
        \end{tabular}
    &  
        \begin{tabular}{@{}l}
        	$\bar{t}^2$\\
        	$\bar{t}^3$\\
        	$\bar{t}^3$\\
        	$\bar{t}^4$\\
        	$\bar{t}^4$\\
        	$\bar{t}^5$\\
        	$\bar{t}^6$
        \end{tabular}
    &  
        \begin{tabular}{@{}l}
        	$1/12$\\
        	$1/4$
        \end{tabular}
    & 
        \begin{tabular}{@{}l}
        	$1/48$\\
        	$1/36$\\
        	$1/36$\\
        	$1/32$\\
        	$1/32$\\
        	$1/30$\\
        	$5/144$
        \end{tabular}
    &
        \begin{tabular}{@{}l}
        	$1/864$\\
        	$1/1944$\\
        	$1/1944$\\
        	$1/3456$\\
        	$1/3456$\\
        	$1/5400$\\
        	$1/7776$
        \end{tabular}
    \\
    \hline
    Model 8
    &
        \hspace{-0.24cm}
        \begin{tabular}{c|c|c|c|c}
            & $U(1)_{f_1}$ 
            & $U(1)_{f_2}$ 
            & $U(1)_R$ 
            & fugacity
            \\
            \hline
            $p_1$ 
            & 1
            & 0 
            & $\frac{1}{\sqrt{3}}$
            & $t_1$
            \\
            $p_2$ 
            & -1/2
            & 1/2
            & $\frac{1}{\sqrt{3}}$
            & $t_2$
            \\
            $p_3$ 
            & -1
            & 0
            & $1-\frac{1}{\sqrt{3}}$
            & $t_3$
            \\
            $p_4$ 
            & 1/2
            & -1/2
            & $1-\frac{1}{\sqrt{3}}$
            & $t_4$
        \end{tabular}
    & 
        \begin{tabular}{@{}l}
        	$p_1^2 p_3$\\
        	$p_1^2 p_2^2$\\
        	$p_1 p_2 p_3 p_4$\\
        	$p_3^2 p_4^2$\\
        	$p_1 p_2^3 p_4$\\
        	$p_2^2 p_3 p_4^2$\\
        	$p_2^4 p_4^2$
        \end{tabular}
    &  
        \begin{tabular}{@{}l}
        	$\bar{t}^3$\\
        	$\bar{t}^4$\\
        	$\bar{t}^4$\\
        	$\bar{t}^4$\\
        	$\bar{t}^5$\\
        	$\bar{t}^5$\\
        	$\bar{t}^6$
        \end{tabular}
    &  
        \begin{tabular}{@{}l}
        	$1/24$\\
        	$1/6$
        \end{tabular}
    & 
        \begin{tabular}{@{}l}
        	$5/432$\\
        	$1/72$\\
        	$1/72$\\
        	$1/72$\\
        	$11/720$\\
        	$11/720$\\
        	$7/432$
        \end{tabular}
    &
        \begin{tabular}{@{}l}
        	$1/3888$\\
        	$1/6912$\\
        	$1/6912$\\
        	$1/6912$\\
        	$1/10800$\\
        	$1/10800$\\
        	$1/15552$
        \end{tabular}
    \\
    \hline
    Model 9
    &  
        \hspace{-0.24cm}
        \begin{tabular}{c|c|c|c|c}
            & $U(1)_{f_1}$ 
            & $U(1)_{f_2}$ 
            & $U(1)_R$ 
            & fugacity
            \\
            \hline
            $p_1$ 
            & -2/5
            & 1/2
            & $2 \left(\sqrt{5}-2\right)$
            & $t_1$
            \\
            $p_2$ 
            & -1/5
            & -1/2
            & $2 \left(\sqrt{5}-2\right)$
            & $t_2$
            \\
            $p_3$ 
            & 2/5
            & 0
            & $2 \left(\sqrt{5}-2\right)$
            & $t_3$
            \\
            $p_4$ 
            & 1/5
            & 0
            & $7-3 \sqrt{5}$
            & $t_4$
            \\
            $p_5$ 
            & 0
            & 0
            & $7-3 \sqrt{5}$
            & $t_5$
        \end{tabular}
    & 
        \begin{tabular}{@{}l}
        	$p_3^2 p_4 p_5$\\
        	$p_1^2 p_3 p_4^2$\\
        	$p_1 p_2 p_3 p_4 p_5$\\
        	$p_2^2 p_3 p_5^2$\\
        	$p_1^3 p_2 p_4^2$\\
        	$p_1^2 p_2^2 p_4 p_5$\\
        	$p_1 p_2^3 p_5^2$
        \end{tabular}
    &  
        \begin{tabular}{@{}l}
        	$\bar{t}^4$\\
        	$\bar{t}^5$\\
        	$\bar{t}^5$\\
        	$\bar{t}^5$\\
        	$\bar{t}^6$\\
        	$\bar{t}^6$\\
        	$\bar{t}^6$
        \end{tabular}
    &  
        \begin{tabular}{@{}l}
        	$1/45$\\
        	$1/9$
        \end{tabular}
    & 
        \begin{tabular}{@{}l}
        	$7/1080$\\
        	$1/135$\\
        	$1/135$\\
        	$1/135$\\
        	$13/1620$\\
        	$13/1620$\\
        	$13/1620$
        \end{tabular}
    &
        \begin{tabular}{@{}l}
        	$1/12960$\\
        	$1/20250$\\
        	$1/20250$\\
        	$1/20250$\\
        	$1/29160$\\
        	$1/29160$\\
        	$1/29160$
        \end{tabular}
    \\
    \hline
    Model 10
    &  
        \hspace{-0.24cm}
        \begin{tabular}{c|c|c|c|c}
            & $U(1)_{f_1}$ 
            & $U(1)_{f_2}$ 
            & $U(1)_R$ 
            & fugacity
            \\
            \hline
            $p_1$ 
            & -1
            & 0 
            & $1/3$
            & $t_1$
            \\
            $p_2$ 
            & -1
            & 1
            & $1/3$
            & $t_2$
            \\
            $p_3$ 
            & 1
            & 0
            & $1/3$
            & $t_3$
            \\
            $p_4$ 
            & 1
            & -1
            & $1/3$
            & $t_4$
            \\
            $p_5$ 
            & 0
            & 0
            & $1/3$
            & $t_5$
            \\
            $p_6$ 
            & 0
            & 0
            & $1/3$
            & $t_6$        
        \end{tabular}
    & 
        \begin{tabular}{@{}l}
        	$p_2^2 p_3^2 p_4 p_5$\\
        	$p_1 p_2 p_3^2 p_5^2$\\
        	$p_2^2 p_3 p_4^2 p_6$\\
        	$p_1 p_2 p_3 p_4 p_5 p_6$\\
        	$p_1^2 p_3 p_5^2 p_6$\\
        	$p_1 p_2 p_4^2 p_6^2$\\
        	$p_1^2 p_4 p_5 p_6^2$
        \end{tabular}
    &  
        \begin{tabular}{@{}l}
        	$\bar{t}^6$\\
        	$\bar{t}^6$\\
        	$\bar{t}^6$\\
        	$\bar{t}^6$\\
        	$\bar{t}^6$\\
        	$\bar{t}^6$\\
        	$\bar{t}^6$
        \end{tabular}
    &  
        \begin{tabular}{@{}l}
        	$1/72$\\
        	$1/12$
        \end{tabular}
    & 
        \begin{tabular}{@{}l}
        	$1/216$\\
        	$1/216$\\
        	$1/216$\\
        	$1/216$\\
        	$1/216$\\
        	$1/216$\\
        	$1/216$
        \end{tabular}
    &
        \begin{tabular}{@{}l}
        	$1/46656$\\
        	$1/46656$\\
        	$1/46656$\\
        	$1/46656$\\
        	$1/46656$\\
        	$1/46656$\\
        	$1/46656$
        \end{tabular}
    \\
    \hline
    \end{tabular}
    }
    }
\caption{The Futaki invariants $F(\mathcal{X}_a, \zeta_p, \eta_h)$ for the toric Calabi-Yau 3-folds $\mathcal{X}_a$ corresponding to the 16 reflexive polygons in $\mathbb{Z}^2$. 
The extremal perfect matchings $p_\alpha$ and the generators in terms of $p_\alpha$ are shown with their global symmetry charges.
\textbf{(Part 1/2)}
}\label{FPtableA}
\end{table}

\begin{table}[H]
    \centering
    \tiny
    \makebox[\textwidth][l]{\hspace*{-0.65cm}
    \resizebox{1.07\hsize}{!}{
    \begin{tabular}{|>{\centering\arraybackslash}p{1.5cm}|p{6.5cm}|p{1.5cm}|p{1.5cm}|p{2cm}|p{2.3cm}|p{2.5cm}|}
    \hline
    Model $a$
    & Global Symmetry 
    & Generators 
    & $t_\alpha = t$
    & $A_0, A_1$
    & $F(\mathcal{X}_a, \zeta_p, \eta_h)$
    & $\|\eta\|^2_p$
    \\
   \hline
    Model 11
    &  
        \hspace{-0.24cm}
        \begin{tabular}{c|c|c|c|c}
            & $U(1)_{f_1}$ 
            & $U(1)_{f_2}$ 
            & $U(1)_R$ 
            & fugacity
            \\
            \hline
            $p_1$ 
            & -1/4
            & -1/3
            & $R_{11, 1}$
            & $t_1$
            \\
            $p_2$ 
            & -1/4
            & 0
            & $R_{11, 2}$
            & $t_2$
            \\
            $p_3$ 
            & 0
            & 2/3
            & $R_{11, 3}$
            & $t_3$
            \\
            $p_4$ 
            & 1/2
            & -1/3
            & $R_{11, 4}$
            & $t_4$
        \end{tabular}
    & 
        \begin{tabular}{@{}l}
        	$p_1^2 p_4$\\
        	$p_3 p_4^2$\\
        	$p_1^3 p_2$\\
        	$p_1 p_2 p_3 p_4$\\
        	$p_1^2 p_2^2 p_3$\\
        	$p_2^2 p_3^2 p_4$\\
        	$p_1 p_2^3 p_3^2$\\
        	$p_2^4 p_3^3$
        \end{tabular}
    &  
        \begin{tabular}{@{}l}
        	$\bar{t}^3$\\
        	$\bar{t}^3$\\
        	$\bar{t}^4$\\
        	$\bar{t}^4$\\
        	$\bar{t}^5$\\
        	$\bar{t}^5$\\
        	$\bar{t}^6$\\
        	$\bar{t}^7$
        \end{tabular}
    &  
        \begin{tabular}{@{}l}
        	$25/504$\\
        	$25/126$
        \end{tabular}
    & 
        \begin{tabular}{@{}l}
        	$125/9072$\\
        	$125/9072$\\
        	$25/1512$\\
        	$25/1512$\\
        	$55/3024$\\
        	$55/3024$\\
        	$25/1296$\\
        	$425/21168$
        \end{tabular}
    &
        \begin{tabular}{@{}l}
        	$25/81648$\\
        	$25/81648$\\
        	$25/145152$\\
        	$25/145152$\\
        	$1/9072$\\
        	$1/9072$\\
        	$25/326592$\\
        	$25/444528$
        \end{tabular}
    \\
    \hline
    Model 12
    &  
        \hspace{-0.24cm}
        \begin{tabular}{c|c|c|c|c}
            & $U(1)_{f_1}$ 
            & $U(1)_{f_2}$ 
            & $U(1)_R$ 
            & fugacity
            \\
            \hline
            $p_1$ 
            & 1/2
            & 0
            & $\frac{1}{16} \left(5 \sqrt{33}-21\right)$
            & $t_1$
            \\
            $p_2$ 
            & -1/2
            & 0
            & $\frac{1}{16} \left(57-9 \sqrt{33}\right)$
            & $t_2$
            \\
            $p_3$ 
            & 0
            & -1/2
            & $\frac{1}{16} \left(57-9 \sqrt{33}\right)$
            & $t_3$
            \\
            $p_4$ 
            & 0
            & 1/2
            & $\frac{1}{16} \left(5 \sqrt{33}-21\right)$
            & $t_4$
            \\
            $p_5$ 
            & 0
            & 0
            & $\frac{1}{2} \left(\sqrt{33}-5\right)$
            & $t_5$
        \end{tabular}
    & 
        \begin{tabular}{@{}l}
        	$p_1^2 p_3 p_4$\\
        	$p_1 p_2 p_4^2$\\
        	$p_1^2 p_3^2 p_5$\\
        	$p_1 p_2 p_3 p_4 p_5$\\
        	$p_2^2 p_4^2 p_5$\\
        	$p_1 p_2 p_3^2 p_5^2$\\
        	$p_2^2 p_3 p_4 p_5^2$\\
        	$p_2^2 p_3^2 p_5^3$
        \end{tabular}
    &  
        \begin{tabular}{@{}l}
        	$\bar{t}^4$\\
        	$\bar{t}^4$\\
        	$\bar{t}^5$\\
        	$\bar{t}^5$\\
        	$\bar{t}^5$\\
        	$\bar{t}^6$\\
        	$\bar{t}^6$\\
        	$\bar{t}^7$
        \end{tabular}
    &  
        \begin{tabular}{@{}l}
        	$31/1120$\\
        	$31/224$
        \end{tabular}
    & 
        \begin{tabular}{@{}l}
        	$31/3840$\\
        	$31/3840$\\
        	$31/3360$\\
        	$31/3360$\\
        	$31/3360$\\
        	$403/40320$\\
        	$403/40320$\\
        	$31/2940$
        \end{tabular}
    &
        \begin{tabular}{@{}l}
        	$31/322560$\\
        	$31/322560$\\
        	$31/504000$\\
        	$31/504000$\\
        	$31/504000$\\
        	$31/725760$\\
        	$31/725760$\\
        	$31/987840$
        \end{tabular}
    \\
    \hline
    Model 13
    &  
        \hspace{-0.24cm}
        \begin{tabular}{c|c|c|c|c}
            & $U(1)_{f}$ 
            & $SU(2)_{x}$ 
            & $U(1)_R$ 
            & fugacity
            \\
            \hline
            $p_1$ 
            & -1/4
            & 1/2
            & $2/3$
            & $t_1$
            \\
            $p_2$ 
            & -1/4
            & -1/2
            & $2/3$
            & $t_2$
            \\
            $p_3$ 
            & 1/2
            & 0
            & $2/3$
            & $t_3$
        \end{tabular}
    & 
        \begin{tabular}{@{}l}
        	$p_3^2$\\
        	$p_1^2 p_3$\\
        	$p_1 p_2 p_3$\\
        	$p_2^2 p_3$\\
        	$p_1^4$\\
        	$p_1^3 p_2$\\
        	$p_1^2 p_2^2$\\
        	$p_1 p_2^3$\\
        	$p_2^4$
        \end{tabular}
    &  
        \begin{tabular}{@{}l}
        	$\bar{t}^2$\\
        	$\bar{t}^3$\\
        	$\bar{t}^3$\\
        	$\bar{t}^3$\\
        	$\bar{t}^4$\\
        	$\bar{t}^4$\\
        	$\bar{t}^4$\\
        	$\bar{t}^4$\\
        	$\bar{t}^4$
        \end{tabular}
    & 
        \begin{tabular}{@{}l}
        	$1/8$\\
        	$3/8$
        \end{tabular}
    & 
        \begin{tabular}{@{}l}
        	$1/32$\\
        	$1/24$\\
        	$1/24$\\
        	$1/24$\\
        	$3/64$\\
        	$3/64$\\
        	$3/64$\\
        	$3/64$\\
        	$3/64$
        \end{tabular}
    &
        \begin{tabular}{@{}l}
        	$1/576$\\
        	$1/1296$\\
        	$1/1296$\\
        	$1/1296$\\
        	$1/2304$\\
        	$1/2304$\\
        	$1/2304$\\
        	$1/2304$\\
        	$1/2304$
        \end{tabular}
    \\
    \hline
    Model 14
    &  
        \hspace{-0.24cm}
        \begin{tabular}{c|c|c|c|c}
            & $U(1)_{f_1}$ 
            & $U(1)_{f_2}$ 
            & $U(1)_R$ 
            & fugacity
            \\
            \hline
            $p_1$ 
            & 1
            & 0
            & $\sqrt{13}-3$
            & $t_1$
            \\
            $p_2$ 
            & 1
            & 1
            & $\frac{1}{3} \left(5 \sqrt{13}-17\right)$
            & $t_2$
            \\
            $p_3$ 
            & -1
            & -1
            & $-\frac{4}{3} \left(\sqrt{13}-4\right)$
            & $t_3$
            \\
            $p_4$ 
            & -1
            & 0
            & $-\frac{4}{3} \left(\sqrt{13}-4\right)$
            & $t_4$            
        \end{tabular}
    & 
        \begin{tabular}{@{}l}
        	$p_1^2 p_3$\\
        	$p_1^2 p_4$\\
        	$p_1 p_2 p_3^2$\\
        	$p_1 p_2 p_3 p_4$\\
        	$p_1 p_2 p_4^2$\\
        	$p_2^2 p_3^3$\\
        	$p_2^2 p_3^2 p_4$\\
        	$p_2^2 p_3 p_4^2$\\
        	$p_2^2 p_4^3$
        \end{tabular}
    &  
        \begin{tabular}{@{}l}
        	$\bar{t}^3$\\
        	$\bar{t}^3$\\
        	$\bar{t}^4$\\
        	$\bar{t}^4$\\
        	$\bar{t}^4$\\
        	$\bar{t}^5$\\
        	$\bar{t}^5$\\
        	$\bar{t}^5$\\
        	$\bar{t}^5$
        \end{tabular}
    &  
        \begin{tabular}{@{}l}
        	$14/225$\\
        	$56/225$
        \end{tabular}
    & 
        \begin{tabular}{@{}l}
        	$7/405$\\
        	$7/405$\\
        	$14/675$\\
        	$14/675$\\
        	$14/675$\\
        	$77/3375$\\
        	$77/3375$\\
        	$77/3375$\\
        	$77/3375$
        \end{tabular}
    &
        \begin{tabular}{@{}l}
        	$7/18225$\\
        	$7/18225$\\
        	$7/32400$\\
        	$7/32400$\\
        	$7/32400$\\
        	$7/50625$\\
        	$7/50625$\\
        	$7/50625$\\
        	$7/50625$
        \end{tabular}
    \\
    \hline
    Model 15
    &  
        \hspace{-0.24cm}
        \begin{tabular}{c|c|c|c|c}
            & $SU(2)_{x_1}$ 
            & $SU(1)_{x_2}$ 
            & $U(1)_R$ 
            & fugacity
            \\
            \hline
            $p_1$ 
            & 1/2
            & 0
            & $1/2$
            & $t_1$
            \\
            $p_2$ 
            & -1/2
            & 0
            & $1/2$
            & $t_2$
            \\
            $p_3$ 
            & 0
            & 1/2
            & $1/2$
            & $t_3$
            \\
            $p_4$ 
            & 0
            & -1/2
            & $1/2$
            & $t_4$            
        \end{tabular}
    & 
        \begin{tabular}{@{}l}
        	$p_1^2 p_3^2$\\
        	$p_1 p_2 p_3^2$\\
        	$p_2^2 p_3^2$\\
        	$p_1^2 p_3 p_4$\\
        	$p_1 p_2 p_3 p_4$\\
        	$p_2^2 p_3 p_4$\\
        	$p_1^2 p_4^2$\\
        	$p_1 p_2 p_4^2$\\
        	$p_2^2 p_4^2$
        \end{tabular}
    &  
        \begin{tabular}{@{}l}
        	$\bar{t}^4$\\
        	$\bar{t}^4$\\
        	$\bar{t}^4$\\
        	$\bar{t}^4$\\
        	$\bar{t}^4$\\
        	$\bar{t}^4$\\
        	$\bar{t}^4$\\
        	$\bar{t}^4$\\
        	$\bar{t}^4$
        \end{tabular}
    &  
        \begin{tabular}{@{}l}
        	$1/16$\\
        	$1/4$
        \end{tabular}
    & 
        \begin{tabular}{@{}l}
        	$1/48$\\
        	$1/48$\\
        	$1/48$\\
        	$1/48$\\
        	$1/48$\\
        	$1/48$\\
        	$1/48$\\
        	$1/48$\\
        	$1/48$
        \end{tabular}
    &
        \begin{tabular}{@{}l}
        	$1/4608$\\
        	$1/4608$\\
        	$1/4608$\\
        	$1/4608$\\
        	$1/4608$\\
        	$1/4608$\\
        	$1/4608$\\
        	$1/4608$\\
        	$1/4608$
        \end{tabular}
    \\
    \hline
    Model 16
    &  
        \hspace{-0.24cm}
        \begin{tabular}{c|c|c|c}
            & $SU(3)_{(x_1, x_2)}$ 
            & $U(1)_R$ 
            & fugacity
            \\
            \hline
            $p_1$ 
            & (-1/3, -1/3)
            & $2/3$
            & $t_1$
            \\
            $p_2$ 
            & (2/3, -1/3)
            & $2/3$
            & $t_2$
            \\
            $p_3$ 
            & (-1/3, 2/3)
            & $2/3$
            & $t_3$     
        \end{tabular}
    & 
        \begin{tabular}{@{}l}
        	$p_1^3$\\
        	$p_1^2 p_2$\\
        	$p_1 p_2^2$\\
        	$p_2^3$\\
        	$p_1^2 p_3$\\
        	$p_1 p_2 p_3$\\
        	$p_2^2 p_3$\\
        	$p_1 p_3^2$\\
        	$p_2 p_3^2$\\
        	$p_3^3$
        \end{tabular}
    &  
        \begin{tabular}{@{}l}
        	$\bar{t}^3$\\
        	$\bar{t}^3$\\
        	$\bar{t}^3$\\
        	$\bar{t}^3$\\
        	$\bar{t}^3$\\
        	$\bar{t}^3$\\
        	$\bar{t}^3$\\
        	$\bar{t}^3$\\
        	$\bar{t}^3$\\
        	$\bar{t}^3$
        \end{tabular}
    &  
        \begin{tabular}{@{}l}
        	$1/6$\\
        	$1/2$
        \end{tabular}
    & 
        \begin{tabular}{@{}l}
        	$1/18$\\
        	$1/18$\\
        	$1/18$\\
        	$1/18$\\
        	$1/18$\\
        	$1/18$\\
        	$1/18$\\
        	$1/18$\\
        	$1/18$\\
        	$1/18$
        \end{tabular}
    &
        \begin{tabular}{@{}l}
        	$1/972$\\
        	$1/972$\\
        	$1/972$\\
        	$1/972$\\
        	$1/972$\\
        	$1/972$\\
        	$1/972$\\
        	$1/972$\\
        	$1/972$\\
        	$1/972$
        \end{tabular}
    \\
    \hline
    \end{tabular}
    }
    }
\caption{The Futaki invariants $F(\mathcal{X}_a, \zeta_p, \eta_h)$ for the toric Calabi-Yau 3-folds $\mathcal{X}_a$ corresponding to the 16 reflexive polygons in $\mathbb{Z}^2$. 
The extremal perfect matchings $p_\alpha$ and the generators in terms of $p_\alpha$ are shown with their global symmetry charges.
\textbf{(Part 2/2)}
}\label{FPtableB}
\end{table}

\subsection{Futaki Invariants $F(\mathcal{X}, \zeta_R, \eta)$ and $F(\mathcal{X}, \zeta_p, \eta)$}\label{twoFuts}

As we have discussed in the sections above, the Futaki invariants of the form $F(\mathcal{X}_a, \zeta_R, \eta)$ are obtained where $\zeta=\zeta_R$ corresponds to the $U(1)_R$ symmetry and the Futaki invariants of the form $F(\mathcal{X}_a, \zeta_p, \eta)$ are obtained where $\zeta=\zeta_p$ imposes a grading on $\mathcal{X}_a$ corresponding to the GLSM field degrees.
Having computed these Futaki invariants for each of the generators $x_h$ of $\mathcal{X}_a$ for the family of toric Calabi-Yau 3-folds $\mathcal{X}_a$ corresponding to the $16$ reflexive polygons in \fref{fig_reflexive}, 
it is natural to ask whether $F(\mathcal{X}_a, \zeta_R, \eta_h)$ and $F(\mathcal{X}_a, \zeta_p, \eta_h)$ form any relationship.

\begin{figure}[h]
    \centering
    \resizebox{0.9\hsize}{!}{\includegraphics[width=1\linewidth]{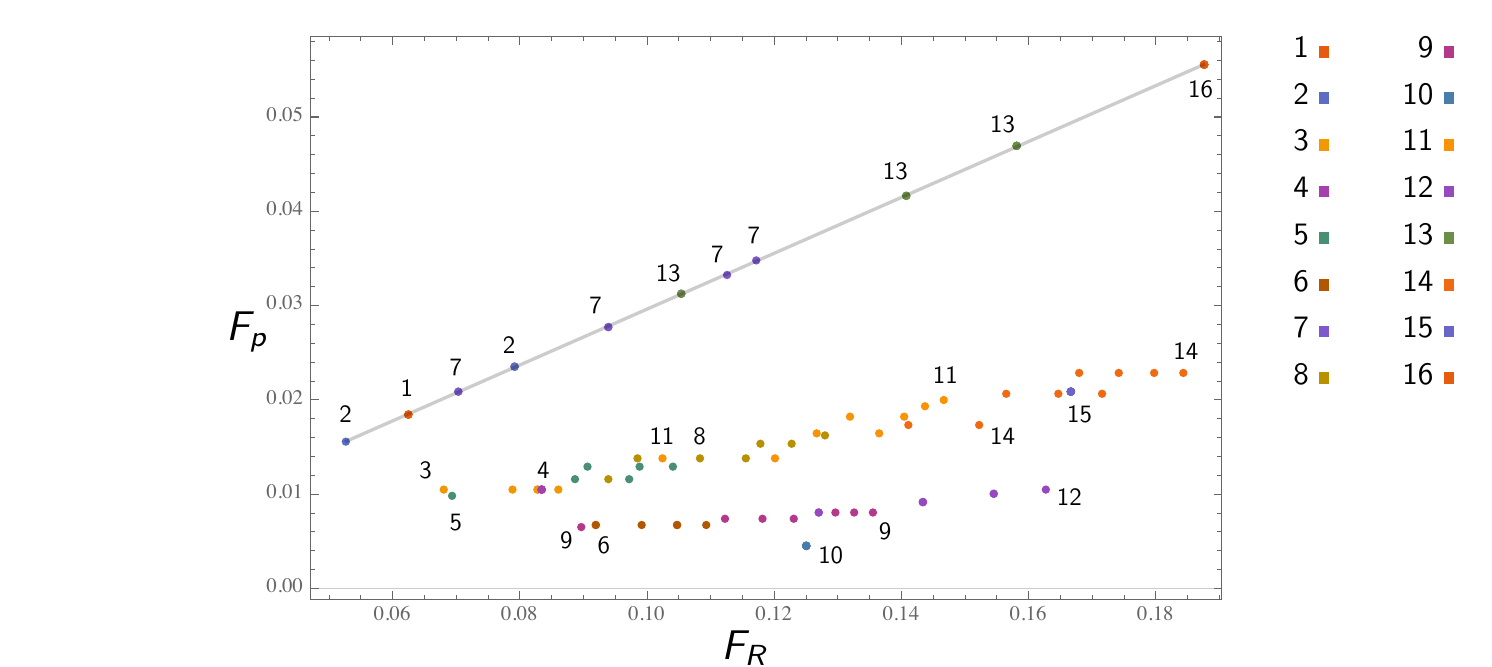}}
    \caption{Futaki invariants $F(\mathcal{X}_a, \zeta_R, \eta_h)$ [$F_R$] against $F(\mathcal{X}_a, \zeta_p, \eta_h)$ $[F_p]$, where $a=1, \dots, 16$ labels the 16 reflexive polygons and their corresponding toric Calabi-Yau 3-folds $\mathcal{X}_a$, and $h=1, \dots, k_a$ labels the generators $x_h$ for a given $\mathcal{X}_a$.}\label{FRvsFp}
\end{figure}

In order to answer this question, let us first plot the Futaki invariants $F(\mathcal{X}_a, \zeta_R, \eta_h)$ against $F(\mathcal{X}_a, \zeta_p, \eta_h)$, where $a=1, \dots, 16$ labels the 16 reflexive polygons and their corresponding toric Calabi-Yau 3-folds $\mathcal{X}_a$, and $h=1, \dots, k_a$ labels the generators $x_h$ for a given $\mathcal{X}_a$. 
The resulting plot is shown in \fref{FRvsFp}.

In particular, we find that there is a clear upper bound based on \fref{FRvsFp}, and \tref{FRtableA} and \tref{FRtableB}.
We observe,
\begin{proposition}
The Futaki invariant $F(\mathcal{X}_a, \zeta_p, \eta_h)$ under a test symmetry $\eta_h$ associated to the $h$-th generator of $\mathcal{X}_a$
has an upper bound in terms of the Futaki invariant $F(\mathcal{X}_a, \zeta_R, \eta_h)$ as follows,
\beal{es04a01}
F(\mathcal{X}_a, \zeta_p, \eta_h) \leq \frac{8}{27} F(\mathcal{X}_a, \zeta_R, \eta_h)
~,~
\eea
where $\mathcal{X}_a$ has a toric diagram, which is one of the 16 reflexive polygons in $\mathbb{Z}^2$.
\end{proposition}
The above can be observed for example in section \sref{example}
on $L_{1, 3, 1}/\mathbb{Z}_{2}$ $(0,1,1,1)$, which corresponds to Model 3
in \fref{fig_reflexive}.
There, when we compare the general expression for the Futaki invariant $F_R$ in \eref{es03a11} with $F_p$ in \eref{es03a18}, we see that the bound in \eref{es04a01} holds for any test symmetry $\eta_h$.

In fact, based on this observation, we conjecture the following, 
\begin{conjecture}
The Futaki invariant $F(\mathcal{X}, \zeta_p, \eta_h)$ 
has an upper bound in terms of the Futaki invariant $F(\mathcal{X}, \zeta_R, \eta_h)$
as given in \eref{es04a01}
for any toric Calabi-Yau 3-fold $\mathcal{X}$, where $\mathcal{X}$ has no factors of $\mathbb{C}$. 
\end{conjecture}

When we consider toric Calabi-Yau 3-folds corresponding to the 16 reflexive polygons in \fref{fig_reflexive},
we note that the bound is saturated as follows, 
\beal{es04a02}
F(\mathcal{X}_{a^*}, \zeta_p, \eta_h) = \frac{8}{27} F(\mathcal{X}_{a^*}, \zeta_R, \eta_h)
~,~
\eea
for a critical subset of toric Calabi-Yau 3-folds $\mathcal{X}_{a^*}$ and for any of their generators.
These critical toric Calabi-Yau 3-folds correspond to Models $1$, $2$, $7$, $13$ and $16$ in \fref{FRvsFp}, which we identify as the abelian orbifolds of the form $\mathbb{C}^3/\mathbb{Z}_3 \times \mathbb{Z}_3$ $(1,0,2)(0,1,2)$, $\mathbb{C}^3/\mathbb{Z}_4 \times \mathbb{Z}_2$ $(1,0,3)(0,1,1)$, $\mathbb{C}^3/\mathbb{Z}_6$ $(1,2,3)$, $\mathbb{C}^3/\mathbb{Z}_4$ $(1,1,2)$ and $\mathbb{C}^3/\mathbb{Z}_3$ $(1,1,1)$, respectively.
The corresponding toric diagrams are all triangles in $\mathbb{Z}^2$ with the origin as the unique internal point, as shown in \fref{FRvsFp}. 

The origin of this upper bound in \eref{es04a02} can be
traced back to the original definitions of the Futaki invariants in \eref{es02a77} and \eref{es02a78}, which are given below, 
\beal{es04a05}
F(\mathcal{X}; \zeta_R, \eta_h) = 
\frac{A_0(\zeta_R)}{2}
- \frac{A_1(\zeta_R)}{6R_h}
~,~
F(\mathcal{X}; \zeta_p, \eta_h) = 
\frac{A_0(\zeta_p)}{2} - \frac{A_1(\zeta_p)}{6 d_h}
~,~
\eea
where $R_h$ and $d_h$ are the $U(1)_R$ charge and the degree in GLSM fields of the generator $x_h$ of $\mathcal{X}$, respectively. 
For abelian orbifolds of $\mathbb{C}^3$, there are only 3 extremal vertices in the toric diagram of $\mathcal{X}$ corresponding to 3 extremal GLSM fields $p_1, p_2, p_3$. 
Each of these GLSM fields have an $U(1)_R$ charge $R(p_\alpha)=2/3$, which implies that the $U(1)_R$ charge of a generator $x_h$ is simply given by 
\beal{es04a06}
R_h = \frac{2}{3} d_h ~.~
\eea
As a result, by setting $F(\mathcal{X}; \zeta_R, \eta_h) = \tilde{c} F(\mathcal{X}; \zeta_R, \eta_h)$, we have
\beal{es04a07}
\frac{A_0(\zeta_p)}{2} - \frac{A_1(\zeta_p)}{6 d_h}
=
\tilde{c}
\left[
\frac{A_0(\zeta_R)}{2} - \frac{A_1(\zeta_R)}{6R_h}
\right]
~,~
\eea
where $\tilde{c}$ is the slope of the bound that we can solve for. 
We note that for abelian orbifolds of $\mathbb{C}^3$, we have $R_h = \frac{2}{3} d_h$, $A_1(\zeta_R)/A_0(\zeta_R)=2$ under $\zeta_R$ and $A_1(\zeta_p)/A_0(\zeta_p)=3$ under $\zeta_p$.
Moreover, we note that the slope $\tilde{c}$ is relating the leading coefficients $A_0(\zeta_p) = \tilde{c}A_0(\zeta_R)$ and $A_1(\zeta_p) = \tilde{c} \frac{3}{2}A_1(\zeta_R)$ for abelian orbifolds of $\mathbb{C}^3$.
Recalling the Laurent expansion in \eref{es02a22} of the Hilbert series $g(t;\mathcal{X}, \zeta)$, we have
\beal{es04a08}
g(\bar{t} = e^{-s}; \mathcal{X}_a, \zeta_p )
&=&
\frac{2 A_0(\zeta_p)}{ s^3} + \dots
~,~
\nn\\
g(t = e^{-(3/2)s}; \mathcal{X}_a, \zeta_R )
&=&
\frac{2 A_0(\zeta_R)}{((3/2)s)^3} + \dots
~,~
\eea
where $\bar{t}$ is the fugacity for $\zeta_p$ and $t$ is the fugacity for $\zeta_R$.
For abelian orbifolds of $\mathbb{C}^3$, we have $t = \bar{t}^{2/3}$ assigned to each GLSM field $p_\alpha$ since $R(p_\alpha)=2/3$. 
According to \eref{es04a08}, we see that for abelian orbifolds of $\mathbb{C}^3$, $A_0(\zeta_p) = \tilde{c}A_0(\zeta_R)$ gives, 
\beal{es04a09}
\tilde{c} = \left(\frac{R_h}{d_h}\right) = \left(\frac{2}{3}\right)^3 = \frac{8}{27}~.~
\eea

In \fref{FRvsFp}, we can also see that
there is also a lower bound provided by a single 
point corresponding to the Futaki invariants associated to Model 10, the Calabi-Yau cone over dP$_3$ in \fref{fig_reflexive}. 
This is another special case because it is the only reflexive polygon in $\mathbb{Z}^2$ with 6 extremal vertices.
The corresponding toric Calabi-Yau 3-fold has generators $x_h$
that have the same $U(1)_R$ charge $R_h=2$ and also have the same degree in GLSM fields $d_h=6$ for all $h=1,\dots,7$. 
Accordingly, we have for Model 10
\beal{es04a10}
R_h = \frac{1}{3} d_h ~,~
\eea
which leads to the lower bound
\beal{es04a11}
F(\mathcal{X}_{10}; \zeta_p, \eta_h)=\frac{1}{27} F(\mathcal{X}_{10}; \zeta_R, \eta_h)
~.~
\eea

In \cite{He:2017gam}, the minimized volumes $V_{min}$ for the Sasaki-Einstein 5-manifolds corresponding to the 16 toric Calabi-Yau 3-folds $\mathcal{X}$ with reflexive toric diagrams $\Delta$ were computed.
As summarized in \eref{es02a40}, it was discovered in \cite{He:2017gam} that these minimum volumes $V_{min}$ 
are bounded by the Euler number $\chi$ defined in \eref{es02a41} and the first Chern number $C_1$ defined in \eref{es02a43}. 
Interestingly, 
the bounds on the minimum volume $V_{min}$ were found to be saturated by reflexive toric diagrams that are triangles and the hexagon -- exactly like what we observe here for the Futaki invariants. 
This leads us to speculate whether the minimized volumes $V_{min}$ of the Sasaki-Einstein 5-manifolds and the topological invariants of the associated toric varieties $X(\Delta)$ form relations and whether such relations
are actually determined by the $U(1)_R$ charges and the degrees in GLSM fields.

In the following subsections, we would like to compare the Futaki invariants with various quantities that arise from the toric Calabi-Yau 3-folds corresponding to the 16 reflexive polygons in \fref{fig_reflexive} and their Sasaki-Einstein base manifolds.
We are going to focus mainly on the Futaki invariants of the form $F(\mathcal{X}; \zeta_R, \eta_h)$, 
where $\zeta_R$ corresponds to the $U(1)_R$ symmetry. 
\\

\subsection{Minimized Volumes and Topological Invariants}\label{Vminandtopinv}

We first compare the minimum volume $V_{min} = V(b^*; Y_a)$ of the Sasaki-Einstein manifold $Y_a$ associated to $\mathcal{X}_a$
with the associated Futaki invariants of the form $F(\mathcal{X}_a; \zeta_R, \eta_h)$, where $\mathcal{X}_a$ are the toric Calabi-Yau 3-folds whose toric diagrams are given by the 16 reflexive polygons in \fref{fig_reflexive}. 
Here, we note that the volume minimization of the original volume function $V(b; Y_a)$ in \eref{es02a33} to $V_{min}$ extremizes the associated central charge $a$-function, giving the superconformal $U(1)_R$ charges as illustrated in \eref{es02a34}.

\begin{figure}[h]
    \centering
    \resizebox{0.9\hsize}{!}{\includegraphics[width=1\linewidth]{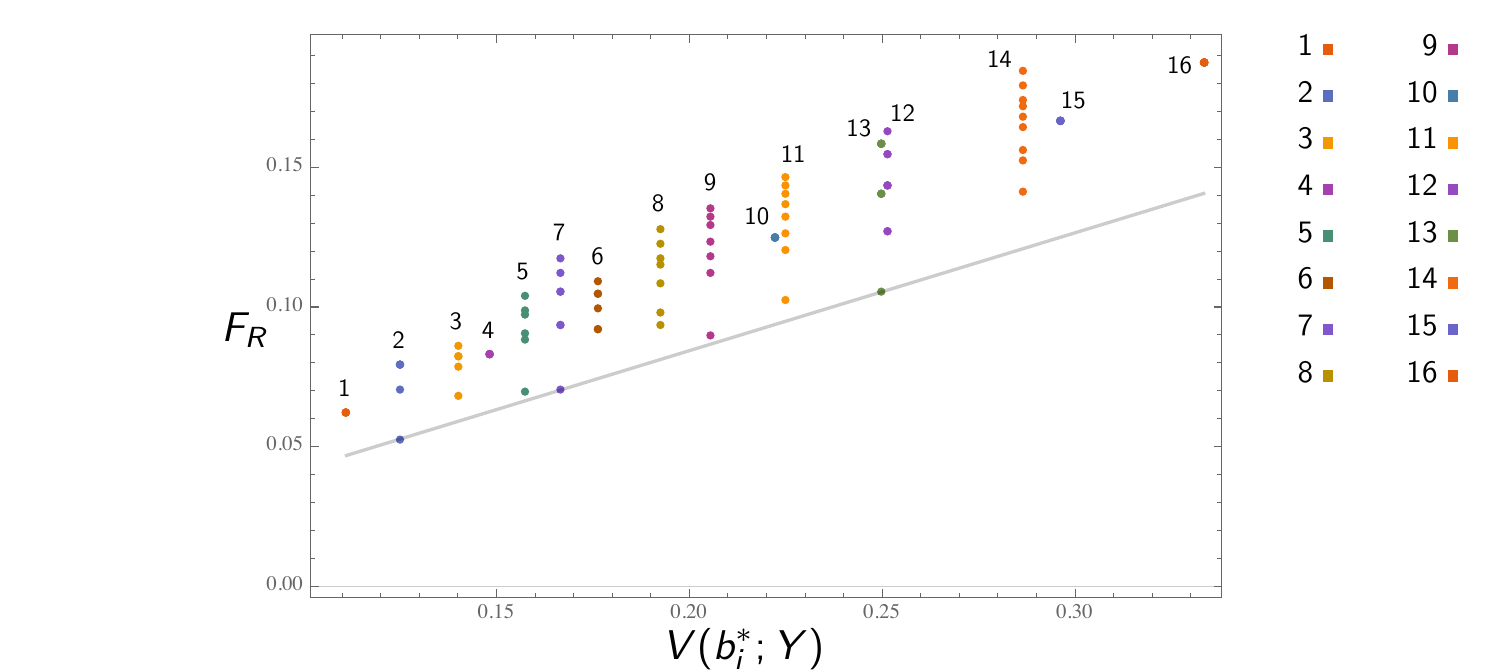}}
    \caption{The minimum volume $V_{min} = V(b_i^*; Y_a)$ of the Sasaki-Einstein $5$-manifold $Y_a$ associated to the toric Calabi-Yau 3-fold $\mathcal{X}_a$ with one of the 16 reflexive polygons as its toric diagram, plotted against the Futaki invariants $F(\mathcal{X}_a; \zeta_R, \eta_h)$ for all generators $x_h$ corresponding to $\mathcal{X}_a$.}\label{VminvsFR}
\end{figure}

\fref{VminvsFR}, based on \tref{FRtableA} and \tref{FRtableB}, shows the Futaki invariants $F(\mathcal{X}_a; \zeta_R, \eta_h)$ for all generators $x_h$ associated to $\mathcal{X}_a$ against the corresponding minimum volume $V_{min} = V(b^*; Y_a)$ corresponding to $Y_a$.
We observe here that the Futaki invariants $F(\mathcal{X}_a; \zeta_R, \eta_h)$ are bounded as follows, 
\begin{proposition}
The Futaki invariant $F(\mathcal{X}_a; \zeta_R, \eta_h)$ under a test symmetry $\eta_h$ associated to the $h$-th generator of $\mathcal{X}_a$ has a lower bound given by the minimum volume $V(b_i^*; Y_a)$ of the corresponding Sasaki-Einstein 5-manifold $Y_a$ as follows,
\beal{es04a20}
F(\mathcal{X}_a; \zeta_R, \eta_h) \geq \frac{27}{64} V(b_i^*; Y_a)
~,~
\eea
where 
$\mathcal{X}_a$
has a toric diagram given by one of the 16 reflexive polygons in $\mathbb{Z}^2$.
\end{proposition}

Based on observations for toric Calabi-Yau 3-folds that do not correspond to reflexive polygons in $\mathbb{Z}^2$, which we summarize in appendix \sref{nonreflexive}, 
we conjecture
\begin{conjecture}
The Futaki invariant $F(\mathcal{X}; \zeta_R, \eta_h)$ has a lower bound given by \eref{es04a20} in terms of the minimum volume $V(b_i^*; Y)$ of the corresponding Sasaki-Einstein 5-manifold $Y$ for any toric Calabi-Yau 3-folds $\mathcal{X}$, where $\mathcal{X}$ has no factors of $\mathbb{C}$. 
\end{conjecture}

When we consider toric Calabi-Yau 3-folds corresponding to the 16 reflexive polygons, 
we can explain the origin of the lower bound on $F(\mathcal{X}_a; \zeta_R, \eta_h)$ by going back to the definition of the Futaki invariant in \eref{es02a76} given by, 
\beal{es04a21}
F(\mathcal{X}_a; \zeta_R, \eta_h) =\frac{A_0(\zeta_R)}{2}-\frac{A_1(\zeta_R)}{6R_h}
~.~
\eea
Because $\zeta_R$ here refers to the $U(1)_R$ symmetry, 
we recall that 
$A_0(\zeta_R)$ is proportional to the minimum volume $V_{min}$ such that $2 \tilde{c} A_{0}(\zeta) = V_{min}$, 
and $A_1(\zeta_R)=2A_0(\zeta_R)$.
Accordingly, we can rewrite the Futaki invariants as follows, 
\beal{es04a22}
F(\mathcal{X}_a; \zeta_R, \eta_h) =
\left(\frac{3}{2}-\frac{1}{R_h}\right)\frac{V_{min}}{6 \tilde{c}}~,~
\eea
where $R_h$ is the $U(1)_R$ charge of the corresponding generator $x_h$ of $\mathcal{X}_a$. 
We can see here that in order to identify the slope of the lower bound on the Futaki invariant $F(\mathcal{X}_a; \zeta_R, \eta_h)$, we have to identify the toric Calabi-Yau 3-fold $\mathcal{X}_a$ with a generator $x_h$ that has the smallest $U(1)_R$ charge $R_h$.
According to the $U(1)_R$ charges collected in \tref{FRtableA} and \tref{FRtableB}, we see that the generators with the lowest $U(1)_R$ charges have $R_h=4/3$ and are part of Models 2, 7 and 13 in \fref{fig_reflexive}, which correspond respectively to the abelian orbifolds of the form $\mathbb{C}^3/\mathbb{Z}_4 \times \mathbb{Z}_2$ $(1,0,3)(0,1,1)$, $\mathbb{C}^3/\mathbb{Z}_6$ $(1,2,3)$ and $\mathbb{C}^3/\mathbb{Z}_4$ $(1,1,2)$.
These 3 abelian orbifolds and their generators with $R_h=4/3$ precisely correspond to the 3 points on the lower bound on the Futaki invariant in \tref{VminvsFR}.

\begin{figure}[H]
    \centering
    \resizebox{0.9\hsize}{!}{\includegraphics[width=1\linewidth]{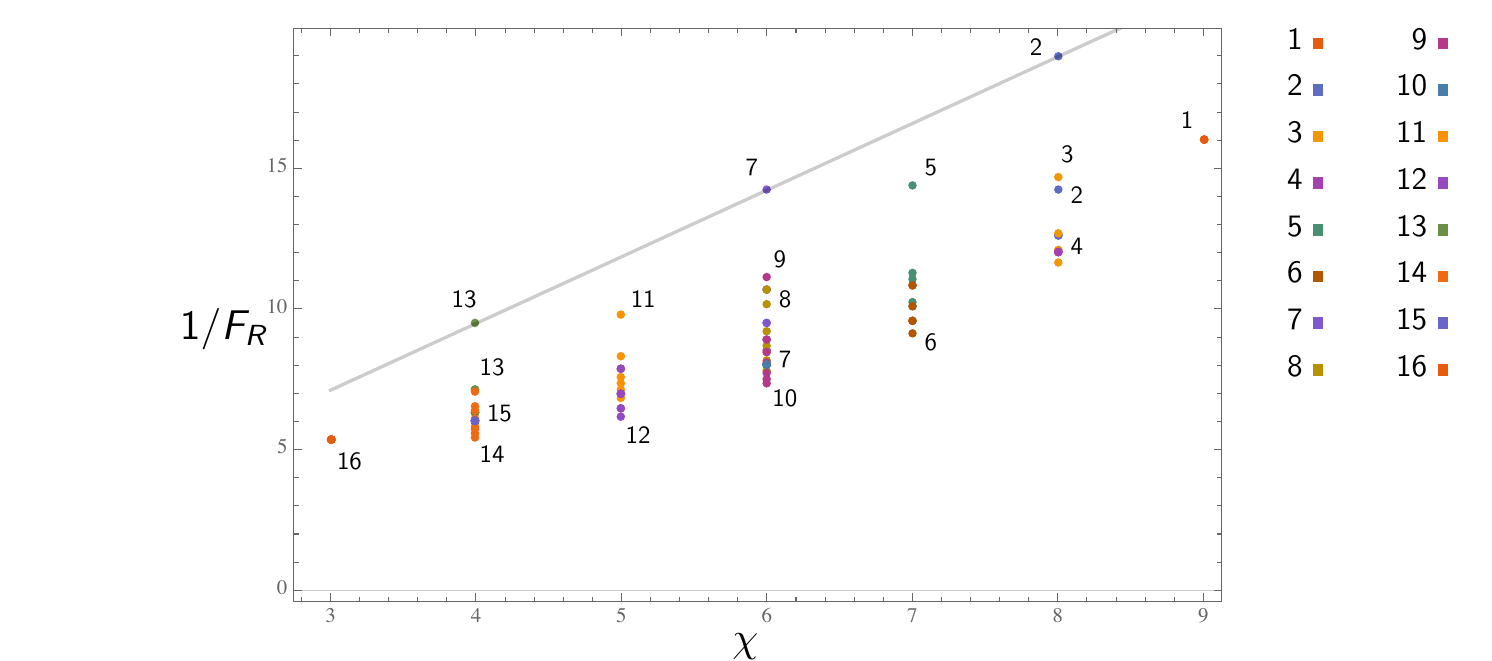}}
    \caption{
    The inverse of the Futaki invariants $F(\mathcal{X}_a; \zeta_R, \eta_h)$ [$F_R$] against the Euler number $\chi$ of the resolved toric varieties $X_a$ corresponding to the toric Calabi-Yau 3-folds $\mathcal{X}_a$ with the 16 reflexive polygons in $\mathbb{Z}^2$ as their toric diagrams.
    }\label{chivsFR}
\end{figure}

In fact, based on \eref{es04a22}, we see that for any toric Calabi-Yau 3-fold with a generator that has a $U(1)_R$ charge $R_h>4/3$, there is a separate line in the plot shown in \fref{VminvsFR} with a slope proportional to $\frac{3}{2}-\frac{1}{R_h}$.
This observation gives insight into the properties of the plot in \fref{VminvsFR}. 
We can draw a straight line for each value of $\frac{3}{2}-\frac{1}{R_h}$ resulting in a bouquet of lines starting at the origin where every point in \fref{VminvsFR} corresponding to a unique value of the Futaki invariant
would lie on one of these lines.
For all the examples we know corresponding to toric Calabi-Yau 3-folds associated to the 16 reflexive polygons in $\mathbb{Z}^2$, $R_h=4/3$ is the lowest possible value for a generator. 
The exception is of course $\mathbb{C}^3$ whose 3 generators have all $U(1)_R$ charge $R_h = 2/3$ according to the associated $4d$ $\mathcal{N}=4$ supersymmetric gauge theory. 
As a result, 
we expect the lower bound in \eref{es04a20} on the Futaki invariants $F(\mathcal{X}_a; \zeta_R, \eta_h)$
to hold for any toric Calabi-Yau 3-folds except for $\mathbb{C}^3$ or toric Calabi-Yau 3-folds that factorize with $\mathbb{C}$ factors.
We check the lower bound in \eref{es04a20} with additional abelian orbifolds of $\mathbb{C}^3$ in appendix \sref{nonreflexive}.
\\

\begin{figure}[H]
    \centering
    \resizebox{0.9\hsize}{!}{\includegraphics[width=1\linewidth]{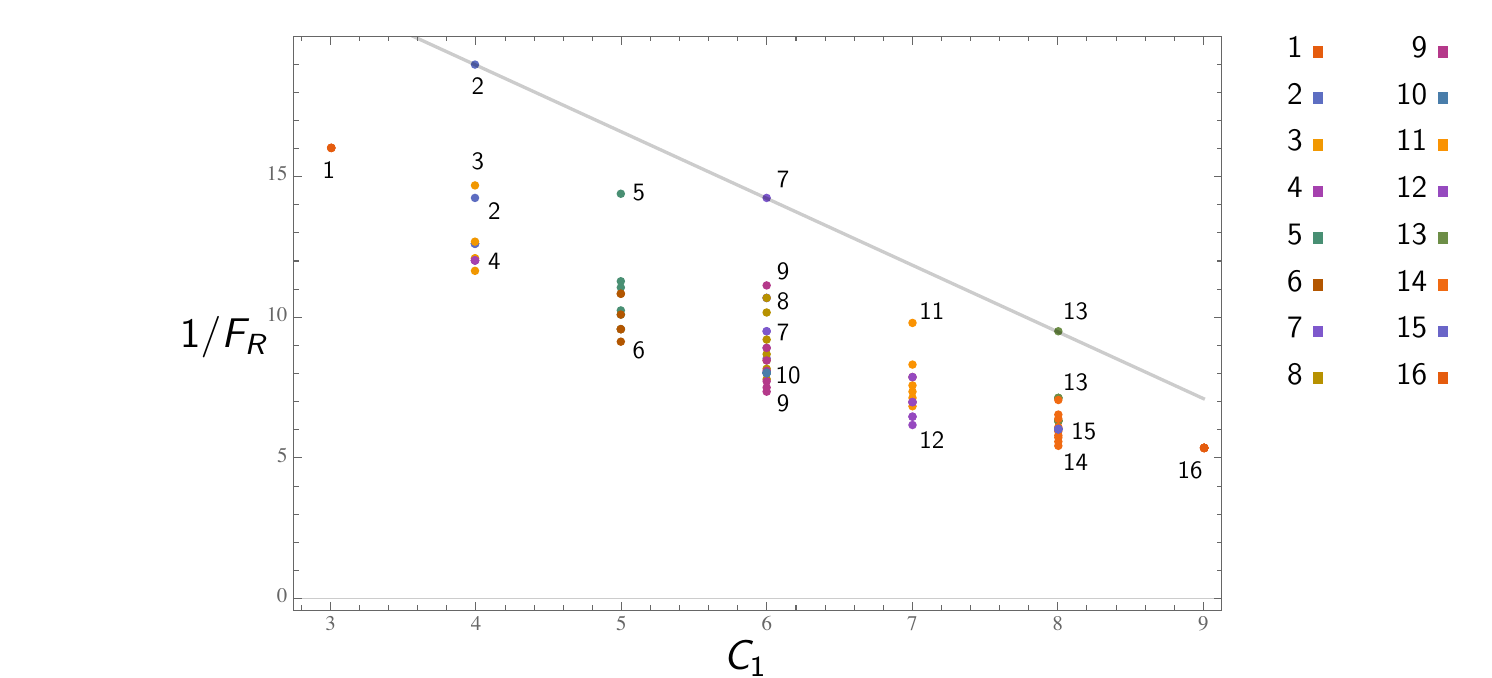}}
    \caption{
    The inverse of the Futaki invariants $F(\mathcal{X}_a; \zeta_R, \eta_h)$ [$F_R$] agains the first Chern number $C_1$ of the resolved toric varieties $X_a$ corresponding to the toric Calabi-Yau 3-folds $\mathcal{X}_a$ with the 16 reflexive polygons in $\mathbb{Z}^2$ as their toric diagrams.}\label{C1vsFR}
\end{figure}

As studied in \cite{He:2017gam}, the topological invariants of the resolved toric varieties $X_a$ built from the reflexive polygons 
can put bounds on the minimized volumes of the Sasaki-Einstein 5-manifolds $Y_a$.
As reviewed in section \sref{minimizedvolandtopinv}, 
the minimum volumes $V_{min} = V(b^*; Y_a)$ were found to satisfy the following bounds,
\beal{es04a30}
1/\chi \leq V_{min} \leq m_3 C_1 ~,~
\eea
where $m_3\sim 3^{-3}$ was found in \cite{He:2017gam} for the 16 reflexive polygons, 
and
$\chi$ is the Euler number and $C_1$ is the first Chern number associated to $X_a$.
Here, the lower bound is saturated for the abelian orbifolds of $\mathbb{C}^3$ where the toric diagrams are triangles and reflexive. 
Accordingly, we expect similar bounds to appear when we plot the Futaki invariants of the form $F(\mathcal{X}_a; \zeta_R, \eta_h)$
for all $\mathcal{X}_a$ corresponding to the 16 reflexive polygons in \fref{fig_reflexive}
against the corresponding Euler number $\chi$ and first Chern numbers $C_1$ of $X_a$,
as shown in \fref{chivsFR} and \fref{C1vsFR}, respectively.

Focusing first on \fref{chivsFR}, we see that similar to the lower bound set by $V_{min}$ in \eref{es04a20}, 
the Futaki invariants $F(\mathcal{X}_a; \zeta_R, \eta_h)$
are bounded in terms of the corresponding Euler numbers $\chi$ as follows, 
\begin{proposition}
The inverse of the Futaki invariant $F(\mathcal{X}_a; \zeta_R, \eta_h)$ under a test symmetry $\eta_h$ associated to the $h$-th generator of $\mathcal{X}_a$
has an upper bound given
by the Euler number $\chi(X_a)$,
\beal{es04a31}
\frac{1}{F(\mathcal{X}_a; \zeta_R, \eta_h)}
\leq \frac{64}{27}~
\chi(X_a)
~,~
\eea
where 
$X_a$
is the toric variety associated to the toric Calabi-Yau 3-fold $\mathcal{X}_a$, whose toric diagram is given by one of the 16 reflexive polygons in $\mathbb{Z}^2$.
\end{proposition}
Here, the slope of the bound $\frac{64}{27}$ corresponds to the inverse of the slope of the bound in \eref{es04a20}.
This is not surprising given that the Euler number $\chi$ sets a lower bound on the minimum volume $V_{min}$ according to \eref{es04a30}.

For the 16 reflexive polygons $\Delta$ in \fref{fig_reflexive}, 
the Euler number $\chi=p$ given by the number of perimeter lattice points of $\Delta$
and 
the first Chern number $C_1=p^\circ$ given by the number of perimeter lattice of the dual polygon $\Delta^\circ$
are not independent to each other 
and satisfy the relationship
\beal{es04a32}
C_1+\chi=12 
~,~
\eea
for all reflexive polygons $\Delta$.
As a result, 
the bound in \eref{es04a31} in terms of the Euler number $\chi$ can be rewritten in terms of the Chern number $C_1$.
Accordingly, we have
\beal{es04a33}
\frac{1}{F(\mathcal{X}_a; \zeta_R, \eta_h)}
\leq \frac{64}{27}~
(12-C_1(X_a))
~,~
\eea
where 
$X_a$
is the toric variety associated to the toric Calabi-Yau 3-fold $\mathcal{X}_a$, whose toric diagram is given by one of the 16 reflexive polygons in $\mathbb{Z}^2$.
We note that the bound in terms of the Chern number $C_1$ is confirmed by the plot in \fref{C1vsFR}.
\\

\subsection{Divisor Volumes}\label{divvol}

\begin{figure}[H]
    \centering
    \resizebox{0.9\hsize}{!}{\includegraphics[width=1\linewidth]{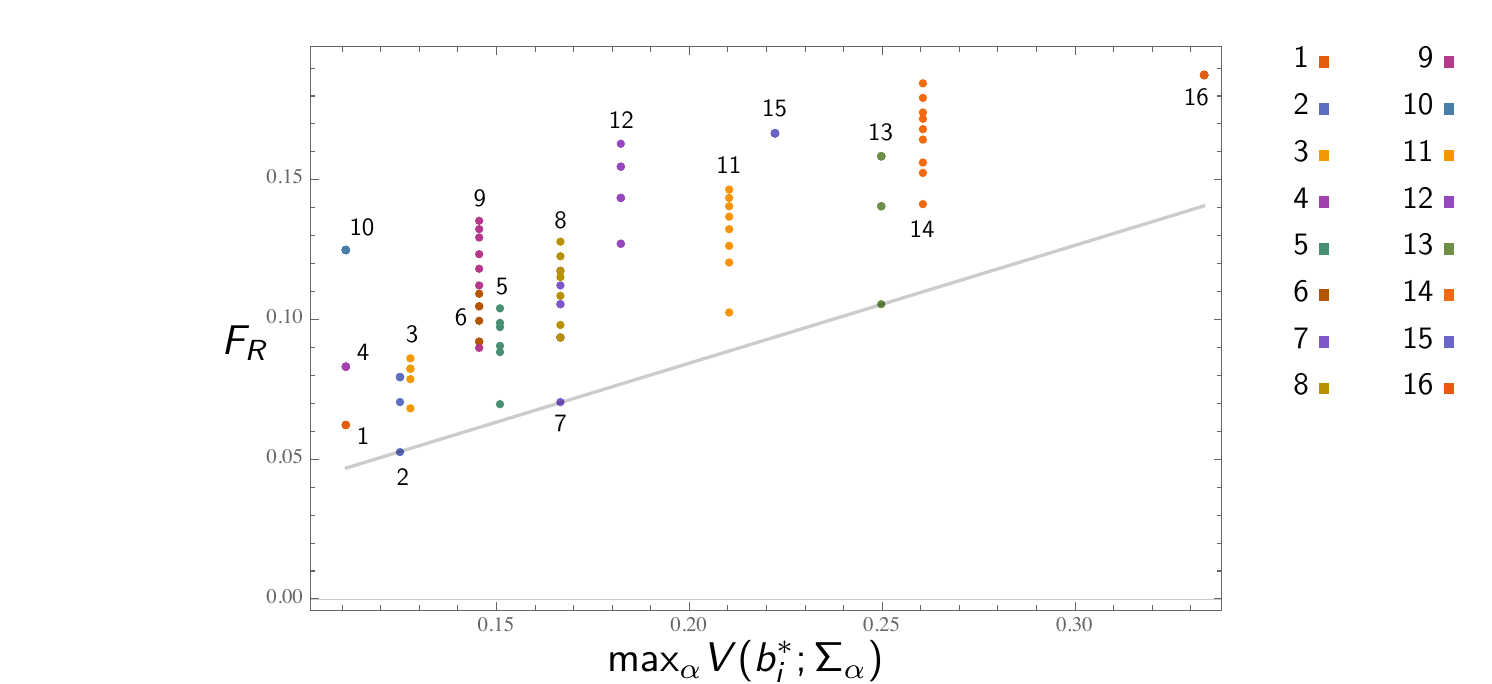}}
    \caption{
    The Futaki invariants $F(\mathcal{X}_a; \zeta_R, \eta_h)$ [$F_R$] against the maximum divisor volume $\max_\alpha V(b^*; \Sigma_{\alpha}^a)$ for the toric Calabi-Yau 3-folds $\mathcal{X}_a$ corresponding to the 16 reflexive polygons in $\mathbb{Z}^2$.
    }\label{divvolmaxvsFR}
\end{figure}

\begin{figure}[H]
    \centering
    \resizebox{0.9\hsize}{!}{\includegraphics[width=1\linewidth]{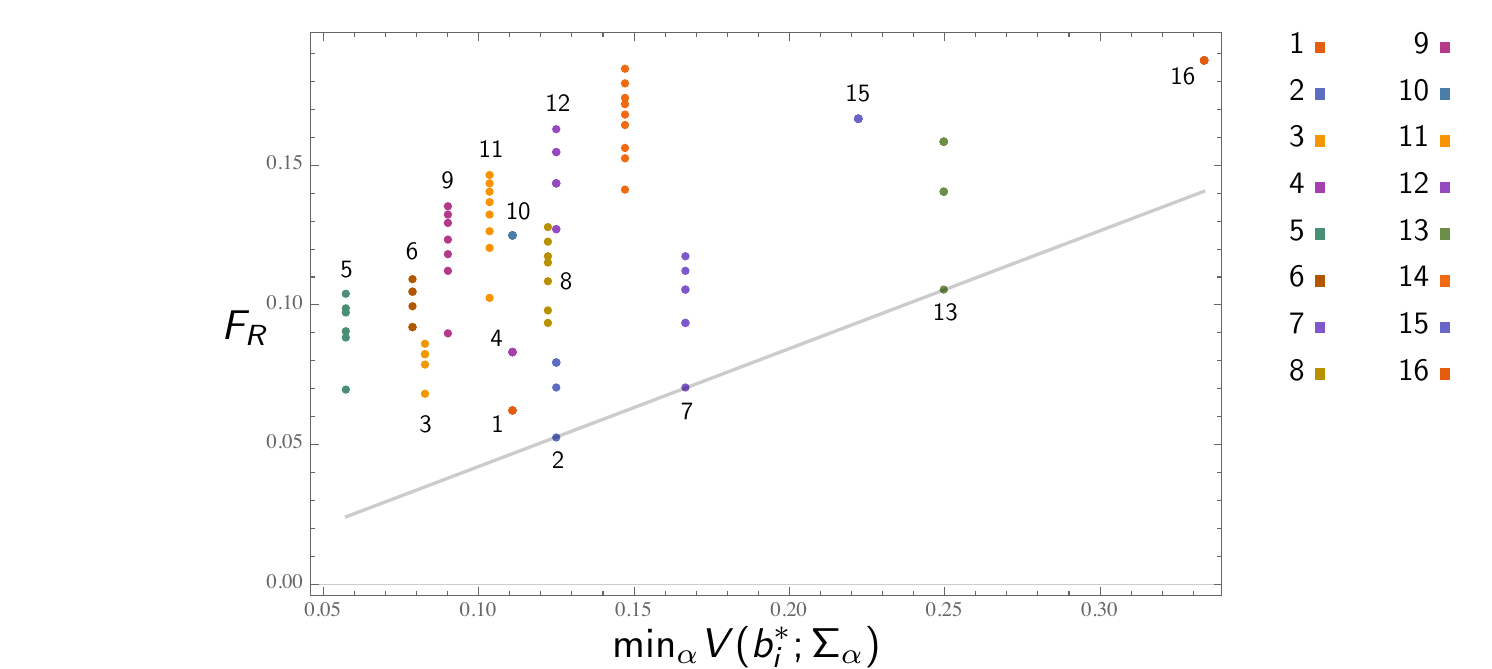}}
    \caption{
    The Futaki invariants $F(\mathcal{X}_a; \zeta_R, \eta_h)$ [$F_R$] against the minimum divisor volume $\min_\alpha V(b^*; \Sigma_{\alpha}^a)$ for the toric Calabi-Yau 3-folds $\mathcal{X}_a$ corresponding to the 16 reflexive polygons in $\mathbb{Z}^2$.
    }\label{divvolminvsFR}
\end{figure}

As discussed in section \sref{divisorvolumes}, 
besides the minimum volume of the Sasaki-Einstein base manifold $Y_a$ of the toric Calabi-Yau 3-folds $\mathcal{X}_a$ that we are considering here, 
we can also obtain volumes associated to the divisors $D_\alpha$
corresponding to the extremal GLSM fields $p_\alpha$ in $\mathcal{X}_a$.
Accordingly, each toric Calabi-Yau 3-fold $\mathcal{X}_a$ with its toric diagram given by one of the 16 reflexive polygons in \fref{fig_reflexive}
is associated to multiple divisors $D_\alpha^a$ with corresponding minimum volumes $V(b^*; \Sigma_{\alpha}^a)$.
In this section, we compare the Futaki invariants of the form $F(\mathcal{X}_a; \zeta_R, \eta_h)$, 
with the divisor volumes $V(b^*; \Sigma_{\alpha}^a)$.
In particular, 
we concentrate on the maximum divisor volume $\max_\alpha V(b^*; \Sigma_{\alpha}^a)$ and the minimum divisor volume $\min_\alpha V(b^*; \Sigma_{\alpha}^a)$ 
for each $\mathcal{X}_a$
and compare it with the corresponding Futaki invariants $F(\mathcal{X}_a; \zeta_R, \eta_h)$ for each generator of $\mathcal{X}_a$.
\fref{divvolmaxvsFR} and \fref{divvolminvsFR}
illustrate the plots of the Futaki invariants $F(\mathcal{X}_a; \zeta_R, \eta_h)$ against $\max_\alpha V(b^*; \Sigma_{\alpha}^a)$ and $\min_\alpha V(b^*; \Sigma_{\alpha}^a)$, respectively.

We observe that the Futaki invariants $F(\mathcal{X}_a; \zeta_R, \eta_h)$ have a lower bound in terms of the maximum divisor volume $\max_\alpha V(b^*; \Sigma_{\alpha}^a)$ and minimum divisor volume $\min_\alpha V(b^*; \Sigma_{\alpha}^a)$ for all toric Calabi-Yau 3-folds $\mathcal{X}_a$ corresponding to the 16 reflexive polygons in \fref{fig_reflexive}.
These lower bounds are given by, 
\beal{es04a40}
F(\mathcal{X}_a; \zeta_R, \eta_h)
\geq
\frac{27}{64}
\max_\alpha V(b^*; \Sigma_{\alpha}^a)
~,~
F(\mathcal{X}_a; \zeta_R, \eta_h)
\geq
\frac{27}{64}
\min_\alpha V(b^*; \Sigma_{\alpha}^a)
~,~
\eea
where the lower bounds are saturated for Models 2, 7 and 13 in \fref{fig_reflexive}
corresponding to the abelian orbifolds of the form 
$\mathbb{C}^3/\mathbb{Z}_4 \times \mathbb{Z}_2$ $(1,0,3)(0,1,1)$, $\mathbb{C}^3/\mathbb{Z}_6$ $(1,2,3)$ and $\mathbb{C}^3/\mathbb{Z}_4$ $(1,1,2)$, 
respectively.
The lower bounds in terms of $\max_{\alpha}V(b^*; \Sigma_{\alpha})$ and $\min_{\alpha}V(b^*; \Sigma_{\alpha})$ in \eref{es04a40}
coincide for these abelian orbifolds of $\mathbb{C}^3$ 
because for abelian orbifolds of $\mathbb{C}^3$ the 3 divisors $D_\alpha$ have all the same minimum volume,
which sets $\max_{\alpha}V(b^*; \Sigma_{\alpha})=\min_{\alpha}V(b^*; \Sigma_{\alpha})$.

We note that the divisor volumes $V(b^*; \Sigma_{\alpha})$ gives the $U(1)_R$ charge $R(p_\alpha)$
of the corresponding GLSM field $p_\alpha$ according to \eref{es02a53}.
Therefore, following the expression for the Futaki invariants $F(\mathcal{X}_a; \zeta_R, \eta_h)$ in \eref{es04a22}, we can derive
\beal{es04a41}
F(\mathcal{X}_a; \zeta_R, \eta_h)
=
\left(\frac{3}{2}-\frac{1}{R_h}\right)\frac{V_{min}}{6 \tilde{c}}
=
\left(\frac{3}{2}-\frac{1}{R_h}\right)\frac{1}{6 \tilde{c}} \frac{V(b_{i}^{*}; \Sigma_{\alpha})}{(3/2) R_{\alpha}}
~.~
\eea
The expression in the parenthesis above is exactly the same factor that we have seen in 
section \sref{Vminandtopinv} for minimized volumes $V_{min}$, 
which is smaller when the generator has a smaller $U(1)_R$ charge. 
Among the 16 reflexive polygons, the smallest possible $U(1)_R$ charge for a generator of $\mathcal{X}_a$ is $4/3$. 
Moreover, compared to the discussions in section \sref{Vminandtopinv}, 
there is an extra factor $\frac{2}{3R_{\alpha}}$ that determines the ratio in \eqref{es04a41}. From the data we have, it turns out that the largest possible $U(1)_R$ charge of a single GLSM field is equal to $2/3$. As a result, the Models 2, 7 and 13 would saturate both the minimal $R_h$ and the maximal $R_{\alpha}$.
Accordingly, 
we can see that the slope of the lower bound in \eref{es04a40}
originates from the lower bound based on the overall minimum volume $V_{min}$ of the Sasaki-Einstein 5-manifolds in \eref{es04a20}. 
Similar to the discussions on the minimized volumes, 
we observe the following, based on \fref{divvolminvsFR}, \tref{FRtableA} and \tref{FRtableB},
\begin{proposition}
The Futaki invariant
$F(\mathcal{X}_a; \zeta_R, \eta_h)$
under a test symmetry $\eta_h$ associated to the $h$-th generator of a toric Calabi-Yau 3-fold $\mathcal{X}_a$
has a lower bound defined by the divisor volumes $V(b^*; \Sigma_{\alpha})$
as follows, 
\beal{es04a42}
F(\mathcal{X}_a; \zeta_R, \eta_h)
\geq
\frac{27}{64}
V(b^*; \Sigma_{\alpha})
~,~
\eea
where $\mathcal{X}_a$ has a toric diagram given by one of the 16 reflexive polygons in $\mathbb{Z}^2$.
\end{proposition}
We expect that the above observation holds more generally, and conjecture
\begin{conjecture}
The Futaki invariant
$F(\mathcal{X}; \zeta_R, \eta_h)$
has a lower bound defined by \eref{es04a42} in terms of the divisor volumes $V(b^*; \Sigma_{\alpha})$
for all toric Calabi-Yau 3-folds $\mathcal{X}$,
where $\mathcal{X}$ has no factors of $\mathbb{C}$.
\end{conjecture}

\subsection{Integrated Curvatures}\label{Riem2}

We note here that higher order terms in the Laurent expansion of the Hilbert series $g(t; \mathcal{X}, \zeta)$, 
\beal{es04a50}
g(t=e^{-s}; \mathcal{X}, \zeta) = \frac{2 A_0(\zeta)}{s^3} + \frac{A_1(\zeta)}{s^{2}} + \frac{A_2(\zeta)}{s} \dots ~,~
\eea
have interpretations with regards to the Sasaki-Einstein base manifold $Y$ of the toric Calabi-Yau fold $\mathcal{X}$. 
For example, 
information about the integrated curvature $\int\text{Riem}^2$ of $Y$
is contained in the coefficient $A_2(\zeta)$ in \eref{es04a50} \cite{herbig2021laurent}.
In this section, we compare the values of the integrated curvatures $\int\text{Riem}^2$ 
for $\zeta=\zeta_R$ being the $U(1)_R$ symmetry
with the Futaki invariants of the form $F(\mathcal{X}_a; \zeta_R, \eta_h)$ for all $\mathcal{X}_a$ corresponding to the 16 reflexive polygons in \fref{fig_reflexive}.

\begin{figure}[H]
    \centering
    \resizebox{0.9\hsize}{!}{\includegraphics[width=1\linewidth]{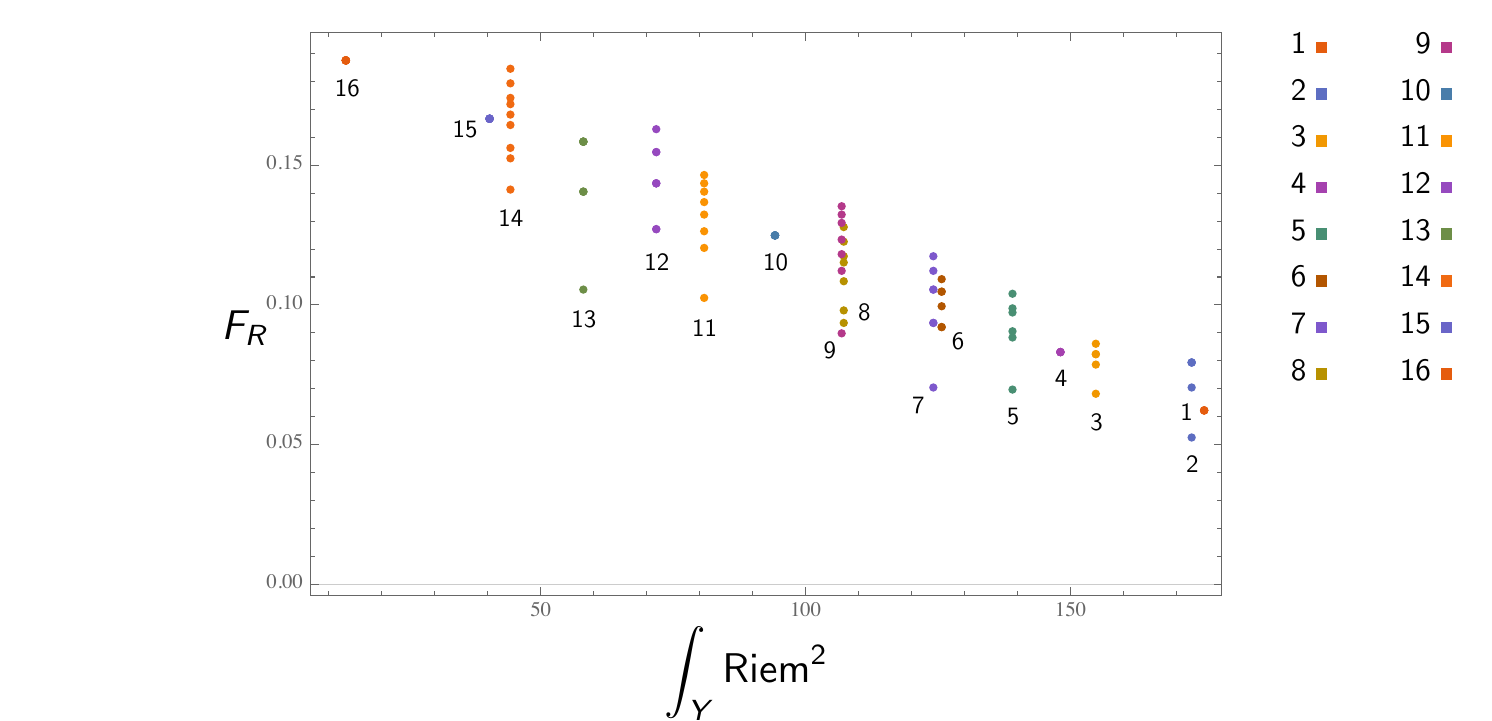}}
    \caption{
    The Futaki invariants $F(\mathcal{X}_a; \zeta_R, \eta_h)$ [$F_R$] against the 
    integrated curvatures $\int_{Y_a} \text{Riem}^2$ for the Sasaki-Einstein 5-manifolds $Y_a$
    corresponding to the 16 reflexive polygons in $\mathbb{Z}^2$.
    }\label{Riem2vsFR}
\end{figure}

\fref{Riem2vsFR} shows the plot 
for the Futaki invariants $F(\mathcal{X}_a; \zeta_R, \eta_h)$ against their corresponding integrated curvatures $\int\text{Riem}^2$ where $\mathcal{X}_a$ corresponds to the 16 reflexive polygons. 
Based on the plot, we see that there does not seem to be an obvious relationship between the integrated curvature and the Futaki invariants. 
This is somewhat not surprising, given that the Futaki invariants $F(\mathcal{X}_a; \zeta_R, \eta_h)$ only depend on the leading coefficients $A_0(\zeta_R)$ and $A_1(\zeta_R)=2A_0(\zeta_R)$
in the Laurent expansion of the Hilbert series in \eref{es04a50}.

Nevertheless, from \cite{herbig2021laurent}, we know that the leading coefficients in the Laurent expansion of the Hilbert series of a Gorenstein ring 
are not all independent. 
Due to the convention of the weights and coefficients adopted in this work, 
our cases do not fit into the conditions in Theorem 1.1 in \cite{herbig2021laurent}. 
However, there should still be some relations among the leading coefficients in the Laurent expansion of the Hilbert series that we want to explore here.

\begin{figure}[H]
    \centering
    \resizebox{0.9\hsize}{!}{\includegraphics[width=1\linewidth]{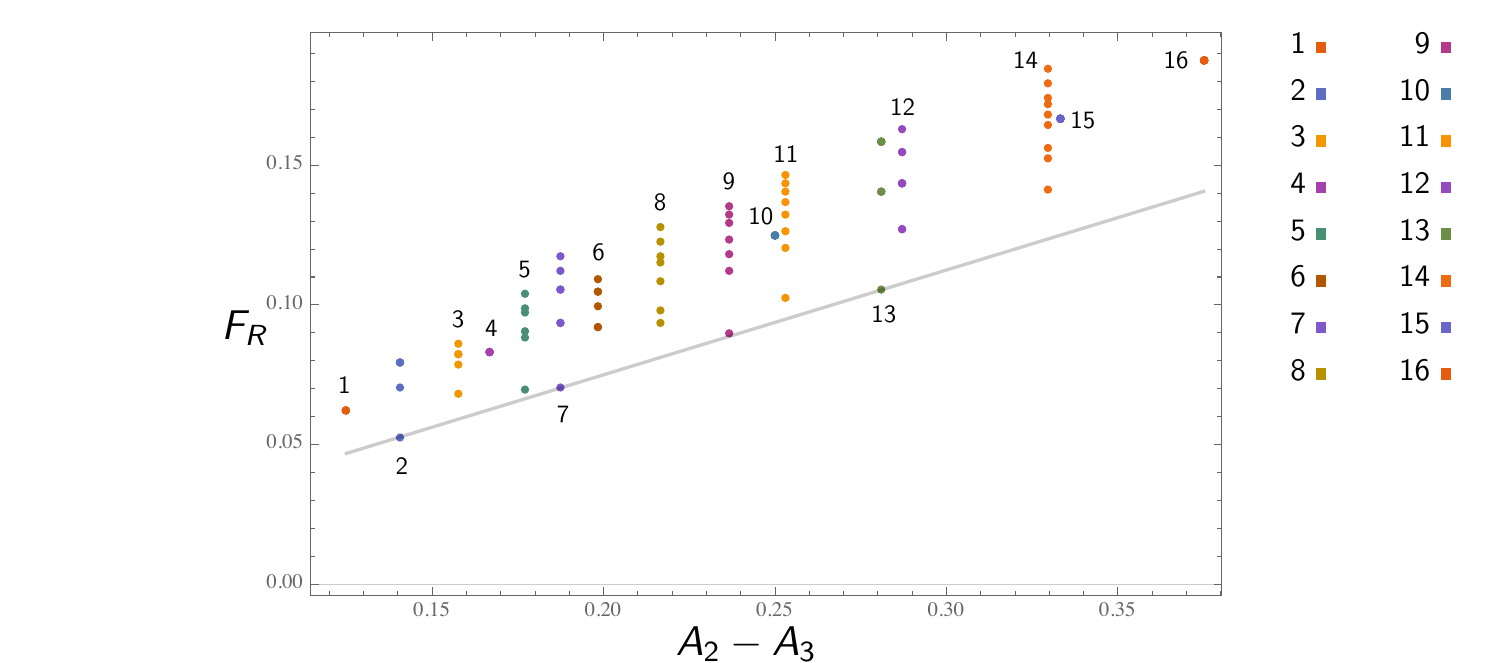}}
    \caption{
    The Futaki invariants $F(\mathcal{X}_a; \zeta_R, \eta_h)$ [$F_R$] against the difference $A_2(\zeta_R)-A_3(\zeta_R)$ for the toric Calabi-Yau 3-folds $\mathcal{X}_a$ corresponding to the 16 reflexive polygons in $\mathbb{Z}^2$.
    }\label{A2A3vsFR}
\end{figure}

The very first relation among the leading coefficients,
\beal{es04a51}
2A_0(\zeta_R)-A_1(\zeta_R)=0,
\eea
which has already been used throughout the paper. 
Motivated by this, let us plot 
the Futaki invariants $F(\mathcal{X}_a; \zeta_R, \eta_h)$ against 
the difference $A_2(\zeta_R)-A_3(\zeta_R)$
for all $\mathcal{X}_a$ corresponding to the 16 reflexive polygons in \fref{fig_reflexive}. 
The resulting plot is shown in \fref{A2A3vsFR}.

We can clearly see in \fref{A2A3vsFR}
that the Futaki invariants $F(\mathcal{X}_a; \zeta_R, \eta_h)$ exhibit a lower bound.
We therefore summarize,
\begin{proposition}
The Futaki invariant $F(\mathcal{X}_a; \zeta_R, \eta_h)$ under a test symmetry $\eta_h$ associated to the $h$-th generator of $\mathcal{X}_a$
has a lower bound in terms of the difference of
coefficients $A_2(\zeta_R)-A_3(\zeta_R)$ as follows,
\beal{es04a51}
F(\mathcal{X}_a; \zeta_R, \eta_h)
\geq
\frac{3}{8}(A_2(\zeta_R)-A_3(\zeta_R))
~,~
\eea
where 
$\mathcal{X}_a$
has a toric diagram given by one of the 16 reflexive polygons in $\mathbb{Z}^2$.
\end{proposition}
We expect this bound to hold more generally and conjecture
\begin{conjecture}
The Futaki invariant $F(\mathcal{X}; \zeta_R, \eta_h)$ 
has a lower bound given by \eref{es04a51} in terms of the difference of
coefficients $A_2(\zeta_R)-A_3(\zeta_R)$
for all toric Calabi-Yau 3-folds $\mathcal{X}$, where $\mathcal{X}$ has no factors of $\mathbb{C}$.
\end{conjecture}

Considering only the toric Calabi-Yau 3-folds $\mathcal{X}_a$ with reflexive polygons as their toric diagrams, 
we observe that
the points lying on the lower bound
correspond to $\mathcal{X}_a$ where some of generators 
have the minimum $U(1)_R$ charge $R_h=4/3$. 
This is precisely the case for Models 2, 7 and 13 in \fref{fig_reflexive},
which correspond to the abelian orbifolds of the form $\mathbb{C}^3/\mathbb{Z}_4 \times \mathbb{Z}_2$ $(1,0,3)(0,1,1)$, $\mathbb{C}^3/\mathbb{Z}_6$ $(1,2,3)$ and $\mathbb{C}^3/\mathbb{Z}_4$ $(1,1,2)$.

Given the definition of the Futaki invariants $F(\mathcal{X}_a; \zeta_R, \eta_h)$ in \eref{es02a77},
when the bound in \eref{es04a51}, we have
\beal{es04a52}
\frac{A_0(\zeta_R)}{2}
- 
\frac{A_0(\zeta_R)}{3 R_h}
=
\frac{3}{8}
(A_2(\zeta_R)-A_3(\zeta_R))
~,~
\eea
with $R_h = 4/3$, 
which then simplifies to
\beal{es04a53}
2 A_0(\zeta_R)
- 3 A_2(\zeta_R)
+ 3 A_3(\zeta_R)
= 0
~.~
\eea
Under \eref{es04a51}, 
we then have
\beal{es04a54}
A_1(\zeta_R)-3A_2(\zeta_R)+3A_3(\zeta_R)=0
~.~
\eea

By considering the higher order coefficients in the Laurent expansion of the Hilbert series 
and by studying their relations with the corresponding Futaki invariants, 
we believe that finding relations of the form in \eref{es04a53} and \eref{es04a54} above
will help us better understand
the K-stability of mesonic moduli spaces of supersymmetric gauge theories. 
Having in mind that the coefficient $A_2(\zeta_R)$ here is related to the
integrated curvature of the Sasaki-Einstein base manifold $Y$, 
we would require more information following coefficients such as 
$A_3(\zeta_R)$ to derive higher order relations amongst the coefficients. 
In fact, we believe that it is possible to introduce notions of generalized K-stabilities of 
the mesonic moduli spaces of supersymmetric gauge theories
that are determined by the higher order coefficients in the Laurent expansion of the associated Hilbert series. 
We leave this analysis for future work.
\\

\section{Comments on K-stability and Discussions \label{KunstableSCFT}} \setall

The K-stability of the moduli spaces of supersymmetric gauge theories has been studied in various contexts \cite{Benvenuti:2017lle, Xie:2019qmw, Amariti:2019pky, Fazzi:2019gvt, Bao:2020ugf}
since the original work in \cite{Collins:2016icw}.
The conjecture in \cite{Collins:2016icw}
has been that when the classical moduli space of a $4d$ $\mathcal{N}=1$ supersymmetric gauge theory
is K-stable, the $4d$ $\mathcal{N}=1$ theory flows in the IR to a $4d$ superconformal field theory.
This is certainly the case for the family of $4d$ $\mathcal{N}=1$ supersymmetric gauge theories that we study in this work,
which are worldvolume theories of D3-branes probing toric Calabi-Yau 3-folds. 
Focusing on toric Calabi-Yau 3-folds corresponding to the 16 reflexive polygons in \fref{fig_reflexive},
we have shown through explicit computations in this work that
the Futaki invariants under the test symmetries associated to each generator of the mesonic moduli spaces
are all positive.

In \cite{Bao:2020ugf}, it was discovered 
that certain $4d$ $\mathcal{N}=1$ supersymmetric gauge theories with moduli spaces that are K-stable
do not have an associated superconformal field theory.
The examples described in \cite{Bao:2020ugf}
involve $4d$ SQCD theories
that are outside the conformal window, even though their 
moduli spaces are K-stable. 
This indicates that K-stability not necessarily implies the existence of a corresponding superconformal field theory for all families of $4d$ supersymmetric gauge theories.
We therefore believe that for certain families of $4d$ supersymmetric gauge theories, the notion of K-stability has to be extended in order to compensate for this discrepancy. 
This might include generalized Futaki invariants as mentioned in section \sref{Riem2}.

In fact, in
\cite{Benvenuti:2017lle}
the notion of stability 
was extended 
for moduli spaces of $4d$ $\mathcal{N}=2$ supersymmetric theories
with associated superconformal field theories. 
An example considered in \cite{Benvenuti:2017lle} is the $A_3$ Argyres-Douglas theory, 
whose combined Higgs and Coulomb branch moduli space can be shown to be unstable under ordinary K-stability, following the computation of Futaki invariants from Hilbert series as outlined in this work.

We hope to investigate similar extensions of K-stability and the introduction of generalized Futaki invariants in future work.
For now, our results in this work summarize how Futaki invariants 
for the mesonic moduli spaces of the family $4d$ $\mathcal{N}=1$ supersymmetric gauge theories corresponding to toric Calabi-Yau 3-folds
form novel relations between geometric and topological invariants such as the Euler number and Chern numbers, the minimum volume of Sasaki-Einstein 5-manifolds
as well as the volumes of divisors.
\\

\section*{Acknowledgments}
J. B. is supported by a JSPS fellowship. Y.-H. H. is supported by STFC grant ST/J00037X/2. R.-K. S. is supported by a Basic Research Grant of the National Research Foundation of Korea (NRF2022R1F1A1073128). He is also supported by a Start-up Research Grant for new faculty at UNIST (1.210139.01)
and a UNIST AI Incubator Grant (1.240022.01). He is also partly supported by the BK21 Program (``Next Generation Education Program for Mathematical Sciences'', 4299990414089) funded by the Ministry of Education in Korea and the National Research Foundation of Korea.

\appendix

\section{Some Exact Values}\label{exactvalues}

Some of the $U(1)_R$ charges $R_\alpha$, 
certain leading coefficients of the Laurent expansion of the Hilbert series,
Futaki invariants $F$ as well as the norm of the test symmetries $\eta$
shown 
in \tref{FRtableA} to \tref{FPtableB}
are algebraic numbers that can be obtained exactly by solving for roots of polynomial equations. 
In the following section, we summarize these polynomials
in \tref{exactR} to \tref{exactnorm},
where dividing by the leading coefficient yields the corresponding minimal polynomials.
In these tables, 
we indicate which $n$-th root of the polynomials corresponds to the quantity in question. 
Here, the ordering of the roots is determined as follows:
any two roots to an equation of the form $z_1=x_1+iy_1$ and $z_2=x_2+iy_2$
are ordered as 
if $x_1\leq x_2$ or 
if $x_1=x_2$ and $y_1\leq y_2$.

\begin{table}[H]
    \centering
    \resizebox{0.8\hsize}{!}{
        \begin{tabular}{|>{\centering\arraybackslash}p{2.7cm}|p{12.5cm}|>{\centering\arraybackslash}p{2cm}|>{\centering\arraybackslash}p{1.5cm}|}
            \hline
            $U(1)_R$ charge & Polynomial & $x_n$ & $n$\\
            \hline
            $r_{5 , 1}$ & $27 x^4 - 162 x^3 + 180 x^2 + 28 x - 48$ & $0.5775$ & 2\\
            $r_{5 , 2}$ & $81 x^4 + 162 x^3 - 36 x^2 - 52 x - 8$ & $0.6398$ & 4\\
            $r_{5 , 3}$ & $81 x^4 - 486 x^3 + 288 x^2 + 448 x - 256$ & $0.5393$ & 2\\
            $r_{5 , 4}$ & $27 x^4 + 54 x^3 - 432 x^2 + 496 x - 96$ & $0.2434$ & 2\\
            
            $r_{6 , 1} = r_{6,5}$ & $3 x^3 - 340 x^2 - 24 x + 72$ &  $0.427$ & 2\\
            $r_{6 , 2} = r_{6,4}$ & $3 x^3 + 206 x^2 - 384 x + 96$ & $0.2978$ & 2\\
            $r_{6 , 3}$ & $3 x^3 + 250 x^2 - 124 x - 8$ & $0.5505$ & 3\\
            
            $r_{11 , 1}$ & $27 x^4 + 126 x^3 + 36 x^2 - 52 x - 16$ & $0.6223$ & 4\\
            $r_{11 , 2}$ & $3 x^4 - 26 x^3 + 4 x^2 + 52 x - 24$ & $0.5016$ & 2\\
            $r_{11 , 3}$ & $27 x^4 + 126 x^3 - 864 x^2 + 1088 x - 256$ & $0.3062$ & 2\\
            $r_{11 , 4}$ & $9 x^4 - 78 x^3 + 112 x^2 + 16 x - 32$ & $0.5698$ & 2\\
            \hline
        \end{tabular}
    }
    \caption{
    The $U(1)_R$ charges $r_{a, \alpha}$ on GLSM fields $p_\alpha$ of Models 5, 6 and 11 in \tref{FRtableA} and \tref{FRtableB} expressed as roots of polynomial equations.
    }\label{exactR}
\end{table}

\begin{table}[H]
    \centering
    \resizebox{0.8\hsize}{!}{
        \begin{tabular}{|>{\centering\arraybackslash}p{2.7cm}|p{12.5cm}|>{\centering\arraybackslash}p{2cm}|>{\centering\arraybackslash}p{1.5cm}|}
            \hline
            $A_0$ & Polynomial & $x_n$ &$n$\\
            \hline
            $A_{5 , 0}$ & $18874368 x^4 + 15597568 x^3 - 580608 x^2 - 1224720 x - 19683$ & $0.2655$ & 4\\
            $A_{6 , 0}$ & $5308416 x^3 + 1999872 x^2 - 1064720 x-27$ & $0.2975$ &3\\
            $A_{11 , 0}$ & $6291456 x^4 + 3276800 x^3 - 1336320 x^2 - 304560 x - 2187$ & $0.3799$ & 4\\
            \hline
        \end{tabular}
    }
    \caption{
    The leading coefficients $A_{a,0}$ in the Laurent expansion of the Hilbert series under $\zeta_R$ for Models 5, 6 and 11 in \tref{FRtableA} and \tref{FRtableB} expressed as roots of polynomial equations.}\label{exactA0}
\end{table}

\begin{table}[H]
    \centering
    \resizebox{0.8\hsize}{!}{
        \begin{tabular}{|>{\centering\arraybackslash}p{2.7cm}|p{12.5cm}|>{\centering\arraybackslash}p{2cm}|>{\centering\arraybackslash}p{1.5cm}|}
            \hline
            $U(1)_R$ & Polynomial & $x_n$ & $n$\\
            \hline
            $R_{5 , 1}$ & $9 x^4 - 90 x^3 + 288 x^2 - 320 x + 96$ & $1.3983$ & 2\\
            $R_{5 , 2}$ & $27 x^4 - 144 x^2 - 80 x + 48$ & $2.4970$ & 4\\
            $R_{5 , 3}$ & $x - 2$ & $2$ &1\\
            $R_{5 , 4}$ & $27 x^4 + 54 x^3 - 288 x^2 - 512 x + 256$ & $3.0986$ & 4\\
            $R_{5 , 5}$ & $9 x^4 - 54 x^3 + 72 x^2 + 32 x - 32$ & $2.6017$ & 3\\
            $R_{5 , 6}$ & $27 x^4 - 378 x^3 + 1728 x^2 - 2944 x + 1536$ & $2.1047$ & 2\\
            \hline
            $R_{6 , 1}$ & $3 x^3 + 26 x^2 - 88 x + 32$ & $2.2527$ & 3\\
            $R_{6 , 2}$ & $3 x^3 - 62 x^2 + 264 x - 288$ & $1.7473$ & 1\\
            $R_{6 , 3}$ & $3 x^3 + 70 x^2 - 108 x - 216$ & $2.5054$ & 3\\
            $R_{6 , 4}$ & $x - 2$ & $2$ &1\\
            $R_{6 , 5}$ & $3 x^3 + 26 x^2 - 88 x + 32$ & $2.2527$ & 3\\
            $R_{6 , 6}$ & $3 x^3 - 62 x^2 + 264 x - 288$ & $1.7473$ &1\\
            \hline
            $R_{11 , 1}$ & $27 x^4 + 18 x^3 - 96 x^2 - 64 x + 32$ & $1.8145$ & 4\\
            $R_{11 , 2}$ & $27 x^4 - 342 x^3 + 1344 x^2 - 1664 x + 512$ & $1.4458$ & 2\\
            $R_{11 , 3}$ & $3 x^4 + 16 x^3 - 16 x^2 - 112 x + 48$ & $2.3686$ & 4\\
            $R_{11 , 4}$ & $x - 2$ & $2$ &1\\
            $R_{11 , 5}$ & $27 x^4 - 90 x^3 - 168 x^2 + 416 x + 384$ & $2.5542$ &3\\
            $R_{11 , 6}$ & $27 x^4 - 450 x^3 + 2712 x^2 - 6944 x + 6304$ & $2.1855$ &1\\
            $R_{11 , 7}$ & $9 x^4 - 108 x^3 + 376 x^2 - 352 x - 144$ & $2.7397$ &2\\
            $R_{11 , 8}$ & $3 x^4 - 62 x^3 + 416 x^2 - 1024 x + 768$ & $2.9252$ &2\\
            \hline
        \end{tabular}
    }
    \caption{
    The $U(1)_R$ charges $R_{a, h}$ on generators $x_h$ of Models 5, 6 and 11 in \tref{FRtableA} and \tref{FRtableB} expressed as roots of polynomial equations.
    }\label{exactk}
\end{table}

\begin{table}[H]
    \centering
    \resizebox{0.8\hsize}{!}{
        \begin{tabular}{|>{\centering\arraybackslash}p{2.7cm}|p{12.5cm}|>{\centering\arraybackslash}p{2.0cm}|>{\centering\arraybackslash}p{1.5cm}|}
            \hline
            $F_{a,h}$ & Polynomial & $x_n$ & $n$\\
            \hline
            $F_{5 , 1}$ & $679477248 x^4 - 54722560 x^3 - 13976064 x^2 + 912708 x + 6561$ & $0.0695$ & 3\\
            $F_{5 , 2}$ & $679477248 x^4 + 262537216 x^3 - 82446336 x^2 + 4304016 x + 59049$ & $0.0973$ & 3\\
            $F_{5 , 3}$ & $56623104 x^4 + 15597568 x^3 - 193536 x^2 - 136080 x - 729$ & $0.0885$ & 4\\
            $F_{5 , 4}$ & $3623878656 x^4 - 437780480 x^3 - 167712768 x^2 + 16428744 x + 177147$ & $0.1042$ & 3\\
            $F_{5 , 5}$ & $452984832 x^4 + 235798528 x^3 + 15510528 x^2 - 4333176 x + 6561$ & $0.0987$ & 4\\
            $F_{5 , 6}$ & $21743271936 x^4 + 6592397312 x^3 - 371589120 x^2 - 34379640 x - 216513$ & $0.0907$ & 4\\
            \hline
            $F_{6 , 1}$ & $95551488 x^3 - 29210112 x^2 + 2010878 x + 27$ & $0.1047$ & 2\\
            $F_{6 , 2}$ & $3439853568 x^3 + 627388416 x^2 - 86818648 x - 1053$ & $0.0920$ & 3\\
            $F_{6 , 3}$ & $322486272 x^3 + 135737856 x^2 - 18659888 x - 243$ & $0.1092$ & 3\\
            $F_{6 , 4}$ & $47775744 x^3 + 5999616 x^2 - 1064720 x - 9$ & $0.0992$ & 3\\
            $F_{6 , 5}$ & $95551488 x^3 - 29210112 x^2 + 2010878 x + 27$ & $0.1047$ & 2\\
            $F_{6 , 6}$ & $3439853568 x^3 + 627388416 x^2 - 86818648 x - 1053$ & $0.0920$ & 3\\
            \hline
            $F_{11 , 1}$ & $75497472 x^4 + 28639232 x^3 - 7856640 x^2 + 381348 x + 2187$ & $0.1202$ & 4\\
            $F_{11 , 2}$ & $4831838208 x^4 - 908066816 x^3 - 19169280 x^2 + 6102864 x + 19683$ & $0.1024$ & 3\\
            $F_{11 , 3}$ & $56623104 x^4 - 14188544 x^3 - 399360 x^2 + 169524 x + 729$ & $0.1365$ & 3\\
            $F_{11 , 4}$ & $18874368 x^4 + 3276800 x^3 - 445440 x^2 - 33840 x - 81$ & $0.1266$ & 4\\
            $F_{11 , 5}$ & $3623878656 x^4 + 993001472 x^3 - 79368192 x^2 - 17626896 x - 115911$ & $0.1404$ & 4\\
            $F_{11 , 6}$ & $29746003968 x^4 + 7597064192 x^3 - 832195584 x^2 - 88849224 x - 282123$ & $0.1320$ & 4\\
            $F_{11 , 7}$ & $339738624 x^4 + 155385856 x^3 + 1391616 x^2 - 4333176 x - 12393$ & $0.1437$ & 4\\
            $F_{11 , 8}$ & $7247757312 x^4 + 1716518912 x^3 - 292761600 x^2 - 16502256 x - 51759$ & $0.1467$ & 4\\
            \hline
        \end{tabular}
    }
    \caption{
    The Futaki invariants $F_{a, h}$ for test symmetry $\eta_h$ of Models 5, 6 and 11 in \tref{FRtableA} and \tref{FRtableB} expressed as roots of polynomial equations.
    }\label{exactF}
\end{table}

\begin{table}[H]
    \centering
    \resizebox{0.8\hsize}{!}{
        \begin{tabular}{|>{\centering\arraybackslash}p{2.2cm}|p{14cm}|>{\centering\arraybackslash}p{1.75cm}|>{\centering\arraybackslash}p{0.75cm}|}
            \hline
            $||\eta_{a,h}||^2$ & Polynomial & $x_n$ & $n$\\
            \hline
            $||\eta_{5 , 1}||^2$ & $25048249270272 x^4 + 4535015702528 x^3 - 31482445824 x^2 - 31072896 x - 2187$ & $0.0075$ &4\\
            $||\eta_{5 , 2}||^2$ & $6262062317568 x^4 + 1031429685248 x^3 + 26005929984 x^2 - 59066496 x - 19683$ & $0.0024$ &4\\
            $||\eta_{5 , 3}||^2$ & $695784701952 x^4 + 7985954816 x^3 - 4128768 x^2 - 120960 x - 27$ & $0.0037$ &4\\
            $||\eta_{5 , 4}||^2$ & $178120883699712 x^4 + 44181254832128 x^3 + 79796109312 x^2 - 214710912 x - 19683$ & $0.0015$ &4\\
            $||\eta_{5 , 5}||^2$ & $2783138807808 x^4 + 115091701760 x^3 + 410517504 x^2 - 466560 x - 2187$ & $0.0022$ &4\\
            $||\eta_{5 , 6}||^2$ & $6412351813189632 x^4 + 112783693709312 x^3 - 318547427328 x^2 - 420743808 x - 19683$ & $0.0033$ &4\\
            \hline
            $||\eta_{6 , 1}||^2$ & $3522410053632 x^3 + 744876933120 x^2 - 2463354496 x-27 $  & $0.0033$ &3\\
            $||\eta_{6 , 2}||^2$ & $285315214344192 x^3 - 711858585600 x^2 - 4508295808 x-27 $  & $0.0054$ &3\\
            $||\eta_{6 , 3}||^2$ & $160489808068608 x^3 + 4009006006272 x^2 - 11668379776 x-27 $  & $0.0026$ &3\\
            $||\eta_{6 , 4}||^2$ & $220150628352 x^3 + 1151926272 x^2 - 8517760 x-3 $  & $0.0041$ &3\\
            $||\eta_{6 , 5}||^2$ & $3522410053632 x^3 + 744876933120 x^2 - 2463354496 x-27 $  & $0.0033$ &3\\
            $||\eta_{6 , 6}||^2$ & $285315214344192 x^3 - 711858585600 x^2 - 4508295808 x-27 $  & $0.0054$ &3\\
            \hline
            $||\eta_{11 , 1}||^2$ & $927712935936 x^4 + 98717138944 x^3 + 1722286080 x^2 - 15002496 x - 2187$ & $0.0064$ &4\\
            $||\eta_{11 , 2}||^2$ & $237494511599616 x^4 + 43097849331712 x^3 - 446449582080 x^2 - 130211712 x - 2187$ & $0.0101$ &4\\
            $||\eta_{11 , 3}||^2$ & $2087354105856 x^4 + 480533020672 x^3 - 1475543040 x^2 - 1354112 x - 27$ & $0.0038$ &4\\
            $||\eta_{11 , 4}||^2$ & $231928233984 x^4 + 1677721600 x^3 - 9502720 x^2 - 30080 x - 3$ & $0.0053$ &4\\
            $||\eta_{11 , 5}||^2$ & $133590662774784 x^4 + 410169376768 x^3 - 1186725888 x^2 - 4302720 x - 2187$ & $0.0032$ &4\\
            $||\eta_{11 , 6}||^2$ & $36003611330740224 x^4 - 78155721211904 x^3 - 317731110912 x^2 - 175996800 x - 2187$ & $0.0044$ &4\\
            $||\eta_{11 , 7}||^2$ & $18786186952704 x^4 + 1907535904768 x^3 - 3661824000 x^2 - 5118336 x - 243$ & $0.0028$ &4\\
            $||\eta_{11 , 8}||^2$ & $534362651099136 x^4 + 1663226085376 x^3 - 6173491200 x^2 - 2900864 x - 27$ & $0.0025$ &4\\
            \hline
        \end{tabular}
    }
    \caption{
    The norms $||\eta_{a, h}||^2$ for test symmetry $\eta_h$ of Models 5, 6 and 11 in \tref{FRtableA} and \tref{FRtableB} expressed as roots of polynomial equations.
    }\label{exactnorm}
\end{table}

\section{Plots for Futaki Invariants $F(\mathcal{X}, \zeta_p, \eta)$}\label{plotsFp}

\begin{figure}[H]
    \centering
    \resizebox{0.9\hsize}{!}{\includegraphics[width=1\linewidth]{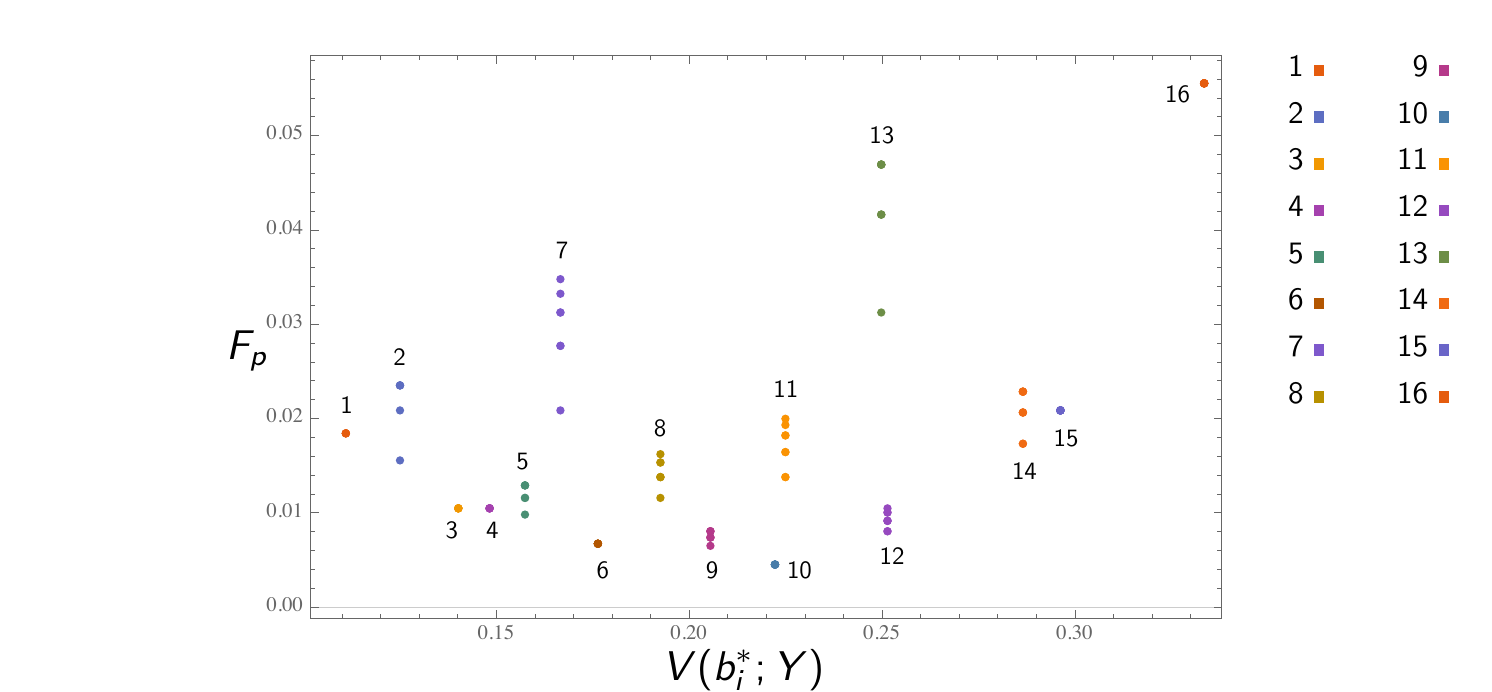}}
    \caption{The minimum volume $V_{min} = V(b_i^*; Y_a)$ of the Sasaki-Einstein $5$-manifold $Y_a$ associated to the toric Calabi-Yau 3-fold $\mathcal{X}_a$ with one of the 16 reflexive polygons as its toric diagram, plotted against the Futaki invariants $F(\mathcal{X}_a; \zeta_p, \eta_h)$ for all generators $x_h$ of $\mathcal{X}_a$.}\label{VminvsFp}
\end{figure}

\begin{figure}[H]
    \centering
    \resizebox{0.9\hsize}{!}{\includegraphics[width=1\linewidth]{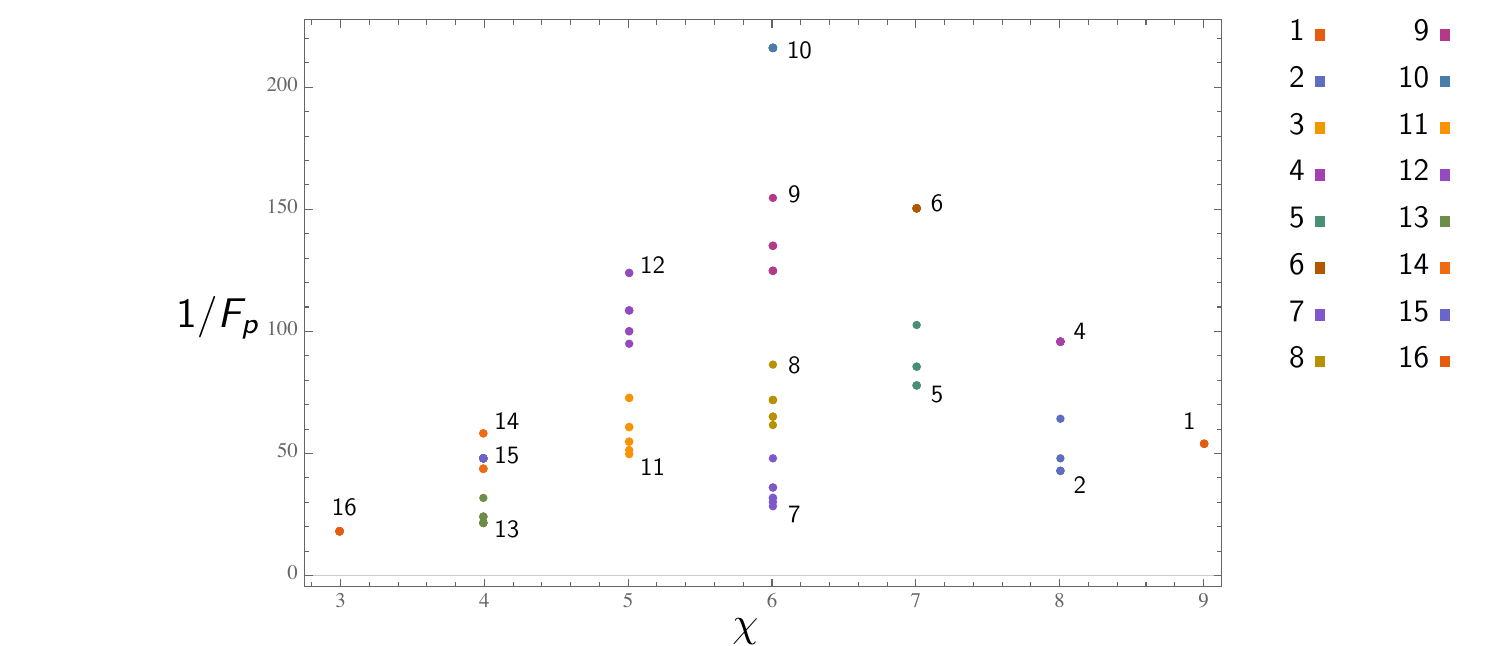}}
    \caption{
    The inverse of the Futaki invariants $F(\mathcal{X}_a; \zeta_p, \eta_h)$ [$F_p$] against the Euler number $\chi$ of the resolved toric varieties $X_a$ corresponding to the toric Calabi-Yau 3-fold $\mathcal{X}_a$ with the 16 reflexive polygons in $\mathbb{Z}^2$ as their toric diagrams.
    }
    \label{chivsFp}
\end{figure}

\begin{figure}[H]
    \centering
    \resizebox{0.9\hsize}{!}{\includegraphics[width=1\linewidth]{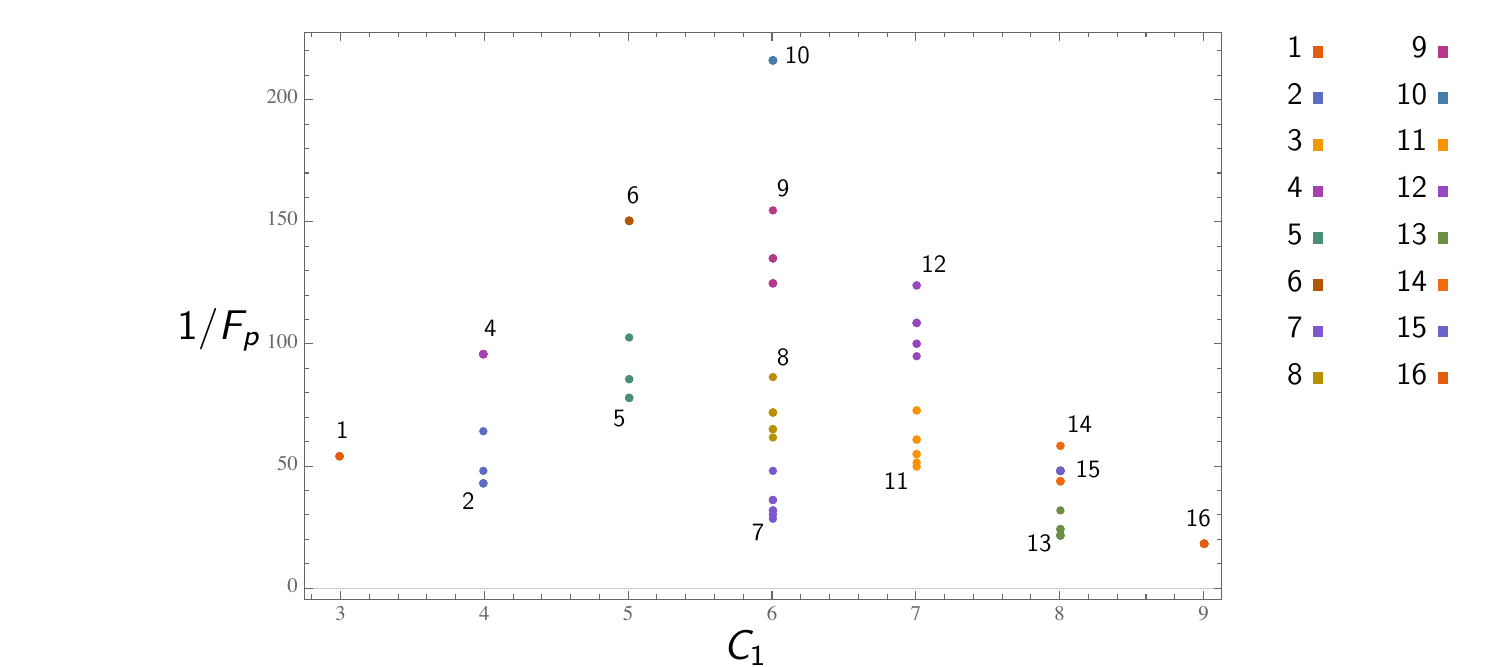}}
    \caption{
    The inverse of the Futaki invariants $F(\mathcal{X}_a; \zeta_p, \eta_h)$ [$F_p$] agains the first Chern number $C_1$ of the resolved toric varieties $X_a$ corresponding to the toric Calabi-Yau 3-fold $\mathcal{X}_a$ with the 16 reflexive polygons in $\mathbb{Z}^2$ as their toric diagrams.
    }\label{C1vsFp}
\end{figure}

\begin{figure}[H]
    \centering
    \resizebox{0.9\hsize}{!}{\includegraphics[width=1\linewidth]{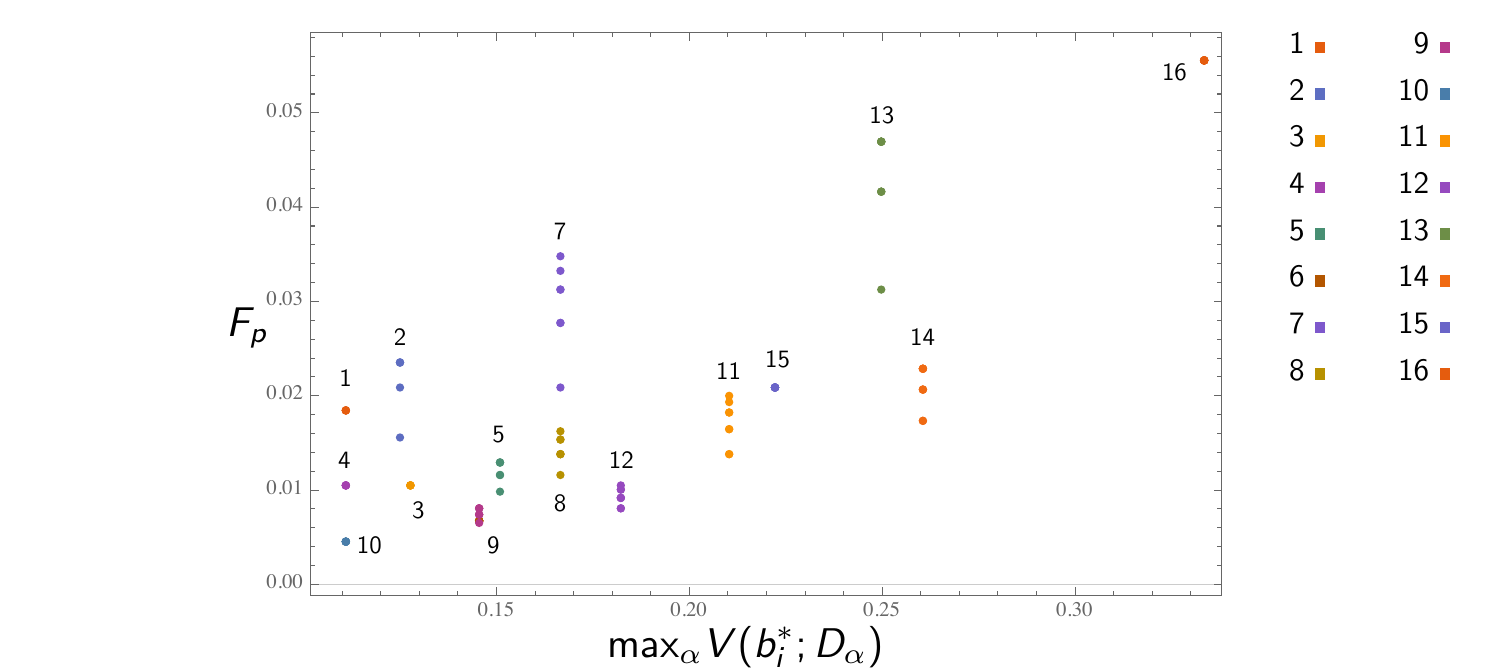}}
    \caption{
    The Futaki invariants $F(\mathcal{X}_a; \zeta_p, \eta_h)$ [$F_p$] against the maximum divisor volume $\max_\alpha V(b^*; \Sigma_\alpha^a)$ for the toric Calabi-Yau 3-folds $\mathcal{X}_a$ corresponding to the 16 reflexive polygons in $\mathbb{Z}^2$.
    }\label{divvolmaxvsFp}
\end{figure}

\begin{figure}[H]
    \centering
    \resizebox{0.9\hsize}{!}{\includegraphics[width=1\linewidth]{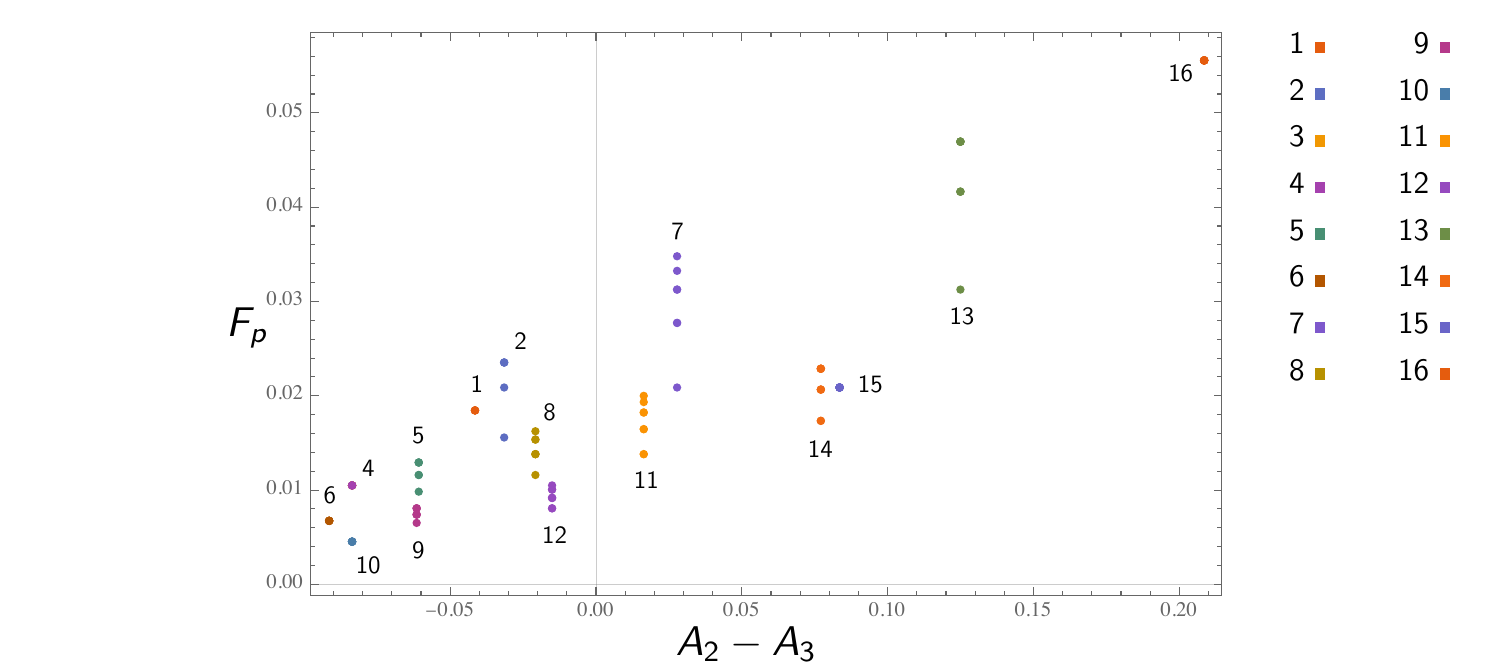}}
    \caption{
    The Futaki invariants $F(\mathcal{X}_a; \zeta_p, \eta_h)$ [$F_p$] against the difference $A_2(\zeta_p)-A_3(\zeta_p)$ for the toric Calabi-Yau 3-folds $\mathcal{X}_a$ corresponding to the 16 reflexive polygons in $\mathbb{Z}^2$.
    }\label{A2A3vsFp}
\end{figure}

In this section, we present plots in \fref{VminvsFp} to \fref{A2A3vsFp} involving the Futaki invariants of the form
$F(\mathcal{X}_a; \zeta_p, \eta_h)$
where the refinement of the Hilbert series of $\mathcal{X}_a$ is under $\zeta_p$ associated to the degrees of GLSM fields.
Here, the toric Calabi-Yau 3-folds $\mathcal{X}_a$ have toric diagrams given by the 16 reflexive polygons in \fref{fig_reflexive}, 
and the test symmetry $\eta_h$ is associated to generator $x_h$ of $\mathcal{X}_a$.
The plots are analogous to the ones shown 
in \fref{VminvsFp} to \fref{A2A3vsFp}, 
corresponding to the Futaki invariants of the form $F(\mathcal{X}_a; \zeta_p, \eta_h)$, where $\zeta_R$ is associated to the $U(1)_R$ symmetry.
\\

\begin{figure}[H]
    \centering
    \resizebox{0.9\hsize}{!}{\includegraphics[width=1\linewidth]{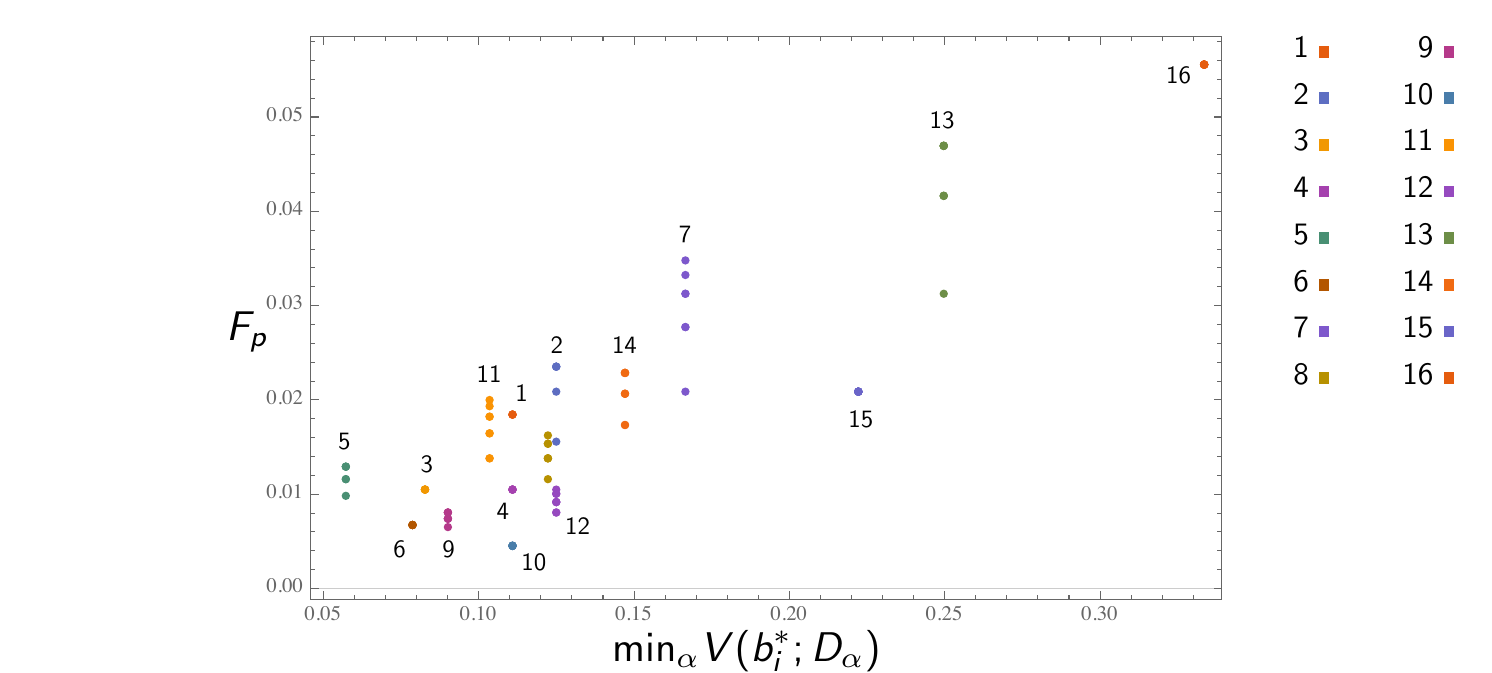}}
    \caption{
    The Futaki invariants $F(\mathcal{X}_a; \zeta_p, \eta_h)$ [$F_p$] against the minimum divisor volume $\min_\alpha V(b^*; \Sigma_\alpha^a)$ for the toric Calabi-Yau 3-folds $\mathcal{X}_a$ corresponding to the 16 reflexive polygons in $\mathbb{Z}^2$.
    }\label{divvolminvsFp}
\end{figure}

\begin{figure}[H]
    \centering
    \resizebox{0.9\hsize}{!}{\includegraphics[width=1\linewidth]{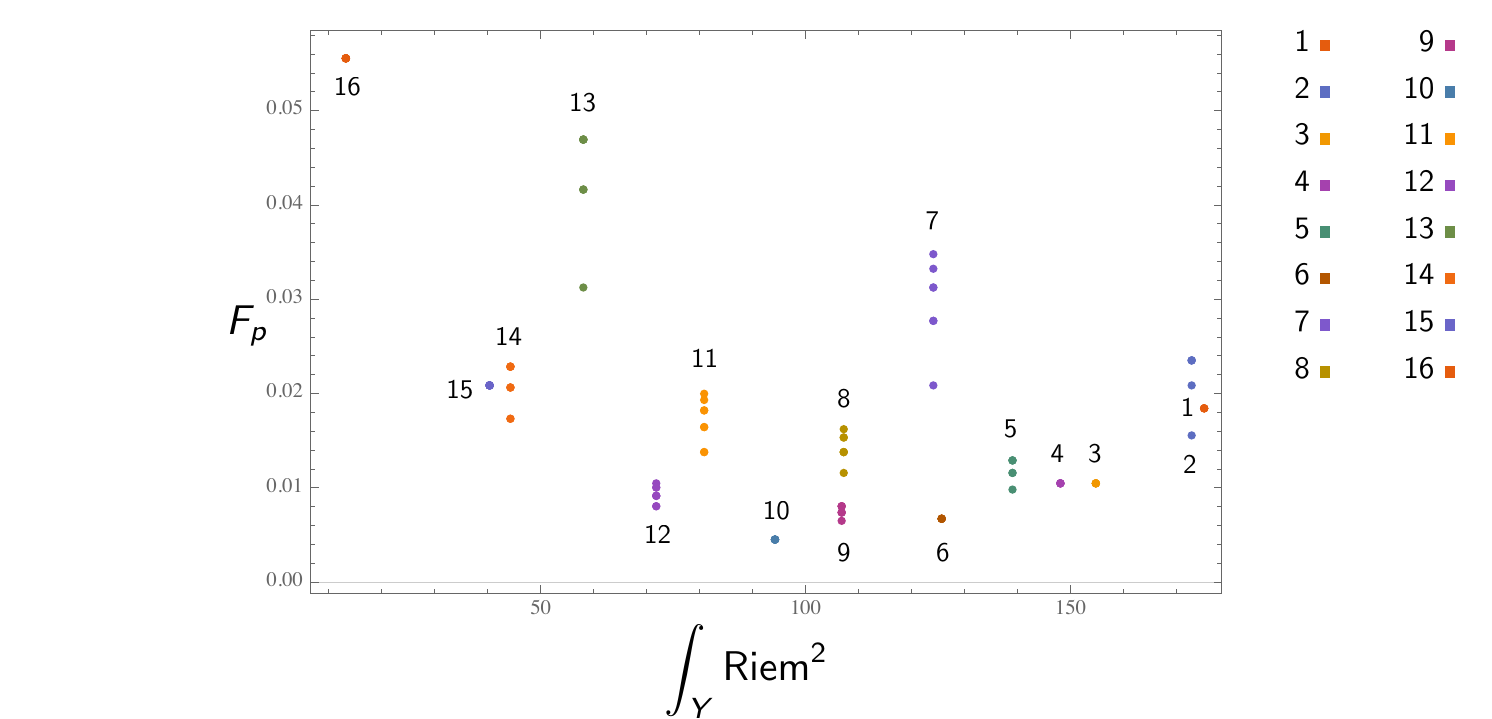}}
    \caption{
    The Futaki invariants $F(\mathcal{X}_a; \zeta_p, \eta_h)$ [$F_p$] against the 
    integrated curvatures $\int_{Y_a} \text{Riem}^2$ for the Sasaki-Einstein 5-manifolds $Y_a$
    corresponding to the 16 reflexive polygons in $\mathbb{Z}^2$.
    }\label{divvolminvsFp}
\end{figure}

In terms of Futaki invariants of the form $F(\mathcal{X}_a; \zeta_p, \eta_h)$, 
where $\zeta_p$ corresponds to the degrees in GLSM fields,
we can see that there are no clear relationships with other geometric and topological invariants associated to $X_a$.
However, given the relationship between 
Futaki invariants of the form $F(\mathcal{X}_a; \zeta_p, \eta_h)$ and $F(\mathcal{X}_a; \zeta_R, \eta_h)$
as studied in section \sref{twoFuts},
we do not completely dismiss the Futaki invariants $F(\mathcal{X}_a; \zeta_p, \eta_h)$ in terms of $\zeta_p$, 
and present them in this section for completeness.
\\

\section{Futaki Invariants and Minimized Volumes for More $\mathbb{C}^3$ Orbifolds}\label{nonreflexive}

In this section, we investigate the behavior of 
Futaki invariants $F(\mathcal{X}_a; \zeta_R, \eta_h)$, where $\zeta_R$ corresponds to the $U(1)_R$ symmetry
for a family of abelian orbifolds of the form
$\mathbb{C}^{3}/(\mathbb{Z}_{n_{1}} \times \mathbb{Z}_{n_{2}})$ with $n_{1} n_{2} = 1, \dots, 12$.
These toric Calabi-Yau 3-folds
have toric diagrams that are not necessarily reflexive.
Investigating the Futaki invariants for this family of toric Calabi-Yau 3-folds
allows us to test whether the bounds identified in section \sref{Vminandtopinv}
on $F(\mathcal{X}_a; \zeta_R, \eta_h)$ associated to the 16 reflexive toric diagrams in \fref{fig_reflexive}
extends beyond these reflexive toric diagrams.
As discussed in section \sref{Vminandtopinv}, 
we investigate here whether the bound on $F(\mathcal{X}_a; \zeta_R, \eta_h)$ in \eref{es04a20} in terms of the minimum volume $V_{min}=V(b^*; Y_a)$ of the Sasaki-Einstein manifolds $Y_a$
still holds for orbifolds of the form $\mathbb{C}^{3}/(\mathbb{Z}_{n_{1}} \times \mathbb{Z}_{n_{2}})$ with $n_{1} n_{2} = 1, \dots, 12$.
\fref{VminvsFROrb}
show the Futaki invariants $F(\mathcal{X}_a; \zeta_R, \eta_h)$ against the minimum volumes $V_{min}=V(b^*; Y_a)$,
where the Sasaki-Einstein manifolds $Y_a$ correspond to the orbifolds of the form $\mathbb{C}^{3}/(\mathbb{Z}_{n_{1}} \times \mathbb{Z}_{n_{2}})$ with $n_{1} n_{2} = 1, \dots, 12$.

\begin{figure}[H]
    \centering
    \resizebox{0.8\hsize}{!}{\includegraphics[width=1\linewidth]{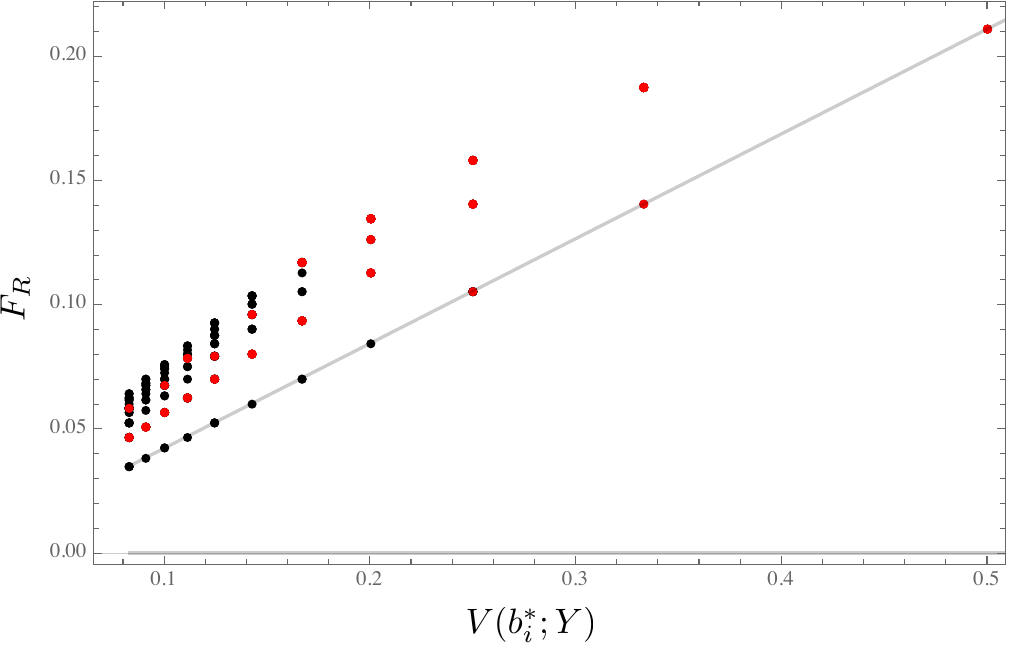}}
    \caption{
    The minimum volume $V_{min} = V(b_i^*; Y_a)$ of the Sasaki-Einstein $5$-manifold $Y_a$ associated to the orbifolds of the form $\mathbb{C}^{3}/(\mathbb{Z}_{n_{1}} \times \mathbb{Z}_{n_{2}})$ with $n_{1} n_{2} = 1, \dots, 12$, plotted against the Futaki invariants $F(\mathcal{X}_a; \zeta_R, \eta_h)$ for all generators $x_h$ corresponding to $\mathbb{C}^{3}/(\mathbb{Z}_{n_{1}} \times \mathbb{Z}_{n_{2}})$.
The red points correspond to all orbifolds of form $\mathbb{{C}}\times\mathbb{C}^{2}/\mathbb{Z}_{n_2}$ while the others are in black.}\label{VminvsFROrb}
\end{figure}

In the family of abelian orbifolds of the form $\mathbb{C}^{3}/(\mathbb{Z}_{n_{1}} \times \mathbb{Z}_{n_{2}})$ with $n_{1} n_{2} = 1, \dots, 12$, 
\fref{VminvsFROrb}
indicates orbifolds of the form 
$\mathbb{{C}}\times\mathbb{C}^{2}/\mathbb{Z}_{n_2}$ with a $\mathbb{C}$ factor with red points. 
Even though these orbifolds with $\mathbb{C}$ factors have clearly toric diagrams that are not reflexive, we can see that their corresponding Futaki invariants 
$F(\mathcal{X}_a; \zeta_R, \eta_h)$ satisfy the bound found in \eref{Vminandtopinv}. 

For orbifolds of the form $\mathbb{{C}}\times\mathbb{C}^{2}/\mathbb{Z}_{n_2}$ with $n_1=1$, 
the corresponding refined Hilbert series can be found to of the following form,
\beal{es90a01}
g(t_\alpha; \mathcal{X}) = \frac{1 - t_2^{n_2} t_3^{n_2}}{(1 - t_1)(1 - t_2 t_3)(1 - t_2^{n_2})(1 - t_3^{n_2})}.
\eea
Using $\zeta=\zeta_R$ to be the $U(1)_R$ symmetry, 
following the calculation in section \sref{Fut}, we can find the Futaki invariants
as, 
\beal{es90a02}
F(\mathcal{X}_a; \zeta_R, \eta_h) = \left(0,\quad \frac{27(n_2-1)}{32n_2^{2}},\quad \frac{27(n_2-1)}{32n_2^{2}},\quad \frac{27}{64n_2}\right)_h
~,~
\eea
where $n_2 = 2, \dots, 12$. 
We can see above that when the toric Calabi-Yau 3-fold has $\mathbb{C}$ factors, the associated generator $x_h$
under test symmetry $\eta_h$
results in a vanishing Futaki invariant
$F(\mathcal{X}_a; \zeta_R, \eta_h)$.
In such a case, 
the corresponding central fibre is isomorphic to the original ring under the test symmetry $\eta_h$, 
and even though the Futaki invariant vanishes it is consistent with the mesonic moduli space being K-stable. 
In Figure \label{VminvsFROrb}, we shall omit these trivial cases, and only plot the Futaki invariants that are non-zero
 $F(\mathcal{X}_a; \zeta_R, \eta_h)>0$. 
\\

\newpage


\bibliographystyle{utphys}
\bibliography{references}

\end{document}